\documentclass[aps,pre,nofootinbib,amsmath,amsfonts,floatfix]{revtex4}

\usepackage{graphicx}
\usepackage{color}

\font\msytw=msbm9 scaled\magstep1

\let\a=\alpha \let\b=\beta  \let\g=\gamma  \let\d=\delta 
  \let\h=\eta   \let\l=\lambda
\let\m=\mu                
\let\s=\sigma \let\t=\tau    
   
\let\G=\Gamma \let\D=\Delta  \let\L=\Lambda 
    \let\Si=\Sigma     
 \let\ee=\epsilon \let\r=\rho 
\let\io=\infty
\def\ie{{i.e. }}\def\eg{{e.g. }}

\def\PP{{\cal P}}\def\EE{{\cal E}}\def\MM{{\cal M}} 
\def\FF{{\cal F}} 
\def\NN{{\cal N}} \def\BB{{\cal B}}
  
\def\DD{{\cal D}}\def\GG{{\cal G}} \def\SS{{\cal S}}
  
\def\ZZ{{\cal Z}}

\def\di{{\partial i}}
\def\dimj{{\partial i \setminus j}}
\def\ij{{\langle i,j\rangle }}

\def\to{\rightarrow}
\def\la{\left\langle}
\def\ra{\right\rangle}

\def\FFF{\hbox{\msytw F}}

\newcommand{\beq}{\begin{equation}}
\newcommand{\eeq}{\end{equation}}

\newcommand{\wt}{\widetilde}

\def\I{\mathbb{I}}

\begin{document}

\title{
Mean field theory of spin glasses
}

\author{Francesco Zamponi}
\affiliation{Laboratoire de Physique Th\'eorique, 
\'Ecole Normale Sup\'erieure,
24 Rue Lhomond, 75231 Paris Cedex 05, France \\
\url{http://www.lpt.ens.fr/~zamponi}
}

\date{\today}

\begin{abstract}
These lecture notes focus on the mean
field theory of spin glasses, with particular emphasis on the presence
of a very large number of metastable states in these systems. 
This phenomenon, and some of its physical consequences, will be discussed in
details for fully-connected models and for models defined on random lattices.
This will be done using the replica and cavity methods.

These notes have been prepared for a course of the PhD
program in Statistical Mechanics at SISSA, Trieste and at the University of Rome ``Sapienza''.
Part of the material is reprinted from other lecture notes, and
when this is done a reference is obviously provided to the original.
I would like to warmly thank 
all the students and colleagues who read these notes, gave
me their feedback and sent me their corrections, that allowed to fix many
errors on the original manuscript.
I also would like to thank SISSA and the University of Rome ``Sapienza'' for
inviting me to give these lectures.
\end{abstract}

\maketitle

\tableofcontents

\clearpage

\section{Introduction}

\subsection{Why study spin glasses?}

Spin glasses have been intensively studied since the seventies \cite{EA75,EA76}. The original motivation
was to describe
a class of magnetic alloys, but
it was realized early on that they are 
representative of a much more general class of disordered systems.
The concept of ``disordered system'' is obviously very generic. Disordered systems in fact display many different
and interesting phenomena whose description might the subject of a collection of monographs.

In these notes
we will limit ourselves to discuss the main physical properties of spin glasses and list
different systems
that share the same properties.
Spin glasses are very interesting for many reasons~\cite{Pa07b}:
\begin{itemize}
\item
Spin glasses are the simplest example of {\it glassy systems}.
Glassy systems are a class of disordered systems that share a common phenomenology,
characterized by a very slow dynamics in the low temperature phase.
In spin glasses, there is a highly non-trivial mean field
approximation where one can study phenomena that have been discovered for the first time in this
context, the most striking one being {\it the existence of many equilibrium states}\footnote{This sentence is too
vague: one should discuss its precise mathematical meaning; although we
will present later a physically reasonable definition, for a careful discussion see Refs. \cite{NS96,Pa98,MPRRZ00,BB04}.}.
\item
The study of spin glasses opens a very important window for studying 
the out-of-equilibrium
behavior of glassy systems, \ie their dynamic evolution when abruptly cooled into the low-temperature
phase starting form high temperature.
In this framework it is possible to derive some of the main properties of generic glassy systems, such as their
history-dependent response \cite{BY86,MPV87,FH91}. This property, in the context of mean field
approximation, is
related to the existence of many equilibrium states.
Aging and the related violations of the equilibrium fluctuation
dissipation relations emerge in a natural way and 
can be studied within this simple setting
\cite{CK93,Cu02,BCKM97}. 
\item
The theoretical concepts and the tools developed in the study of spin glasses are based on two
logically equivalent although very different, methods: the algebraic broken replica symmetry method and the probabilistic cavity
approach \cite{MPV87}. Both 
have a wide domain of applications. Some of the properties that appear in the
mean field approximation, like ultrametricity, are unexpected and counterintuitive. 
\item
Spin glasses also provide a testing ground for a more mathematically inclined probabilistic approach: the
rigorous proof of the correctness of the solution of the mean field model came out after twenty years of
efforts where new ideas, \eg new variational principles~\cite{Gu03},
were at the basis of a recent rigorous proof (see \cite{Ta03} and \cite{FT06} for a concise explanation of the main
ideas) of the correctness of the mean field
approximation in the case of the infinite range Sherrington-Kirkpatrick and $p$-spin models that we will introduce below.
\end{itemize}

\subsection{Physical systems}

Many physical systems have been described using methods and
ideas borrowed from the spin glass physics.
We list some of them below in order to illustrate
the wide variety of physical situations. Details can be found in the references.
\begin{itemize}
\item Real spin glasses: these are typically metallic materials hosting magnetic impurities
located at random positions. The spin polarization of the electrons around the magnetic impurity 
is oscillating at large distance,
\beq
S_{\mathrm{ind}}(r) \sim \frac{\cos 2 k_F r}{r^3} \ ;
\eeq
these are called {\it Friedel oscillations} and are related to the existence of the Fermi surface
(see chapter 2 of \cite{FH91} for more details). Therefore, the coupling between two spins
has random sign and intensity, because the distance between two spins is a random variable. 
The simplest idealization of the interaction between two spins $S_i$ and $S_j$ 
is a coupling term $-J_{ij} S_i \cdot S_j$ in the Hamiltonian, and $J_{ij}$ is taken to be
a random variable.
In these materials the {\it disorder} (\ie the values of the $J_{ij}$) 
is due to the doping with impurities, so in some sense it
is put there ``by hand'' when preparing the sample (it is called {\it quenched disorder} in the literature).
\item Glass-forming liquids~\cite{LN07,Ca09,BB09}: many liquids, when cooled fast enough, freeze in an amorphous state or {\it glass}.
One may think for example to a system of hard spheres, or of point-like particles interacting via a
Lennard-Jones potential.
In a glass, the density profile is not uniform and the particles prefer to stay close to a set of sites that is
not periodic, unlike in a crystal. In the glass 
case, the local environment of each particle is different.
Yet no disorder is 
present in the original Hamiltonian because the interactions between the
particles are deterministic. We may say instead that the disorder is {\it self-generated} by the system.
In addition to hard-spheres and Lennard-Jones systems, many more complicated liquids like molecular liquids
and polymeric liquids display a glass transition.
\item Colloidal dispersions are typically made by mesoscopic particles dispersed in water or other solvents.
The interaction between the particles can be tuned by adding different components to the solutions, and a wide
variety of potentials, ranging from purely hard-core to long range interactions have been created.
These systems usually display, at high enough concentration of particles, a dynamical arrest very similar to
the one observed in glass forming liquids, Some ideas from glass physics have been applied to the study of these systems, mainly
of their dynamics~\cite{DFSTZ01}. Note that, due to the wide variety of potentials that can be engineered, 
in some cases the underlying microscopic phenomenon 
might be very different, leading to different arrested phases such as gels, stripe phases, etc.
\item Quantum glasses: there are many quantum systems that exhibit a glassy behavior. Obviously, one can
consider spin glass materials in which quantum effects are important. But in addition, people have found
an {\it electron (or Coulomb) glass} (see \eg \cite{MI04} and references therein) 
by considering a system of electrons close to the
metal-insulator transition. The observation of an electronic glass phase has also been reported in
high-$T_c$ superconducting cuprates \cite{Ko07}. A particular glass phase (called {\it valence bond 
glass}) is present in a model of hopping electrons on a frustrated lattice \cite{TB08}.
Glass phases are expected also in systems of interacting bosons like cold atoms, 
see \eg \cite{FFI08} and references therein. Recently, the observation of a superfluid-like response in
disordered solid He$^4$ has motivated the study of superfluid glassy ({\it superglass}) phases~\cite{BC08}.
\item Random lasers: one can consider a cavity filled with a disordered system, \eg a solution of mesoscopic
particles in a liquid or glassy matrix with different refractive index. The particles act as a disordered set of
scatterers for light; therefore the modes of the electromagnetic field into
the cavity are disordered. If an amplifying molecule is present in the solution, coherent amplification of the 
modes can be realized and a lasing phenomenon is observed; many modes can be excited at the same time.
The dynamics of the phases of the lasing modes can be described by an equation that closely resemble that of
a spin glass model; this led to the prediction of a glassy phase for the system, manifested in a locking
of the phases of the modes to random position. This phenomenon has been called {\it random mode locking} and is
the disordered version of the standard mode-locking phenomenon that is observed in multimode laser cavities~\cite{ACRZ06}.
\item Granular materials: these are ensemble of macroscopic particles, that, unlike in colloidal systems, are
not dispersed in a solvent. Their mass is so large that the potential energy due to gravity ($\sim m g d$, with $d$
the diameter of one particle) is much bigger than $k_B T$. Therefore in these systems thermal fluctuations are
irrelevant, and their physics is dominated by gravity and frictional forces between the grains. 
A typical example is rice in a silos. In addition, the
system might be agitated by an external force, like for nuts transported in a truck. 
The dynamics of these systems under
external forces is very important for industrial applications.
The configurations of the grains are typically amorphous, and some concepts borrowed from the physics of glasses have
been applied to describe them, see \eg \cite{Gran94}.
\item Biological systems: spin glass models have been used for a long time to describe several different biological systems. 
Probably the most successful example is that of neural networks~\cite{Am89};
in addition, other phenomena such as the folding of proteins have been studied
using these methods~\cite{BW87,TW97,Wo05}. Another important application of spin glass models is the inference
of correlations hidden in biological data~\cite{CLM09,SM09,WWSHH09,MWBC10,HRLR09}. This field is now growing very quickly
and there are many other applications that can be listed here.
\end{itemize}

\subsection{Optimization problems}

In addition to the above long list of physical systems, 
spin glass techniques have been applied to a large class of computer science problems,
called ``optimization problems'': this connection dates back from twenty years at least
\cite{MPV87,Mo07}. In an optimization problem, one looks for a
configuration of parameters minimizing some cost function (the length
of a tour in the traveling salesman problem (TSP), the number of violated
constraints in constrained satisfaction problems, etc.) \cite{PS98}.

As an example, consider a linear system of Boolean equations~\cite{Mo07}: it is given
a set of $N$ Boolean variables $x_i$ with indices $i=1,\ldots,N$. 
Any variable shall be False (F) or True (T). The sum of two variables,
denoted by $+$, corresponds to the logical exclusive OR between these
variables defined through,
\begin{eqnarray} \label{sumrule}
F + T &=& T + F = T \quad , \nonumber \\ 
F + F &=& T + T = F  \quad .
\end{eqnarray}
In the following we shall use an alternative representation of the above 
sum rule. Variables will be equal to 0 or 1 instead of $F$ or $T$, 
respectively. The $+$ operation then corresponds to adding
integer numbers, modulo two. 

A linear equation involving three variables is for example
$x_1 + x_2 + x_3 = 1$.
Four among the $2^3=8$ assignments of $(x_1,x_2,x_3)$ satisfy 
the equation: $(1,0,0)$, $(0,1,0)$, $(0,0,1)$ and $(1,1,1)$.
A Boolean system of equations is a set of Boolean equations
that have to be satisfied together. For instance,
the following Boolean system involving four variables 
\begin{equation} \label{xoreq02}
\left\{ \begin{array} {l}
x_1 + x_2 + x_3 = 1 \\
x_2 + x_4  = 0 \\
x_1 + x_4 = 1 
\end{array} \right.
\end{equation}  
has two solutions: $(x_1,x_2,x_3,x_4)=(1,0,0,0)$ and $(0,1,0,1)$. 
A system with one or more solutions is called satisfiable. 
Determining whether a Boolean system admits 
an assignment of the Boolean variables satisfying all the
equations constitutes the XORSAT 
(exclusive OR Satisfaction) {\it decision} problem.

In spin language the problem can be reformulated as follows.
We associate to each variable $x_i=0,1$ a spin $S_i=(-1)^{x_i}$.
The equation $x_1 + x_2 = a$, where $a=0,1$, can be rewritten
as $J S_1 S_2 = 1$, where $J=(-1)^a$, and similarly for equations
involving more variables\footnote{An equation of length $K$, $x_1 + \cdots + x_K = a$,
is equivalent to $J S_1 \cdots S_K = 1$.}.
The system (\ref{xoreq02})
can be rewritten as
\beq
-S_1 S_2 S_3 + S_2 S_4 - S_1 S_4 = 3 \ .
\eeq
Consider for simplicity a system of $M$ equations, each involving two variables;
it is equivalent to
\beq\label{XORSATH}
H =- \sum_{(i,j)} J_{ij} S_i S_j = -M \ ,
\eeq
where the sum is over all the pairs $(i,j)$ that appear in one of the $M$ 
equations\footnote{For a generic system of $M$ equations labeled by $a=1,\cdots,M$
one has
$H = -\sum_{a=1}^M J_a S_{i_1^a} \cdots S_{i_{K_a}^a} = M$,
where $K_a$ is the length of equation $a$ and $i_1^a, \cdots, i_{K_a}^a$ is
the set of variables belonging to equation $a$.
}.
The XORSAT decision problem is therefore equivalent to the following question:
is the ground state energy of $H$ equal to $-M$ or not?

In the decision problem one is asked to determine whether a given set of constraints can
be satisfied or not. One can also consider the {\it optimization} version of the XORSAT problem, that consists in
finding the ground state of $H$, or in other words of finding the maximum possible number
of equations that can be simultaneously satisfied.
The connection between (zero temperature) statistical mechanics and optimization should be
clear from this example, and as we will see the Hamiltonian (\ref{XORSATH}) is a typical
spin glass Hamiltonian.

In spin glasses the $J_{ij}$ are random variables. For instance, in the example
(\ref{XORSATH}), we can decide that each variable appears in {\it exactly} in $z$ equations;
the total number of equation is then $M= Nz/2$. Consider a graph such that each variable
is a vertex and a link $(i,j)$ correspond to an equation involving $S_i$ and $S_j$.
Then with this choice the model is defined on a graph such that each vertex has exactly
$z$ neighbors. We give equal probability to all graphs satisfying this constraint.
For each equation (link) the corresponding coupling $J_{ij}$ is taken as
a random variable.

In computer science, random distribution of instances, such as the one we introduced above, 
have been used as a benchmark to
test the behavior of {\it search algorithms}, \ie algorithms that
try to find a ground state of $H$ (a solution of the problem).
In physical language, a search algorithm correspond in some cases 
to a dynamical rule that, starting from a configuration of the spins,
attempts to explore the configuration space while looking for the ground
state. A typical example is Monte Carlo dynamics at very low temperature.
The presence of a low-temperature ``glassy'' phase in the model, associated to slow
dynamics, is clearly important for the performances of these algorithms. This is 
an important motivation to study the spin glass phases of such Hamiltonians.

Note that, despite the beautiful studies of the average
properties of the TSP, Graph partitioning, Matching, etc., based on 
spin glass methods \cite{MPV87},
a methodological gap between the field of statistical physics and that
of computer science is far from being bridged. 
In statistical physics statements are usually made on the 
properties of samples that are typical with respect to some 
disorder distribution (\ie distribution of the $J_{ij}$).
In optimization, however, one is interested in solving one (or
several) particular instances of a problem, and needs efficient
ways to do so, that is, requiring a computational effort growing
not too quickly with the number of data defining the instance.
Knowing precisely the typical properties for a given
distribution of instances might not help much to
solve practical cases. Unfortunately, statistical mechanics is for
the moment unable to tell us precise properties for a given sample,
\ie for a given realization of the couplings.

Finally, note that the recent developments in quantum computing triggered
some efforts to study the performances of quantum algorithms to solve
these problems (see \eg~\cite{ST06} for a review). 
Therefore, recently quantum versions of the problems
(in which Ising spins are replaced by Pauli matrices and a transverse field
is added) have been considered~\cite{Fa01}. Understanding the properties of quantum
spin glasses may also 
be important 
in this respect.

\subsection{Models and universality classes}

The simplest spin glass Hamiltonian has 
the form:
\beq H=\sum_{i,k}^{1,N}J_{ik} S_{i}S_{k}\, ,  \label{SIMPLEST}
\eeq
where the $J$'s are {\it quenched} (i.e. time independent) random variables located on the links  connecting
two points of the lattice and
the $S$'s are Ising variables (i.e. they are equal to $\pm 1$).  The total number of points is denoted with $N$ and
it goes to infinity in the thermodynamic limit. We will always assume that $J_{ii}=0$, obviously, and
$J_{ij}=J_{ji}$.

We can consider four models, whose solution is increasingly difficult to obtain~\cite{Pa07b}:
\begin{itemize}
\item
The Sherrington-Kirkpatrick (SK, or {\it fully connected}) model \cite{SK75,MPV87}: 
All $J$'s are random and different from zero, with a Gaussian or a bimodal
distribution with variance $N^{-1/2}$.  The coordination number 
$z = N-1$ 
goes to infinity with
$N$.  In this case a mean field theory is valid in the infinite $N$ limit \cite{MPV87}.
\item
The Bethe lattice model \cite{VB85,cavity,cavity_T0}: The spins live on a random lattice such that each variable has $z$
neighbors, therefore only $Nz/2$ $J$'s are
different from zero: they have finite variance, it is convenient to choose $z^{-1/2}$ in order to have a good limit
$z\to\io$. In this case a modified mean field theory is valid. Note that this model correspond exactly to the XORSAT
problem (\ref{XORSATH}) if the distribution of the $J$ is bimodal (up to a rescaling of $H$).
\item
The large range model \cite{FL03}: The spins belong to a finite dimensional lattice
of dimension $D$.  Only nearest spins at a distance less than $R$ interact and the variance of the $J$'s is
proportional to $1/R^{D/2}$.  If  $R$ is large,   the corrections to mean field theory are small for
thermodynamic quantities. 
They
may, however, change the large distance behavior of the correlations functions and the nature of the phase transition,
which may even disappear.
\item
The Edwards-Anderson ({\it finite dimensional}) model \cite{EA75,EA76}: The spins belong to a finite dimensional lattice of dimension
$D$: only nearest neighbor interactions are different from zero and their variance is $D^{-1/2}$.
In this case finite corrections to mean field theory are present, that are certainly very large in
one or two dimensions, where no transition is expected. The Edwards-Anderson model  corresponds to the limit $R=1$ of 
the large range Edwards-Anderson model; both models are expected to belong to the same universality class. The
large range Edwards-Anderson model provides a systematic way to interpolate between the mean field results and
the short range model.
\end{itemize}

As far as the free energy is concerned, one can prove the following  rigorous results:
\beq\label{limiti}
\begin{split}
&\lim_{z \to \infty}\mbox{Bethe}(z)=\mbox{SK} \ , \\
&\lim_{R \to \infty}\mbox{Large range}(R)=\mbox{SK} \  , \\
&\lim_{D \to \infty}\mbox{Edwards-Anderson}(D)=\mbox{SK} \  , 
\end{split}\eeq

The Sherrington-Kirkpatrick model is thus also a good starting point for studying the finite-dimensional case with short-range interactions, which 
is both the most realistic and the most difficult case to study. This starting
point becomes worse and worse when the dimension decreases; for instance, it is not of any 
use in the limit 
where $D=1$.

In the following we will mostly focus on the mean field theory of spin glasses, which gives the correct solution
of the fully connected (SK) and Bethe lattice models. This theory is very complex and 
has already been the subject of
several books and review papers~\cite{BY86,MPV87,CC05}.
Giving a complete account of the mean field theory of glasses is already a task that goes beyond the aim of these notes.

One of the main results of the theory is the existence of two distinct classes of models displaying a very different
phenomenology:
\begin{enumerate}
\item
The models defined above (except possibly the finite dimensional version)
belong to a class of models called ``full replica symmetry
breaking'' (fRSB). At high temperature they are in a paramagnetic phase akin to that of the ferromagnetic Ising model. 
Upon lowering the temperature, however, they undergo a 
transition to a spin glass phase. For these
models, it is a second-order phase transition, to which are 
associated a diverging correlation
length and power-law singularities controlled by critical exponents.
Even in the
mean field description, 
the low temperature phase is very
complex. The equilibrium states are organized in an intricate 
hierarchical way, and 
the order parameter 
is a function. The mean field theory of fRSB models is reviewed in \cite{MPV87,BY86}.
\item
A class of simpler models exist, where the (many) equilibrium states are organized in a much
simpler way: different states are simply uncorrelated, in a sense that we will be made
precise below. 
These models are called ``one-step replica symmetry
breaking'' models (1RSB). The transition to the spin glass phase, in these models,
is quite different from the fRSB case: although it is still second order from the
thermodynamic point of view, the order parameter jumps at the transition, making it
first order in some sense. In this case, the identification of a diverging correlation
length and associated critical exponents is not evident.
The simplest representative of this class of models
is the spin glass Hamiltonian
\beq
H[S] = \sum_{i,j,k}^{1,N} J_{ijk} S_i S_j S_k \ ,
\eeq
which is called 3-spin glass. Again, we assume that $J_{ijk}$ is zero when two or more
indexes are equal, and that they are symmetric under permutations of the indexes.
More generally one can consider $p>2$ spin
interactions, hence the name $p$-spin glass. As in the previous case, one can
consider the fully connected, Bethe lattice, large range and finite
dimensional versions of this model, and the relations (\ref{limiti}) hold also in
this case for the free energy. The mean field theory of 1RSB models is reviewed
in \cite{CC05}.
\end{enumerate}

The analysis of 1RSB models is fortunately much simpler than that of fRSB ones, and 
their phenomenology is also 
quite interesting and rich. 
Additionally, many
interesting systems like fragile glasses and many optimization problems 
are conjectured to belong to this class.
We will therefore start our analysis by 
studying $p$-spin glasses,
and then describe (shortly) the solution of the more complicated SK model.

\subsection{Frustration and quenched disorder}

There are two common ingredients in all the models and physical systems we discussed
above: {\it disorder} and {\it frustration}. The disorder, in some cases, is built in
the Hamiltonian (the coupling $J_{ij}$ are random); it represents for instance the
random position of the impurities. Clearly, the impurities might diffuse throughout 
the sample, so $J_{ij}$ should formally be considered as dynamical variables. Because the time
scale of this evolution is much larger than any interesting time scale in the glass problem, however, 
we can
consider the $J_{ij}$ as essentially constant \ie {\it quenched}.

When computing the partition function, we thus 
keep the $J$'s fixed,
\beq
Z_J(\b) = \sum_S e^{-\b H[S]} \ ,
\eeq
and from the partition function we can compute observables such as the energy, entropy,
free energy, magnetization, etc. 
It is {\it these} quantities that should then be averaged over
the distribution of the $J$'s. The average free energy must therefore be defined 
by a so-called {\it quenched} average over the disorder:
\beq
f = \overline{f_J} = -T \overline{\frac1N \log Z_J } \ .
\eeq
In this way, the usual thermodynamic identities are satisfied: for instance, the average
entropy is
\beq
s = -\frac{d f}{d T} = \overline{-\frac{d f_J}{d T}} = \overline{s_J} \ .
\eeq
Ideally, we would like to know the properties of the system for each given realization
of the $J$'s, that corresponds to a given physical sample. Fortunately, one can show that
{\it intensive} quantities such as $f$, $s$, etc., are {\it self-averaging}, which 
means
that in the large volume limit they converge with probability one (with respect to the
distribution of the $J$'s) to the average defined above.
As far as such observables
are concerned, the average over the disorder is representative of the behavior of the typical
sample. 
Yet, as discussed above, in some applications (mainly in computer science) one 
would like to know the properties of rare samples corresponding to particular choices of
the $J$'s, or to have bounds that hold for any choice of the $J$'s. Unfortunately, these problems cannot
be tackled using the methods
described here. 

In other cases, the disorder is self-generated by the system, as for glass-forming
liquids. In these cases, clearly, the explicit average over the disorder is not needed.

Another crucial ingredient is {\it frustration}. In the examples above, this is due to the fact
that the $J$'s have random signs. Therefore, a given spin is subject to fields due to their neighbors
that have different signs. Some wants it to point up and others to point down. For this reason
finding the ground state is not trivial and as we will see many degenerate ground states may be present.
Note that this would not happen if all the $J$'s were negative: in this case the ground state, even in presence
of disorder, would simply be a 
configuration where all spins are equal.

\subsection{What is missing in these notes}

It will be impossible to cover all the relevant issues about the complex physics of spin glasses.
In the following, we will try to review some aspects of this problem, by alternating general discussions
with some more technical sections where methods and techniques of general importance will be introduced.

We will focus more on {\it equilibrium} properties of {\it mean field} spin glasses, and
a detailed investigation of the {\it dynamics} of spin glasses will not be done. Still, dynamics is very
important and is probably the most relevant aspect for making contact with experiments.
Excellent reviews
can be consulted by readers who wish to dig 
deeper into this important subject~\cite{Cu02,CC05}.
The notes are divided in two parts: the first is devoted to fully-connected models and the replica method,
while the second to Bethe lattice models and the cavity method. We will not discuss finite dimensional models
since the extension of these concepts to finite dimensional models is still debated;
reviews and further references 
can be found in \cite{Ca09,BB09} for 1RSB models and \cite{MPRRZ00} for fRSB ones.
Some exercises are proposed at the end of each section. 
Ideally they should be done while reading the notes; the appropriate moments are marked in the text
by {\bf $\Rightarrow$ Ex. N.n}.


\clearpage

\section{Fully connected models}
\label{sec:fully}

We said in the introduction that
frustration causes the existence of many thermodynamic
states and that this is the main interesting property of glassy systems
and in particular of spin glasses.
The aim of this section is to make this statement more precise, by looking to the
exact solution of fully connected models: the SK model belongs 
to the fRSB class
and the spherical $p$-spin model belongs 
to the 1RSB class.
Our aim here is to understand the nature of the transition and of the low temperature
phase; to identify the symmetry that is broken (if any) and a correct order parameter;
and to discuss what are the relevant susceptibilities that diverge signaling the transition.

\subsection{Free energy functional}

\subsubsection{The fully connected Ising ferromagnet}
\label{sec:FCIM}

Before turning to the more complicated case of spin glasses, we will here review
very briefly the concept of {\it metastable state} for the familiar Ising ferromagnet.
We will limit ourselves to the fully connected case where the definition is much simpler;
for a general discussion see~\cite{La69}.

Let us then consider the fully connected Ising model, whose Hamiltonian is given by
\beq\label{HFCIM}
H[S] = -\frac{J}{2 N} \sum_{i,j}^{1,N} S_i S_j - \BB \sum_{i=1}^N S_i = -\frac{N J}{2} (m[S])^2 - N \BB m[S]  \  ,
\eeq
where we defined $m[S] = \sum_{i=1}^N S_i /N$ as the magnetization per spin with $S_i=\pm 1$.  
The Hamiltonian $H[S]$, and consequently the Gibbs probability $P[S] \propto \exp -\b H[S]$, depend only on the 
magnetization $m[S]$. 
The total probability that the system has magnetization $m[S]=m$ can thus be written
as the product of the probability of a given configuration with $m[S]=m$ times the number of such configurations;
the latter is a combinatorial factor counting the number of ways one can choose $N_+ = N (1+m)/2$ spins (out of $N$)
to be equal to $+1$. We obtain
\beq\label{PmIFC}
P(m) \propto e^{N [ \b J m^2 /2 + \b \BB m]} \binom{N}{N_+} \sim e^{N [ \b J m^2 /2 + \b \BB m + s_0(m)]} = e^{-\beta N f(m)} \ ,
\eeq
where
\beq\label{S0}
s_0(m) = \lim_{N\to\io} \frac{1}{N} \log \binom{N}{N (1+m)/2} = 
- \frac{1+m}2 \log\frac{1+m}2 -
\frac{1-m}2 \log\frac{1-m}2 \  .
\eeq
The function $f(m)$ defined in Eq.~(\ref{PmIFC}) is the {\it large deviation function} associated to the magnetization $m$.
If plotted as a function of $m$, it has a familiar form: at high temperature it is a convex function with a single minimum in $m=0$, while
below a {\it critical temperature} (in this case $T=1$)
there are two minima at $m=\pm m^*$ and a maximum in $m=0$ and $f(m)$ is no more convex, see figure~\ref{fig:fm}.
In presence of a non-zero external field $\BB$, one of the two minima has a lower $f(m)$ (a higher probability).

The function $f(m)$ is related to the probability, {\it at equilibrium}, to find the system in a configuration with magnetization $m$.
This means that at low temperatures, there is high probability of finding the system with magnetization $\pm m^*$ (one of the two values
will be preferred for $\BB \neq 0$), while there is a low probability of finding an intermediate value of $m$, in particular $m\sim 0$.
In other words, the system spends a lot of time close to configurations with $m = \pm m^*$, and much less time close to configurations with
$m \sim 0$. 
Yet in order to go from $-m^*$ to $+m^*$, the magnetization must cross $m\sim 0$. The number of such transitions
must 
be very small, or  
the probability of $m\sim 0$ would otherwise be large. The only possible solution is
that the system stays for a long time close to
$- m^*$, then performs a fast jump to $m^*$, stays a long time there, then performs a fast jump in the other direction, and so on.

To be more precise we should introduce a model of the dynamics and analyze it in details ({\bf $\Rightarrow$ Ex.\ref{sec:fully}.\ref{ex:1.1}}). 
A nice and detailed discussion of this aspect by R.~Monasson
can be downloaded from {\tt http://www.phys.ens.fr/$\sim$monasson/Appunti/ising.ps}.
It turns out that the system stays close to $\pm m^*$ 
for large time intervals, whose length 
scales
as $\t_\pm \sim \exp N t_\pm$, while the rare jumps between these two states take a time that increases only polynomially
with $N$, $\t_{jump} \sim N^\a$.
In other words, if prepared close to one of the two minima of the free energy, the system remains there with high probability, for a time
that scales exponentially with $N$. This description is true for both minima, and in particular for the one with higher $f(m)$, which makes it
less probable. This minimum
is thus a classic example of a metastable state.

We are led to identify metastable states with the minima of a suitable free energy function $f(m)$.  Before turning to a more general definition,
it is useful
to highlight some of its properties that will be important in the following.
We know that $f(m)$ is an analytic function of
{\it both} $m$ and $\b$. It does not show any singularity at the critical temperature. Yet
we also know that for $\BB=0$ there is a phase
transition at $T=1$. 
The average magnetization is zero above $T=1$ and non-zero below.
The total free energy of the system is indeed given by
\beq
f = -\frac{T}{N} \log Z = -\frac{T}{N} \int dm e^{-\b N f(m)} = \min_m f(m) \ .
\eeq
The singularity of the thermodynamic observables (energy, magnetization, etc.) at the phase transition 
comes {\it from the bifurcation of the minima of $f(m)$}, \ie by the
minimization involved in the computation of $f$, and not by a singularity of $f(m)$ itself. This 
very peculiar property is characteristic of fully connected models.
We will see that in more realistic models the situation is completely different. 
The fact that $f(m)$ is analytic in $\b$ at all $\b$ nonetheless suggests that we can
compute it by a series expansion for small $\b$ (actually, in this
case $\b f$ is just a linear function of $\b \,$!).
We will follow exactly this strategy in the next sections.

\subsubsection{Metastable states in fully connected models}

A general result of statistical mechanics (see \eg \cite{MPV87,Ga00}) states 
that it is always possible to decompose the equilibrium 
probability distribution as a sum 
over {\it pure} states.
In finite dimensional systems, pure states are defined by taking the thermodynamic
limit with a given boundary condition~\cite{Ga00}. If there are many pure states, one can 
select one of them by adding to the system
a small field: the probability distribution of the pure state can be thought as the limit of zero field
of the Gibbs measure in presence of the field.

In a fully-connected system such as the Ising model defined in (\ref{HFCIM}), however, there is
no space notion since all the spins interact with all others. 
There is therefore no boundary, and no 
boundary conditions can be applied to the system. The
pure states can only be selected by using an external field.
In this way we are able to define the Gibbs distribution restricted to one pure
state, $P^\a(S_1,\cdots,S_N)$. We can write the decomposition of the Gibbs measure as
\beq
P(S_1,\cdots,S_N) = \frac{e^{-\b H[S]}}{Z} =  \sum_\a w_\a P^\a(S_1,\cdots,S_N) \ ,
\eeq
where $\a$ is an index labeling the states and $w_\a$ is the weight of
each state, $\sum_\a w_\a = 1$. 
In general, a pure state is specified by $P^\a$, or equivalently by the full set of correlation
functions $\la S_{i_1} \cdots S_{i_n} \ra_\a$.
A very important property of the probability distributions of pure
states is the {\it clustering} property, \ie the fact that connected
correlation vanish at large distance~\cite{Pa98}. Since for a fully connected
model there is no space notion, the clustering property reads simply
\beq
P^\a(S_1,\cdots,S_N) = \prod_{i=1}^N P^\a_i(S_i) \ .
\eeq
In other words, spins are completely decorrelated within one state.
The single-spin probability distribution is specified by the average
magnetization of the spin $S_i$, $m_i^\a=\sum_S S P^\a_i(S)$; in fact
for Ising spins we have
\beq
P^\a_i(S) = \frac{1+m_i^\a S}2 \  .
\eeq
Thus, in a fully connected spin model,
a {\it pure state} $\a$ is completely determined by the set of local
magnetizations $m^\a_i$, $i=1,\cdots,N$. Note that this result is valid
{\it only} for these very special models.

\subsubsection{The general definition of the free energy functional}

In the case of the Ising model the two states are characterized by a uniform
magnetization, $m^\a_i \equiv m^\a$.
In a disordered system, for a given sample (realization of the $J$'s) 
each state is characterized by an amorphous magnetization or density
profile. Therefore a good starting point to compute the properties of these states
is the free energy as a functional of the 
magnetization/density profile\footnote{In field theory it is the generating
  function of the irreducible correlation functions.}~\cite{TAP77,MPV87,DG06}.
It is a standard object in statistical mechanics, but it is useful to review
here its definition and fix some notations.
As we said above, in the general case the magnetization profile is not enough to determine
a state since one should specify all the set of correlation function; however, the knowledge of
$m_i^\a$ is already a good approximation. Therefore the definition of free energy functional
that we will give in the following is an useful concept also for finite dimensional systems.

Consider a system of spins\footnote{For a system of particles replace the
  magnetic field by an external (chemical) potential.}
in which an external magnetic field\footnote{We will use the letter
$b$ to denote external magnetic fields and the letter $h$ to denote internal
fields due
to the other spins of the system.}
 $b_i$ acts on the spin $S_i$; the free
energy is
\beq
-\b F[b] = \log \sum_S e^{-\b H[S] + \b \sum_i b_i S_i} \ .
\eeq
Here and in the following we omit the explicit dependence 
of the free energy on $\b$. Note that $F[b]$ is extensive, \ie proportional to $N$;
we will use capital letters for extensive quantities.
The local magnetization in presence of these fields is
\beq\label{mdib}
m_i[b] = \la S_i \ra_b = -\frac{d}{db_i} F[b] \ ,
\eeq
and the susceptibility is
\beq\label{chiij}
\chi_{ij} = \frac{d m_i}{d b_j} = -\frac{d^2 F[b]}{db_i db_j} = \b \la
(S_i-m_i) (S_j-m_j) \ra_b \ .
\eeq
Note that $\chi_{ij}$ is
a positive matrix\footnote{We can write $\chi_{ij}=\la \d S_i \d S_j \ra$ or in matrix notation
$\chi = \la \d S \d S^T \ra$. Then, for any vector $v$, we have $v^T \chi v =  \la (\d S \cdot v)^2 \ra \geq 0$.
This result holds in particular for the eigenvectors of $\chi$, and therefore the eigenvalues are all positive.
}.

The free energy functional $\G[m]$ is defined as the
Legendre transform of $F[b]$:
\beq\label{Fdim}
-\b \G[m] = -\b \max_{b} \left[ F[b] + \sum_i b_i m_i\right] =
\min_{b} \left[ \log \sum_S e^{-\b H[S] + \b \sum_i b_i (S_i - m_i)} \right] \ ;
\eeq
in this way $b=b[m]$ is a solution of (\ref{mdib}), \ie it is the set of local
fields $b_i[m]$ that are needed to enforce the magnetizations $m_i$.
The maximum condition comes from the fact that the susceptibility (\ref{chiij}) is positive,
hence the second derivative of $F[b]+\sum_i b_i m_i$ is negative.

Define, for fixed $b_i$ and $m_i$, the average
\beq
\la O \ra = \frac{\sum_S O[S] e^{-\b H[S] + \b \sum_i b_i (S_i - m_i)}}
{\sum_S e^{-\b H[S] + \b \sum_i b_i (S_i - m_i)}} \ ;
\eeq
the field $b$ is determined by
the condition that the derivative with respect to $b_i$ of the last expression
in Eq.~(\ref{Fdim}) vanishes. This condition can be written using the above definition as
\beq
\la S_i - m_i \ra = 0 \ . 
\eeq
The solution $b[m]$ is the derivative of $\G[m]$:
\beq\label{bdim}
b_i = \frac{d}{dm_i} \G[m] \ , \hskip2cm \frac{d b_i}{d m_j} =  \frac{d^2
  \G[m]}{dm_i dm_j} = (\chi^{-1})_{ij} \geq 0 \ .
\eeq
and the free energy $F[b]$ is the inverse Legendre transform of $\G[m]$:
\beq
-\b F[b] = -\b  \min_{m}  \left[ \G[m] - \sum_i b_i m_i \right] \ ;
\eeq
the stationarity implies that $m[b]$ is a solution of (\ref{bdim}), and it
must be a minimum since the second derivative of $\G[m]$ is positive. 
This result leads to an important observation: if there are no external fields, $b_i=0$,
the free energy of the system is
\beq
F = -T \log \sum_S e^{-\b H[S]} = \min_{m} \G[m] \ .
\eeq 

\begin{figure}
\includegraphics[width=8cm]{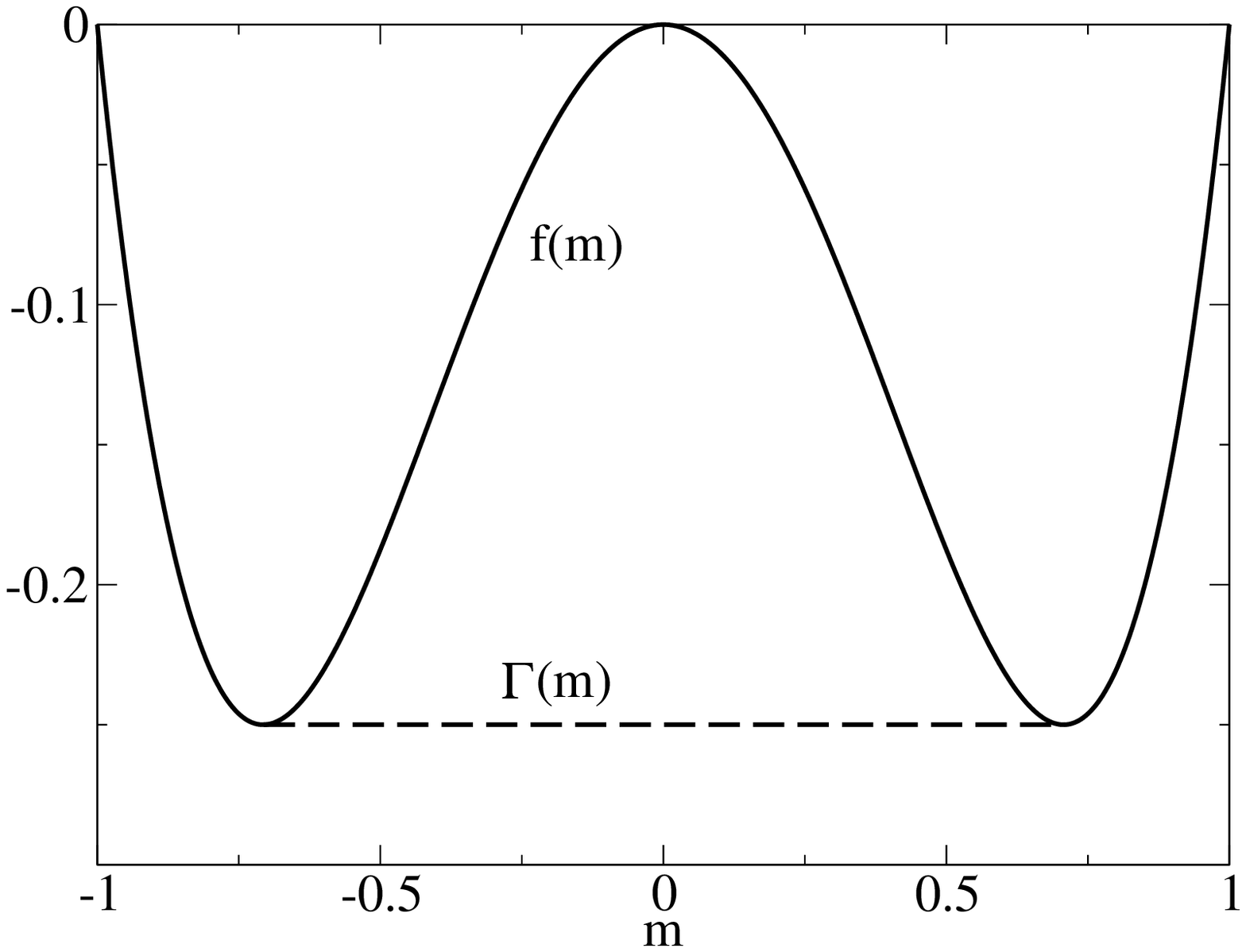}
\caption{The functions $f(m)$ and $\G(m)/N$ for the fully connected Ising ferromagnet
at $\BB=0$ and $T < 1$.}
\label{fig:fm}
\end{figure}

But we still have a problem: $\G[m]$ is a convex function,
hence it cannot have local minima, which is a problem if we
want to use $\G[m]$ to define metastable states. What is wrong?
The problem can be easily understood by computing
$F[b]$ and $\G[m]$ for the fully connected Ising model. We consider a uniform
field, $b_i = b$. Using the
definition of $F[b]$ and the results of section~\ref{sec:FCIM}:
\beq
F(b) = -T \log \int dm e^{-\b N [f(m) - b m]} = N \min_m [f(m) - b m] \ .
\eeq
Hence $F(b)$ is convex even if $f(m)$ is not convex; inverting the Legendre
transform, we obtain that $\G(m)$ is the convex envelope of $f(m)$, see figure~\ref{fig:fm}.

We are therefore interested in $F(m)=N f(m)$ and not really in $\G(m)$. How can we define in general
a non-convex functional $F[m]$, such that its minima are the metastable states?
A way out of this problem is to compute the high temperature expansion of $\G[m]$ defined above. The reason
is that at $\b=0$ there is no interaction and $F[m] = \G[m]$. When expanding around this 
limit, the convexity can be lost if metastable state appear. One can check explicitly that this gives the correct
result for the Ising ferromagnet, which
we will do
at the end of the computation.
Note that for finite dimensional systems, it is not possible to give a general definition of $F[m]$, and
metastable states are more difficult to define. Still, expressions of $F[m]$ based on high temperature or low
density expansions are often used to define metastable states in an approximate way.
In the following, we will denote by $F[m]$ the functional obtained by a high temperature expansion
of $\G[m]$ defined in Eq.~(\ref{Fdim}).

\subsubsection{The Georges-Yedidia expansion}

We will now derive a high temperature-small coupling expansion of this
functional following the strategy of \cite{GY91}; 
we will see that in fully connected models, where individual
couplings vanish for $N\to\io$, the expansion can be
truncated after a finite number of terms, yielding a non-convex free energy functional
$F[m]$ whose minima can be identified with the metastable states of the system. 

To simplify the notation we
define $A^\b[m] = -\b F[m]$
and $\l^\b_i = \b b_i$, and then
\beq
A^\b[m] = \log \sum_S e^{-\b H[S] + \sum_i \l^\b_i (S_i - m_i)} \ ,
\eeq
where as discussed above $\l^\b$ is determined by setting $\la S_i-m_i\ra =0$ (the average is on the measure defining $A^\b$
for fixed $\l^\b$ and $m$)
and thus $\l^\b_i = -\partial_{m_i} A^\b[m]$.
For $\b=0$ we can easily compute $A^0[m]$ because there is no interaction
among the spins. We 
give explicit examples below. 

We wish now to compute
the derivatives of $A^\b[m]$ at
$\b=0$.
By introducing the ``observable''
\beq\label{U}
U[S] = H[S]-\la H \ra - \sum_i \partial_\b \l_i^\b (S_i-m_i) \ 
\eeq
and recalling that $m_i = \la S_i \ra$ for all $\b$, we get
\beq\begin{split}
&\la U \ra = 0  \ , \\
&\frac{d}{d\b} \la O \ra = \la \frac{\partial O}{\partial \b} \ra - \la O U \ra \ , \\
&0=\frac{dm_i}{d\b} =\frac{d\la S_i \ra}{d\b} = -\la S_i U \ra = - \la (S_i - m_i) U \ra \ .
\end{split}
\eeq
We then obtain
\beq\begin{split}
&\frac{d}{d\b} A^\b[m] = \la -H[S] + \sum_i \partial_\b \l_i^\b (S_i-m_i) \ra =
-\la H\ra \ , \\
&\frac{d^2}{d\b^2} A^\b[m] = \la H U \ra = \la U^2 \ra \ , \\
& \frac{d^3}{d\b^3} A^\b[m] = - \la U^3 \ra + \la 2 U \frac{\partial
  U}{\partial \b}\ra = -\la U^3 \ra \ .
\end{split}\eeq
and so on. To compute these derivatives at $\b=0$, we need to know $\partial_\beta
\l^\b_i(\b=0)$ that enters in $U$, and for the higher order derivatives, higher derivatives of $\l^\b$ also appear. 
The derivatives of $\l^\b$ at $\b=0$ can be computed recalling that
$\l^\b_i = -\partial_{m_i} A^\b[m]$, so
\beq
\frac{d^n}{d\b^n} \l^\b_i =  -\frac{\partial}{\partial m_i}\frac{\partial^n A^\b[m]}{ \partial
  \b^n} \ .
\eeq
For instance,
\beq
\begin{split}
&\frac{d}{d\b} \l^\b_i(\b=0) = \frac{d}{dm_i} \la H\ra_0 \ , \\
&\frac{d^2}{d\b^2} \l^\b_i(\b=0) = -\frac{d}{dm_i} \la U^2 \ra_0 \ ,
\end{split}\eeq
and so on.

\subsubsection{Free energy functional for a generic Ising model}

As an example, we consider a model of Ising spins with Hamiltonian
\beq
H[S] = -\frac12 \sum_{i\neq j} J_{ij} S_i S_j - \BB \sum_i S_i \ .
\eeq
First we need to compute the zeroth order term:
\beq\begin{split}
& A^0[m] = \log \sum_S e^{ \sum_i \l^0_i (S_i - m_i)} = \sum_i \left[- \l^0_i m_i + \log
(2 \cosh \l^0_i) \right] \ , \\
& \frac{d A^0[m]}{d\l^0_i} = \la S_i - m_i \ra_0 = \tanh(\l^0_i) - m_i =0 \ .
\end{split}\eeq
Expressing $\l^0_i$ as a function of $m_i$ we get
\beq
A^0[m] = \sum_i s_0(m_i) = -\sum_i \left( \frac{1+m_i}2 \log\frac{1+m_i}2 +
\frac{1-m_i}2 \log\frac{1-m_i}2 \right) \ .
\eeq

Note that at $\b=0$ the spins are uncorrelated,
$\la S_i S_j \ra_0 =
m_i m_j$, $\la S_i S_j S_k \ra_0 = m_i m_j m_k$, and so on, which
allows us to compute
\beq\begin{split}
&\left. \frac{d}{d\b} A^\b[m] \right|_{\b=0} = -\la H \ra_0 = \frac12
\sum_{i\neq j} J_{ij} m_i m_j + \BB\sum_i m_i \ , \\
&\left.\frac{d}{d\b} \l^\b_i \right|_{\b=0} = \frac{d}{dm_i} \la H\ra_0 = -\sum_{j(\neq i)}
J_{ij} m_j - \BB \ .
\end{split}\eeq

Plugging the last equation into (\ref{U}) we obtain
\beq
U_0 = -\frac12 \sum_{i \neq j} J_{ij} (S_i -m_i)(S_j-m_j) \ ,
\eeq
which allows us to compute the second and third derivatives of $A^\b$.
The result is
\beq\begin{split}
&\left.\frac{d^2}{d\b^2} A^\b[m]\right|_{\b=0} = \la U_0^2 \ra_0 = \frac12
\sum_{i\neq j} J_{ij}^2 (1-m_i^2)(1-m_j^2) \ , \\
& \left.\frac{d^2}{d\b^2} \l^\b_i \right|_{\b=0} = 2 m_i \sum_{j(\neq i)}
J_{ij}^2 (1-m_j^2) \ , \\
&\left.\frac{d^3}{d\b^3} A^\b[m]\right|_{\b=0} = -\la U_0^3 \ra_0 = 2
\sum_{i\neq j} J_{ij}^3 m_i (1-m_i^2) m_j (1-m_j^2) 
+\sum_{i\neq j \neq k} J_{ij} J_{ik} J_{jk} (1-m_i^2) (1-m_j^2) (1-m_k^2) \ . 
\end{split}\eeq

Collecting these results, and going back to the original notation, we obtain
the final result
\beq\begin{split}
-\b F[m] &=-\sum_i \left( \frac{1+m_i}2 \log\frac{1+m_i}2 +
\frac{1-m_i}2 \log\frac{1-m_i}2 \right) + \b \frac12
\sum_{i\neq j} J_{ij} m_i m_j + \b \BB\sum_i m_i \\
&+ \frac{\b^2}4
\sum_{i\neq j} J_{ij}^2 (1-m_i^2)(1-m_j^2)
+ \frac{\b^3}6 \left[ 2
\sum_{i\neq j} J_{ij}^3 m_i (1-m_i^2) m_j (1-m_j^2) \right. \\ 
&+ \left. \sum_{i\neq j \neq k} J_{ij} J_{ik} J_{jk} (1-m_i^2) (1-m_j^2) (1-m_k^2)
\right] + O(\b^4) \ , \\
\b b_i[m] &= \text{atanh}(m_i) - \b \left[ \sum_{j(\neq i)}
J_{ij} m_j + \BB \right] + \b^2 m_i  \sum_{j(\neq i)}
J_{ij}^2 (1-m_j^2) + O(\b^3) \ .
\end{split}\eeq
In absence of external fields $b_i=0$, the expressions simplify to the so-called Thouless-Anderson-Palmer (TAP) equations,
\beq\label{TAP_SK}
m_i = \tanh \b h_i \ , \hskip2cm h_i = \BB + \sum_{j(\neq i)}
J_{ij} m_j - \b m_i  \sum_{j(\neq i)}
J_{ij}^2 (1-m_j^2) + O(\b^2) \ ,
\eeq
where $h_i$ is the effective magnetic field provided by the neighboring spins at site $i$.

\subsubsection{Back to the fully connected ferromagnet}

Before going to the more complicated SK model, it is instructive to
look at the fully connected ferromagnet, where
$J_{ij}=\frac{1}{N}$ for all $ij$. In this case the spins are all equivalent, so
we expect that at the free energy minimum the magnetizations
are all equal, $m_i = m$. 
Then it is easy to see that the terms of order $\b^2$ and $\b^3$ in the free energy expansion vanish for 
$N\to\io$. 

Keeping only the $O(\b)$ term, the TAP equations simplify to
the familiar mean-field result
\beq
m = \tanh[\b (\BB+ m)] \ ,
\eeq
and the free energy is
\beq\label{ferro}
f(m) = N^{-1} F[m] = -T s_0(m) - \frac12 m^2 - \BB m \ ,
\eeq
which is the correct result, as we anticipated in Eq.~(\ref{PmIFC}).
At this point it should be clear that $f(m)$ is not convex because
we obtained it by neglecting higher order terms in $\b$. The true function $\G(m)$ is convex
but non-analytic at low temperature (see Figure~\ref{fig:fm}), 
so we cannot obtain it from
a high-temperature expansion.
This example suggests that the truncation of the high-temperature expansion has the effect
of making the existence of {\it metastable states} manifest in $F[m]$.
Note that obviously only the local minima of the TAP free energy, \ie the
solutions of the TAP equations with positive Hessian $\frac{d^2 F}{dm_i
  dm_j}$, can be interpreted as metastable states.

\subsubsection{TAP equations for the SK model}

In the SK model we set $\BB=0$ and the $J_{ij}$ are Gaussian random variables with zero mean and
variance $\overline{J_{ij}^2} = \frac{1}{N}$. The $J_{ij}$ are thus typically
of order of $1/\sqrt{N}$, so the terms $O(\b^2)$ in the TAP free energy
are now relevant. The terms $O(\b^3)$ can, however, still be neglected because they vanish
for $N\to\io$.

Note that the term $\sum_{i\neq j} J_{ij} m_i m_j$
is a sum of a large number of terms; the signs of $J_{ij}$ and $m_i$ are
random, but we expect the sign of $m_i$ to be correlated with the sign of
$h_i = \sum_{j(\neq i)} J_{ij} m_j$. This last quantity is the sum of $N$ terms,
each of order $1/\sqrt{N}$, and is therefore finite for large $N$. As the sign
of $m_i$ and $h_i$ are correlated, we expect $\sum_{i\neq j} J_{ij} m_i m_j = \sum_i m_i
h_i \sim N$.

Conversely, the term $O(\b^2)$ in the free energy is a sum over a large number
of terms, all of them positive. In this case fluctuations are therefore
less important, so
we can replace, for large $N$, 
$J_{ij}^2$ with its average value.
Defining $q=\frac{1}N \sum_i m_i^2$, we get
\beq\label{TAPSK}
\begin{split}
-\b F[m] &= \sum_i s_0(m_i) + \b \frac12 \sum_{i\neq j} J_{ij} m_i m_j +
N \frac{\b^2}4 (1-q)^2 \ , \\
h_i &= \sum_{j(\neq i)}
J_{ij} m_j + \b m_i (1-q)  \ 
\end{split}\eeq
the TAP equations for the SK model.

At high temperatures these equations have only the paramagnetic
solution $m_i=0$.
We can study the stability of this solution upon lowering the temperature.
The stability matrix for the paramagnet is obtained from (\ref{TAPSK}):
\beq
\left. \frac{d^2 F[m]}{dm_i dm_j} \right|_{m_i=0} = (\b + \b^{-1})\d_{ij} - J_{ij} \ ;
\eeq
the stability condition is that all the eigenvalues should be positive.
The spectrum of the matrix $J_{ij}$ is known to be the Wigner semicircle defined 
in the interval $[-2,2]$. For $T>1$, one has $\b+\b^{-1} > 2$, and the paramagnet is stable.
At $T=1$, however, the spectrum touches zero, hence zero modes appear suggesting
that below $T=1$ the paramagnet becomes unstable. 

Yet for $T<1$, again $\b+\b^{-1} > 2$, so it seems that the paramagnet is stable
for all temperatures. This strange result is in fact incorrect: the paramagnet is indeed
unstable at low temperatures. A
clear signature of this fact is obtained by considering its total free energy,
$f_\mathrm{para} = F[m=0]/N = - \b /4 - T \log 2$, and computing from it
the entropy,
$s_\mathrm{para} = -df_\mathrm{para}/dT = \log 2 - \b^2 / 4$. 
This last quantity becomes negative for 
$T \leq 1/(2 \sqrt{\log 2}) \sim 0.911$. This behavior is nonsensical because
we are dealing with Ising spins,
the states of the system are discrete and the entropy must be positive.
The paramagnet actually becomes unstable
at $T=1$, but
it is missed by the TAP equations (\ref{TAPSK}) because of the approximations
we made. There are different ways to understand this. For instance, one can look at the 
leading terms in the small $\b$ expansion; the resummation of these terms is divergent
when $1-\b^2 J^2 (1-q)^2 > 0$, which shows that the TAP equations (\ref{TAPSK}) do not
make sense for the paramagnet ($q=0$) at $T<1$~\cite{DG06}.
Alternatively, one can derive the same condition using 
the cavity method that we will discuss in the following \cite{Pl02}.
This fact points out that
the approximations we made in neglecting higher order terms and substituting 
others with their average are not completely harmless. In fact, while they are correct
for stable states, they are not for unstable states and doing them blindly might 
stabilize solutions that are otherwise unstable. 

What happens, then, below $T=1$?
At low temperatures the TAP equations have many solutions with $m_i \neq 0$ that we
would like to interpret as (stable or metastable) thermodynamic states.
The solution of
the SK model is, however, rather complex, so we
first investigate a much simpler model, the spherical $p$-spin model.

\subsubsection{Spherical $p$-spin model}

We will now compute the TAP free energy for the spherical $p$-spin model~\cite{CS92}, which
we will study in details.
In order to obain the $p$-spin model, we
replace the Ising spins by real continuous variables $\s_i$, and include
the constraint
$\sum_i \s_i^2 = N$, \ie the spins live on the $N$-dimensional sphere of radius $\sqrt{N}$. For
this reason this is called a {\it spherical model}. 
Although this simplification is very convenient for analytical calculations, it
is not useful for the SK model because the spherical version of the SK model
is simply equivalent to a ferromagnet~\cite{CD95}.
The $p$-spin model is therefore defined by the Hamiltonian
\beq\label{Hpspin}
H[\s] = -\frac{1}{p!} \sum_{i_1 \cdots i_p} J_{i_1 \cdots i_p} \s_{i_1} \cdots
\s_{i_p} =- \sum_{i_1 < i_2 < \cdots < i_p} J_{i_1 \cdots i_p} \s_{i_1} \cdots
\s_{i_p}  \ ,
\eeq
where the coupling constants $J$ are again Gaussian random variables with zero mean and average
\beq
\overline{J_{i_1 \cdots i_p}^2} = \frac{p!}{2 N^{p-1}} \ .
\eeq

Using the integral representation of the delta function, the zeroth order term is given by
\beq\label{eq:delta}\begin{split}
e^{A^0[m]} & = \int d\s \d\left(\sum_i \s_i^2- N\right) e^{\sum_i \l^0_i (\s_i-m_i)} =
\int_{-i \io}^{i\io} \frac{d\mu}{2\pi} \int d\s e^{-\mu \sum_i \s^2_i + \mu N + \sum_i \l^0_i
  (\s_i-m_i)} \\
& = \int_{-i \io}^{i\io}  \frac{d\mu}{2\pi} \exp \left[N \mu + \frac{N}2 \log\left(\frac{\pi}{\mu}\right)
+\frac{1}{4\mu}\sum_i (\l_i^0)^2 - \sum_i \l^0_i m_i \right] \ .
\end{split}\eeq
For large $N$ we can evaluate the integral via a saddle point, and
the
stationarity condition for $\l_i^0$ gives $\l_i^0 = 2\mu m_i$, so
\beq
A^0[m] = N \text{st}_{\mu} 
\left[ \mu (1-q) + \frac{1}2 \log\left(\frac{\pi}{\mu}\right) \right] \ .
\eeq
The stationarity condition for $\mu$ further gives $\mu = \frac{1}{2(1-q)}$, and finally
\beq
A^0[m] = N \frac12 \log (1-q) \ ,
\eeq
up to an irrelevant constant. By a similar computation one can show that the
spins are uncorrelated up to $1/N$ corrections, \ie $\la \s_i \s_j \ra = m_i m_j$.
Note also that $\la \s_i^2 \ra = 1$ due to the spherical constraint.
The $O(\b)$ term is then simply $\frac{1}{p!} \sum_{i_1 \cdots i_p} J_{i_1 \cdots i_p} m_{i_1} \cdots
m_{i_p}$. The operator $U_0$ is therefore
\beq
U_0 = -\frac{1}{p!} \sum_{i_1 \cdots i_p} J_{i_1 \cdots i_p} [ \s_{i_1} \cdots
\s_{i_p}- m_{i_1} \cdots m_{i_p} - p (\s_{i_1} - m_{i_1}) m_{i_2} \cdots m_{i_p} ] \ .
\eeq

To compute the $O(\b^2)$, we assume as in the SK case that we can replace
$J^2$ by its average. In computing the average of $U_0^2$ we must therefore only keep  the terms with the
same coupling, \ie such that the indices $i_1 \cdots i_p$ are equal up to a
permutation. Otherwise, the two $J$ have random sign and the contribution is
of subleading order for large $N$. 
It is also useful to recall the relation $\la U (\s_i - m_i) \ra = 0$.
We then get
\beq
\left.\frac{d^2}{d\b^2} A^\b[m]\right|_{\b=0} = \la U_0^2 \ra_0 =
\frac{N}2 [ 1 - p q^{p-1} + q^p (p-1) ] \ ,
\eeq
and the TAP free energy is~\cite{CC05}
\beq\label{TAPpspin}
f[m] = \frac{1}N F[m] = -\frac{1}{2\b} \log(1-q) - 
\frac{1}{p! N} \sum_{i_1 \cdots i_p} J_{i_1 \cdots i_p} m_{i_1} \cdots
m_{i_p} - \frac{\b}4 [ 1 - p q^{p-1} + q^p (p-1) ] \ .
\eeq

\subsubsection{Summary and remarks}

In this section we computed the free energy functional for some spin glass models that we
will discuss in the following sections. It is thus useful to summarize some important remarks that emerged
during the discussion: 
\begin{enumerate}
\item In performing the high-temperature expansion, we did not really define $F[m]$
by taking the maximum as in (\ref{Fdim}). Instead, we continued the $\b=0$ solution.
This is not completely correct since the convexity of $F[m]$ is lost this way.
\item The ``advantage'' is that the local minima of $F[m]$ can be considered
as {\it metastable states}, as we discussed in the case of the ferromagnet in external
field. We will go back to this issue in the next section.
\item The approximations made in deriving $F[m]$ for disordered models can have important
effects on the stability of the solutions, in particular they can stabilize solutions 
(\eg the paramagnet) that are otherwise unstable.
\end{enumerate}
Given these remarks, we now turn to the analysis of the solution of the TAP equations
for the simplest case of the spherical $p$-spin.

\subsection{Metastable states and complexity}

\subsubsection{The simplest example: the spherical $p$-spin glass model}

In the last section we defined the spherical $p$-spin model~\cite{CS92}.
It is the simplest
spin glass for reasons that will soon be clear, and is thus a good
starting point to understand the physics of spin glasses~\cite{CC05}.

As we discussed above, the Gibbs measure can be decomposed in a set
of pure states, that in fully connected models are
completely determined by the set of local
magnetizations $m^\a_i$, $i=1,\cdots,N$.
Note that the same holds for spherical spins because the distribution of a single
spin is Gaussian with $\la \s_i^2 \ra =1$ due to the spherical
constraint, and therefore the only free parameter is the average $m_i^\a$.
The local magnetizations of pure states are the minima of
the TAP free energy functional $F[m]$ in Eq.~(\ref{TAPpspin})~\cite{MPV87,TAP77}. 
The weight $w_\a$ of state $\a$ is proportional to $\exp [-\b N f_\a]$, where
$f_\a = F[m_i^\a]/N$. In general the TAP free energy $F[m_i]$ depends explicitly
on temperature, so the whole structure of the states may also depend strongly
on temperature.

We derived the expression (\ref{TAPpspin}) 
of the TAP free energy for the fully connected $p$-spin models, 
from which the distribution of states
can be computed. Here we will not perform the full computation, but it
can be found
in~\cite{CC05}. We will instead explain the result, and
in the next section present
a simpler method to obtain the same result.

It is convenient to start the description from very low temperature.
In fact, a peculiar property of the spherical $p$-spin model --that
greatly simplifies
the description of the results of the TAP computation-- is that
the dependence of the free energy functional on $T$ is very simple.
The spin glass states (those with $m_i \neq 0$) are labeled by their intensive energy $e$ at $T=0$. The number of
states of energy $e$ is $\Omega(e)=\exp N\Si_0(e)$, where the function $\Si_0(e)$ is called 
{\it complexity}. It is a concave function that vanishes continuously at the 
ground state energy $e_\mathrm{min}$ and goes discontinuously to $0$ above some value 
$e_\mathrm{th}$. 

\begin{figure}[t]
\centering
\includegraphics[width=.6\textwidth]{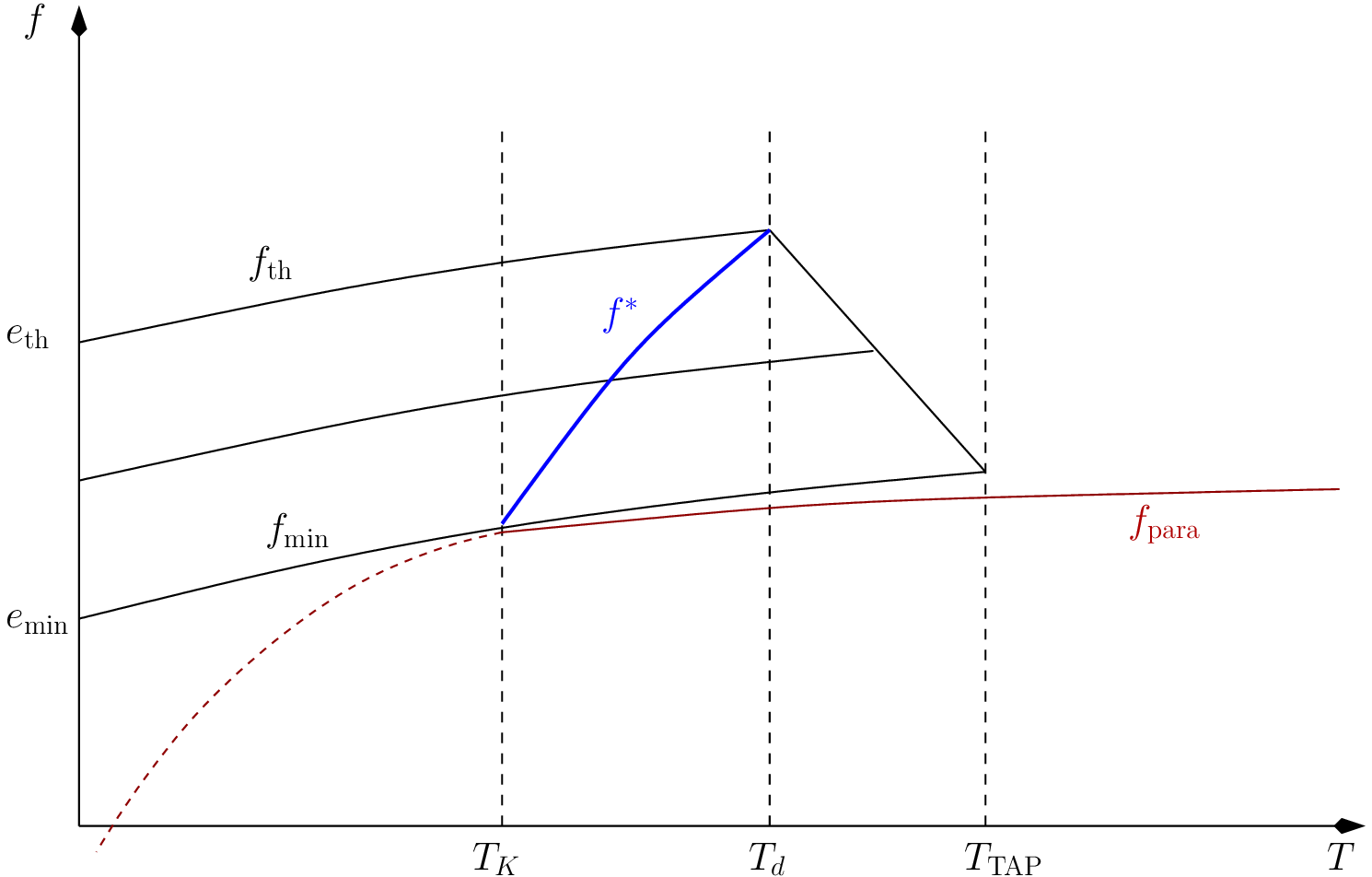}
\caption[Evolution of the TAP states of the spherical $p$-spin model]
{
Sketch of the evolution in temperature\footnote{Note 
that in the spherical $p$-spin model, because of the continuous variables,
the entropy is {\it negative}, hence $df/dT = -s >0$ and the free energy increases
with temperature; in most models (and in all discrete models) instead the entropy is positive
and the free energy decreases with temperature. This fact does not change the qualitative
shape of the figure.} 
of the TAP
states for the spherical $p$-spin model~\cite{BCKM97}. 
Each group of TAP states of 
free energy $f$ can be
followed in temperature until it becomes unstable and disappears. The
complexity vanishes continuously at the ground state $f_\mathrm{min}$ 
and goes abruptly to 0 above the maximum free energy $f_\mathrm{th}$.
The blue line is the free energy $f^*$ of the states that dominate the partition function. 
Between $T_d$ and $T_K$, the red line represents the equilibrium free energy $f^*-T\Si(f^*)$ that takes into account 
the entropic contribution of the degeneracy of the states and is equal to the free energy
of the paramagnet.
}
\label{fig1:TAP}
\end{figure}

At finite temperatures, the minima are ``dressed'' by thermal 
fluctuations but they maintain their identity and one can follow their
evolution at $T>0$. At some temperature $T_\mathrm{max}(e)$, thermal fluctuations
are so large that the states with energy $e$ become unstable and disappear,
until, at high enough temperature $T>T_{\rm TAP}$, the last states (the ones with $f=f_{\rm min}$) disappear.
The temperature evolution of the states is sketched in Fig.~\ref{fig1:TAP}.
At finite temperature, the number of spin glass states of a given free energy
density $f$ is $\Omega(f)=\sum_\a \d(f -f_\a) =\exp N \Si(f)$, where $\Si(f)=\Si_0(e(f))$ and $e(f)$ 
is the $T=0$ energy of the states of free energy $f$. Like its zero-temperature counterpart, the function $\Si(f)$
vanishes continuously at $f=f_\mathrm{min}$ and drops to zero above $f=f_\mathrm{th}$.
A qualitative plot of $\Si(f)$ is reported in Fig.~\ref{fig1:Scqualit}.

In addition to the spin glass states with $m_i \neq 0$, the TAP equations always admit a paramagnetic
solution with $m_i\equiv 0$. However, the paramagnetic state can be considered a pure state only for
$T > T_d$. In fact, if one studies the equilibrium dynamics inside the paramagnetic state~\cite{CC05},
one realizes that spin-spin dynamical correlation functions decay to their long-time value only for $T>T_d$.
When $T<T_d$, the correlation remain larger than its equilibrium value even for infinitely long times. This
means that starting from some equilibrium configuration in the paramagnetic state, the system cannot
explore the whole paramagnetic state, which is therefore not a well defined ergodic component.
Unfortunately for the moment one can realize this problem only through a dynamic calculation. One would
like to see that $m_i=0$ is not a pure state directly from the TAP computation, but how to do this is not clear
for the moment. The dynamical TAP approach of~\cite{Bi99} could help to clarify this issue.

In summary, $p$-spin models are characterized by the existence of a paramagnetic state for $T>T_d$, and
by an {\it exponential number} of spin glass states for low 
temperatures $T<T_{\rm TAP}$. The transition between the paramagnetic and the spin glass regime happens
through a series of very interesting phase transitions that we now discuss.

\subsubsection{The partition function}

The partition function $Z_\a = e^{-\b N f_\a}$ of a pure state $\a$ can be thought as the contribution of this
state to the total partition function. Therefore,
we can write the total contribution of the spin glass states to the partition function $Z$ 
in the following way:
\beq
\label{Zm1}
\begin{split}
Z & = e^{-\b N f_\mathrm{tot}(T)} \sim \sum_\a e^{-\b N f_\a} = \int df \sum_\a \d(f -f_\a) e^{-\b N f} \\
& = \int df \Omega(f) e^{-\b N f}
= \int_{f_\mathrm{min}}^{f_\mathrm{th}}df \, e^{N [\Si(f)-\b f]}
\sim e^{N [\Si(f^*)-\b f^*]} \ ,
\end{split}
\eeq
where $f^* \in [f_\mathrm{min},f_\mathrm{th}]$ is such that $f - T \Si(f)$ 
is minimum, \ie it is the solution of
\beq\label{saddleSi}
\frac{d\Si}{df} = \frac{1}{T} \ ,
\eeq
provided that it belongs to the interval $[f_\mathrm{min},f_\mathrm{th}]$.
Starting from high temperatures, one encounters three distinct temperature regions:
\begin{itemize}
\item For $T > T_d$, the free energy density of the paramagnetic state is
smaller than $f - T\Si(f)$ for any $f\in [f_\mathrm{min},f_\mathrm{th}]$, so the paramagnetic
state dominates and coincides with the Gibbs state. Spin glass states exist for $T<T_{\rm TAP}$ but 
they are irrelevant as their contribution to the total partition function is exponentially smaller than the one
of the paramagnetic state.
\item For $T_d\geq T \geq T_K$, a value $f^* \in [f_\mathrm{min},f_\mathrm{th}]$ is found, such that
$f^* - T \Si(f^*)$ is exactly equal to $f_\mathrm{para}$. In other words, the total free energy of all spin
glass states coincides with the analytic continuation of the free energy of the paramagnetic state below $T_d$.
This can be interpreted as follows. The paramagnetic state
is obtained from the superposition of an
{\it exponential number} of spin glass states of {\it higher} individual free energy
density $f^*$. The Gibbs measure is split on this exponential number of
contributions: however, no phase transition happens at $T_d$ because of the
equality $f^* - T \Si(f^*)=f_\mathrm{para}$ which guarantees that the free energy is
analytic on crossing~$T_d$. 
\item For $T < T_K$, the partition function is dominated by the lowest free 
energy states, $f^* = f_\mathrm{min}$, with $\Si(f_\mathrm{min})=0$ and 
$f_\mathrm{tot}(T)=f_\mathrm{min} - T \Si(f_\mathrm{min}) = f_\mathrm{min}$. At $T_K$ a phase transition occurs;
the free energy and its first derivatives are continuous but the second derivative
of $f_\mathrm{tot}$ with respect to $T$ (the specific heat) has a jump.
\end{itemize}
As we already discussed, the paramagnetic solution $m_i=0$ disappears for $T<T_d$. 
In the range of temperatures $T_d > T > T_K$, the paramagnetic state is replaced by a strange state, in
which the phase space of the model
is disconnected in an exponentially large number of states, giving a contribution
$\Si(T) \equiv \Si(f^*(T))$ to the total entropy of the system.
The entropy $s(T)$ for $T_d > T > T_K$ can thus be written as
\beq
s(T) = \Si(T) + s_\mathrm{vib}(T) \ ,
\eeq
where $s_\mathrm{vib}(T)$ is the individual entropy of a state of free energy $f^*$. 
The phase transition at $T_K$ is signaled by the vanishing of the contribution $\Si(T)$.
A similar ``entropy crisis'' scenario
is realized in a very simple
completely solvable model, the so-called Random Energy Model ({\bf $\Rightarrow$ Ex.\ref{sec:fully}.\ref{ex:1.2}}).


\subsubsection{A method to compute the complexity}
\label{sec:realreplica}

We presented the above picture as the result of a TAP computation that we did not describe explicitly.
How can we compute the properties of the metastable states, for example
the density of states $\Omega(f)$, without solving explicitly the TAP equations?
For systems that present a structure of the free energy landscape similar
to $p$-spin glasses, a general method to compute the complexity as a
function of the free energy of the states has been proposed in~\cite{Mo95}. 
The aim of this section is to present this method
in some details.

\begin{figure}[t]
\centering
\includegraphics[width=7cm,angle=0]{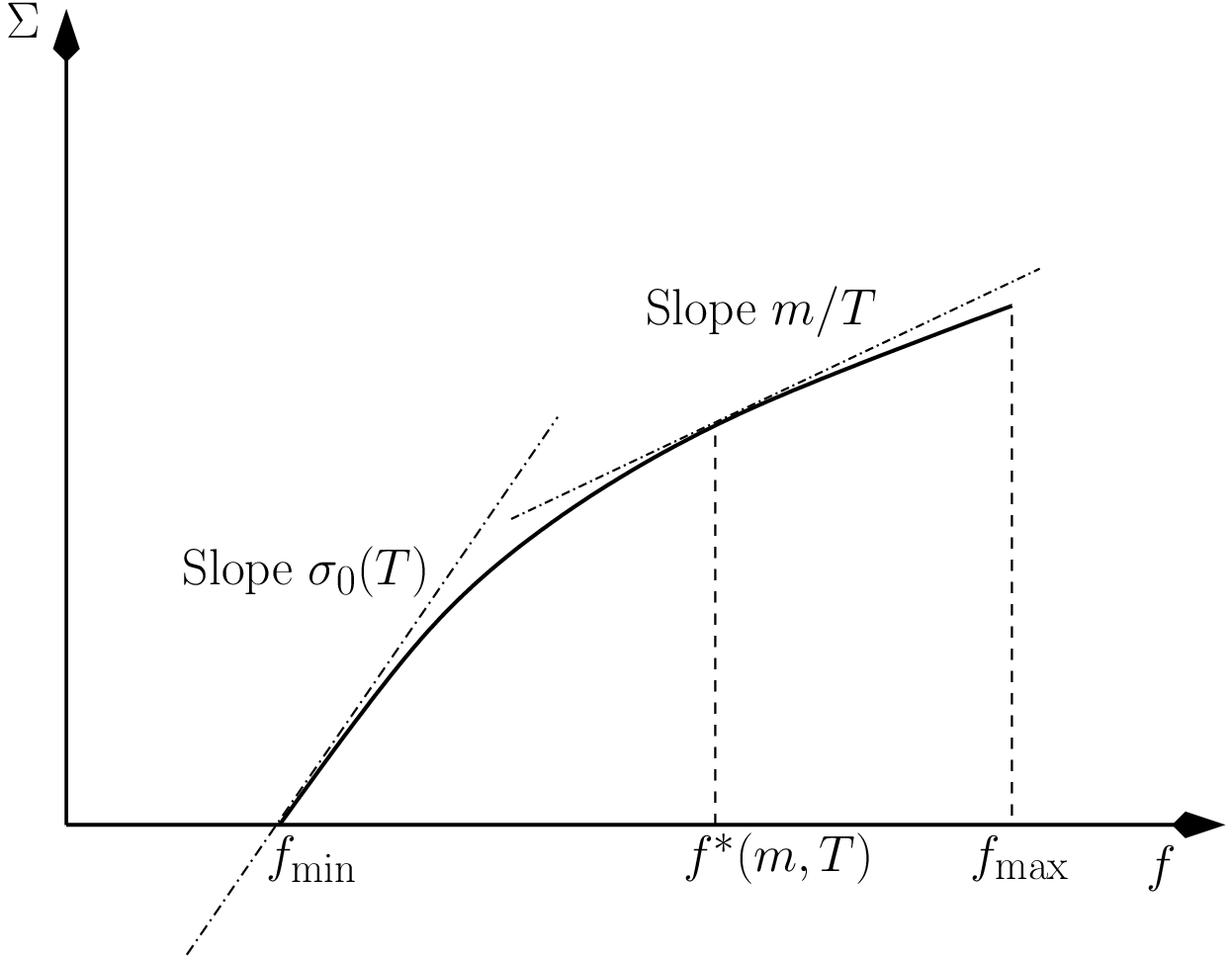}
\caption[Qualitative behavior of the complexity]
{A sketch of the complexity as a function of the
free energy density for systems belonging to the $p$-spin class. The value
$f^*(m,T)$, solution of $\frac{d\Si}{df}=\frac{m}{T}$, is also reported.
}
\label{fig1:Scqualit}
\end{figure}

The main problem we have to face is that, unlike in a ferromagnet, the
states cannot be classified according to symmetry. In fact, the local
magnetizations $m_i^\a$ are different and amorphous in each state.
In principle, we should put an infinitesimal local magnetic field that is 
{\it different for each spin} in order to select a state. However,
the local magnetizations of the states depend on the couplings $J$, so
we must apply
the small field {\it before} taking the average over the disorder
and that field is itself
correlated with the disorder. We therefore cannot 
study the states by selecting them according to an external field, as is usually done for
standard phase transitions.

The idea of \cite{Mo95} is to bypass this problem by considering 
$m$ copies of the original system that are coupled by
a small attractive term.
As we will show below, 
for $T<T_d$, it is possible to choose the small attractive coupling in such a way that
{\it i)} the $m$ copies are constrained to be in the same TAP state, and
{\it ii)} they are uncorrelated within the TAP state.
In this situation, the free energy of the $m$ copies inside a TAP state is just $m$ times
$f_\a$, because the $m$ copies are independent in that state.
Then, at low enough temperatures, the partition function of the replicated system is 
the sum over all states of the contribution of a single TAP state which is $e^{-\b N m f_\a}$:
\beq
\label{Zm}
Z_m  \sim \sum_\a e^{-\b N m f_\a}
= \int_{f_\mathrm{min}}^{f_\mathrm{th}}df \, 
e^{N [\Si(f)-\b m f]}
\sim  e^{N [\Si(f^*)-\b m f^*]} \ ,
\eeq
where now $f^*(m,T)$ is such that $m f - T \Si(f)$ is minimum and
satisfies the equation
\beq\label{saddleSim}
\frac{d\Si}{df} = \frac{m}{T} \ .
\eeq
As a result, we see that an additional weight $m$ has been given to the
term $-\b f$ in (\ref{Zm}).
This is a crucial result: because of this, the full complexity function 
can be computed from the knowledge
of the free energy of the replicated system. Defining
\beq
 \Phi(m,T) = -\frac{T}{N} \log Z_m = \min_f [ m f - T \Si(f) ] = m f^*(m,T) - T \Si(f^*(m,T)) \ ,
\eeq
it is indeed straightforward
to show that
\beq
\label{mcomplexity}
\begin{split}
&f^*(m,T) = \frac{\partial \, \Phi(m,T)}{\partial m} \ , \\
&\Si(m,T) = \Si(f^*(m,T)) = m^2 \frac{\partial \,[ m^{-1} \b \Phi(m,T)]}{\partial m} = 
m \b f^*(m,T) - \b \Phi(m,T) \ .
\end{split}
\eeq
The function $\Si(f)$ can then be reconstructed from the parametric plot of $f^*(m,T)$ 
and $\Si(m,T)$ by varying $m$ at fixed temperature.
From the knowledge of $\Si(f)$ all the information on the TAP states contained in Fig.~\ref{fig1:TAP} can be reconstructed.

Let us expand the complexity at low free energy as
\beq
\Si(f) = \Si(f_\mathrm{min}) + \s_0(T) (f - f_\mathrm{min}) + \cdots \ .
\eeq
Hence $\s_0(T)$ is the slope of $\Si(f)$ at $f_\mathrm{min}$ and we made its temperature dependence explicit.
For $m=1$, the glass transition happens when $\b$ equals $\s_0(T)$, 
or in other words $T_K$ is the solution of $\s_0(T)=1/T$. 
However, if $m$ is allowed to assume real values by an analytical continuation and if $m < 1$, 
this condition is replaced by $\s_0(T) = m/T$.
Because the temperature dependence of $\s_0(T)$ is usually mild,
the glass transition is shifted to lower
temperatures for $m<1$, see Fig.~\ref{fig1:Scqualit}. 
In other words, for any $T<T_K$ there
exists a value $m_s(T) < 1$, such that for $m < m_s(T)$ the system is in the ``high temperature'' phase (where $f > f_{\rm min}$ and $\Si > 0$), 
while for $m> m_s(T)$ it is in the ``low temperature'' spin glass phase (where $f = f_{\rm min}$).
The line $m_s(T)$ is defined by the condition $\s_0(T) = m_s(T)/T$, hence $m_s(T) = T \s_0(T)$.
Because the free energy is always continuous and
is {\it independent} of $m$ in the spin glass phase (being simply the value $f_\mathrm{min}(T)$,
such that $\Si(f_\mathrm{min})=0$), one can compute the free energy of the glass below $T_K$
simply as $f_\mathrm{glass}(T)=f_\mathrm{min}(T)=\Phi(m_s(T),T)/m_s(T)$. In this way we can compute the thermodynamic
properties of the system. Note that by definition, at $T_K$ we have $m_s(T_K)=T_K \s_0(T_K) = 1$.

In summary,
this method allows us to compute the complexity $\Si(f)$ at any temperature,
provided we are able to compute
the free energy of $m$ copies of the original system constrained to be in the same
TAP state and to perform the analytical continuation to real $m$.
In~\cite{Me99} this method was applied to the spherical $p$-spin 
system and it 
was shown that it reproduces the results obtained from the
explicit TAP computation (see also \cite{CC05} for a detailed discussion). 
In the next section we discuss this computation in detail.


\subsubsection{Replicated free energy of the spherical $p$-spin model}
\label{sec:realreplicapspin}

In order to show explicitly how the method works, we perform here the explicit computation for the $p$-spin model following~\cite{Me99}.

We wish to compute the free energy of $m$ copies of the original system,
coupled by an attractive term, the role of which will be discussed after the computation has been performed. 
To fix ideas, we might choose the following Hamiltonian for the $m$ replicas:
\beq
H = H_J[\s_1] + \cdots + H_J[\s_m] - \ee \sum_{a,b}^{1,m} \sum_{i=1}^N \s_i^a \s_i^b \ .
\eeq
Note that the $m$ copies all have the {\it same} couplings
$J$.
In principle we should compute $\Phi(m,T)$ for a given set of couplings, however, thanks to the self-averaging property,
we can equivalently (in the thermodynamic limit) take the average over $J$ of the free energy of the total
system:
\beq
\Phi(m,T)= \overline{ -\frac{T}{N} \log Z_m } =
\overline{-\frac{T}{N} \log \int D\s_1 \cdots D\s_m e^{-\b ( H_J[\s_1] + \cdots + H_J[\s_m] )  +\b  \ee \sum_{a,b}^{1,m} \sum_{i=1}^N \s_i^a \s_i^b    }} \ .
\eeq
To lighten the notation we include the spherical constraint in the integration measure, which
we now denote $D \s$, \ie 
$D\s = \left(\prod_i d\s_i \right) \delta(\sum_i \s_i^2 = N)$.

Let's forget for a moment about the coupling term and set $\ee=0$.
The main problem 
is that we have to perform the average of the logarithm of the partition function,
which is not an easy task. 
Using the 
identity 
\beq
\log x = \lim_{n\to 0} \partial_n x^n \ , \hskip2cm \Phi(m,T)=-\frac{T}{N} \lim_{n\to 0} \partial_n  \overline{(Z_m)^n }
\eeq
transforms the problem into that of calculating
the average over the disorder of $(Z_m)^n$. Although this problem is difficult for
real $n$, it can be solved
for integer $n$. In fact, for integer $n$ (and $m$) we get
\beq\label{Zrepmn}
\overline{(Z_m)^n} =\overline{ \int D\s_1 \cdots D\s_{nm} e^{-\b ( H_J[\s_1] + \cdots + H_J[\s_{nm}] )}} \ ,
\eeq
where we now have a system of $m\times n$ copies or {\it replicas}. By convention we assume that replicas 
$\{1,\cdots, m\}$ are
the original ones (coupled), and $\{m+1, \cdots, 2m\}$ are a copy of them (coupled), and so on. Note that there is
no coupling between replicas belonging to different blocks $\{1+\ell m, \cdots, (\ell+1) m \}$. 
Labeling replicas by an index $a=1,\cdots,nm$, and dropping irrelevant 
normalization constants, we get
\beq\label{eq:Zmnapp}
\begin{split}
\overline{(Z_m)^n} &\propto
\int D\sigma_i^a \prod_{i_1<\cdots< i_p} \int dJ_{i_1\cdots i_p}\;
\exp\left[- J_{i_1\cdots i_p}^2 \frac{N^{p-1}}{p!} + 
\b J_{i_1\cdots i_p} \sum_{a=1}^{mn} \sigma_{i_1}^a \cdots \sigma_{i_p}^a\right] \\
&\propto \int D\sigma_i^a \prod_{i_1<\cdots< i_p}  \exp\left[\frac{\beta^2p!}{4N^{p-1}} \sum_{a,b}^{1,mn} 
\sigma_{i_1}^a\sigma_{i_1}^b \cdots \sigma_{i_p}^a\sigma_{i_p}^b\right] \\ &=
 \int D\sigma_i^a \; \exp\left[\frac{\beta^2}{4N^{p-1}} \sum_{a,b}^{1,mn} \left(\sum_i^N\sigma_i^a\sigma_i^b\right)^p\right]
= \int D\sigma_i^a \; \exp\left[N \frac{\beta^2}{4} \sum_{a,b}^{1,mn} \left(\frac{1}{N}\sum_i \sigma_i^a\sigma_i^b\right)^p \right]
 \ .
\end{split}\eeq 
Note that when taking the average over the $J$'s we had to take into account that they are symmetric under
permutations, so the average must only be taken over $i_1 < \cdots <i_p$.

We now see why the replica trick is useful. In fact, taking the average over the disorder for the replicated system,
we eliminated the couplings $J$ and introduced a coupling between replicas. 
The advantage is that once the disorder is eliminated, the replicated (and coupled) system becomes
translationally invariant. Then, for a fully-connected model, the Hamiltonian becomes a function of a global quantity\footnote{
We have already seen an example of the same property in the study of the Curie-Weiss model.
},
which in this case is the overlap between two different replicas of the system:
\beq\label{Qab}
Q(\s^a,\s^b) = \frac{1}{N}\sum_i \sigma_i^a\sigma_i^b \ .
\eeq
Note that $Q(\s^a,\s^a)=1$ due to the spherical constraint. 
This quantity measures 
the extent to which the configurations of replica $a$ and replica $b$ are different.

A way to compute the replicated partition function starting from Eq.~\eqref{eq:Zmnapp} is reviewed in~\cite{CC05}.
Here we present a slightly different route, just to provide the reader with two different ways of doing the same computation.
From Eq.~\eqref{eq:Zmnapp} we can write
\beq\label{eq:Zmnapp2}
\begin{split}
\overline{(Z_m)^n} &\propto  \int D\sigma_i^a \int \prod_{a<b}^{1,mn} \left\{ dQ_{ab} \, \d\left( Q_{ab} -   \frac{1}{N}\sum_i \sigma_i^a\sigma_i^b \right) \right\}
 \; \exp\left[N \frac{\beta^2}{4} \sum_{a,b}^{1,mn} Q_{ab}^p \right] \\
 &= \int dQ \; \exp\left[N \frac{\beta^2}{4} \sum_{a,b}^{1,mn} Q_{ab}^p \right]
   \int d\sigma_i^a \, \prod_{a\leq b}^{1,mn} \d\left( N Q_{ab} -  \sum_i \sigma_i^a\sigma_i^b \right) 
   = \int dQ \; \exp\left[N \frac{\beta^2}{4} \sum_{a,b}^{1,mn} Q_{ab}^p \right] J(Q) \ ,
\end{split}\eeq
where $dQ = \prod_{a<b} dQ_{ab}$ and in the second line we grouped the delta function that imposes the spherical constraint in $D\s$ with the other
delta functions, hence adding the diagonal term for $a=b$ with $Q_{aa}=1$.

We thus have to compute the Jacobian
\beq
J(Q) =  \int d\sigma_i^a \, \prod_{a\leq b}^{1,mn} \d\left( N Q_{ab} - \sum_i \sigma_i^a\sigma_i^b \right) =
\int d\vec\s^a \d(N Q_{ab} - \vec \s^a \cdot \vec \s^b) \ ,
\eeq
where we think to $\s_i^a$ as a $N$ dimensional vector $\vec\s^a$ with $\vec\s^a\cdot \vec\s^b = \sum_i \s_i^a \s_i^b$.
We note that $Q$ is a symmetric real matrix and it can thus be diagonalized\footnote{
This way of performing the computation has been suggested by Pierfrancesco Urbani~\cite{KPUZ13}.
}. We have $Q = \L^T D \L$ where
$\L^{-1} = \L^T$ is a $mn \times mn$ orthogonal matrix with $\det \L=1$, and $D$ is a diagonal matrix, $D_{aa} = \l_a$ being the eigenvalues
of $Q$. 
Calling $\s$ a matrix composed by the collection of $\vec\s^a$, $\s = \{ \vec\s^1 \cdots \vec\s^{mn} \}$, we define
$\mu = \s \L^T$ and $\mu^T = \L \s^T$, and we have
\beq
J(Q) = 
\int d\s \, \d(N  \L^T D \L - \s^T \s) =
\int d\s \, \d(N D - \L \s^T \s \L^T) =
\int d\m \, \d(N D - \m^T \m)
\ ,
\eeq
where all the manipulations above are possible because $\L$ is an orthogonal matrix and $\det\L=1$.
We obtain
\beq
J(Q) = \int d\vec\mu^a \prod_{a=1}^{mn} \d( N \l_a - \vec\mu^a \cdot \vec\mu^a ) \prod_{a<b}^{1,mn} \d( \vec\mu^a \cdot \vec\mu^b )
\propto \int d\vec\mu^a \prod_{a=1}^{mn} \d( \l_a - \vec\mu^a \cdot \vec\mu^a ) \prod_{a<b}^{1,mn} \d( \vec\mu^a \cdot \vec\mu^b ) \ ,
\eeq
where the factor $N$ can be eliminated by rescaling the vectors $\vec\mu^a$, giving a proportionality constant that we neglect.
We can now introduce polar coordinates where $\mu^a = | \vec\m^a |$ and $\hat \mu^a$ is a unit $N$-dimensional vector
that encodes the orientation of $\vec\mu^a = \mu^a \hat\mu^a$. We obtain
\beq
J(Q) \propto \int \prod_{a=1}^{mn} d\hat\mu^a d\mu^a(\mu^a)^{N-1}  \prod_{a=1}^{mn} \frac{1}{\sqrt{\l_a}}\d\left( \sqrt{\l_a} - \mu^a \right) 
\prod_{a<b}^{1,mn}  \frac{1}{\sqrt{\l_a \l_b}}\d( \hat\mu^a \cdot \hat\mu^b ) \ .
\eeq
The angular integration gives a constant independent of the matrix $Q$ (see~\cite{KPUZ13} for an explicit computation) and we finally obtain
\beq\label{eq:JQfinal}
J(Q) \propto \prod_{a=1}^{mn} (\l_a)^{(N-1)/2}  \prod_{a=1}^{mn} \frac{1}{\sqrt{\l_a}} 
\prod_{a<b}^{1,mn}  \frac{1}{\sqrt{\l_a \l_b}} = \prod_{a=1}^{mn} (\l_a)^{(N-mn-1)/2} = (\det Q)^{(N-mn-1)/2} \sim (\det Q)^{N/2}
\ ,
\eeq
where the last result holds for large $N$.

Plugging Eq.~\eqref{eq:JQfinal} in Eq.~\eqref{eq:Zmnapp2}
we get\footnote{
The proportionality constant in Eq.~\eqref{reppart} should be discussed a bit more carefully. At the leading exponential order in $N$, it
gives a contribution of the form $\exp( m n N C)$ where $C$ is a numerical constant. This can be argued because when $n m =0$ the
proportionality constant must be 1. This term shifts the free energy Eq.~\eqref{eq:freeapp} by a multiple of $mT$, which corresponds to a
constant shift of the entropy and can therefore be neglected. Other subleading corrections disappear in the large $N$ limit. We conclude that
the proportionality constant can really be neglected.
}
\begin{equation}\label{reppart}
\begin{split}
&\overline{(Z_m)^n} \propto \int d Q_{ab} \ e^{N X(Q)} \ , \\
&X(Q)=\frac{\beta^2}{4} \sum_{ab} Q_{ab}^p+ \frac{1}{2} \log \det Q \ .
\end{split}
\end{equation}
The advantage of this form of the integral is that we can use the saddle point (or Laplace, or steepest-descent) method, 
to solve it in the limit $N\to\infty$. 
This simplification is a consequence of the mean-field structure of the model, and 
results from
decoupling
the sites
using
the replica trick.
The saddle-point method states that in the limit $N\to\infty$ the integral (\ref{reppart}) is concentrated in the 
maximum of the integrand. 

Note, however, that we have to be careful for  two reasons. First, the free energy is 
given by
\beq\label{eq:freeapp}
\Phi(m,T) = -T
\lim_{N\to\infty}\frac{1}{N} \lim_{n\to 0}\partial_n  \int dQ_{ab}\; \exp\left[N X(Q)\right] 
\eeq
and thus we should {\it first} take the limit $n\to 0$, and then $N\to\infty$. Unfortunately, we are unable to do so because
$X$ is not an explicit function of $n$ and because we first need to send $N\to\infty$ to solve the integral.
As a conclusion, we need to exchange the order of the two limits, solve the integral, find a parametrization of the
matrix $Q_{ab}$, and finally take the $n\to 0$ limit. This operation is clearly quite dangerous from a strict mathematical
point of view.

Second, we must
pay attention to what is meant
by the ``maximum'' of $X$. The problem is that
the number of independent elements of $Q_{ab}$, $n(n-1)/2$, 
becomes negative is the limit $n\to 0$. 
It is hard to say what is the maximum of a function with a negative number of variables.
There is, however, a criterion we can use
to select the correct saddle point. The corrections to the saddle point result are given by the Gaussian integration
around the saddle point itself. As a result, this integration gives the square root of the determinant of the second
derivative matrix of $X$. 
In order to have a sensible result, we must have 
that the analytic continuation of all the eigenvalues of this
matrix are negative. 
In other words, we have to select saddle points with a negative-defined second derivative of $X$~\cite{AT78}.

\subsubsection{1-step replica symmetry breaking}
\label{sec:1rsbpspin}

The saddle point equation for $Q$ is complicated and cannot be solved in general. We need a simple 
{\it ansatz} on the form of the matrix $Q$ in order to find the solution.
How can we guess the form of $Q$? 
At this point we have to reintroduce the coupling term.
A quick look to the previous computation reveals that in presence of the coupling
the $nm$ replicas are divided in blocks of
$m$ replicas, such that each block is coupled while replicas in different blocks are not coupled.
Let us label by $B$ the blocks.
Because the coupling term only depends on $Q_{ab}$, we can carry out the same computation and we find
\beq\label{XQeps}
X(Q)=\frac{\beta^2}{4} \sum_{ab} Q_{ab}^p+ \frac{1}{2} \log \det Q + \b\ee \sum_B \sum_{ab \in B} Q_{ab} \ .
\eeq
We see that the coupling term {\it breaks explicitly} the replica symmetry.

The matrix element $Q_{ab}$, according to (\ref{Qab}), measures the overlap of replicas $a$ and $b$, \ie how
much the two replicas are close to each other. Therefore, it is very natural to assume that the replicas within a
same block have high overlap due to the external coupling, while replicas in different blocks
have much smaller overlap: in practice we can assume that their overlap is zero, as it would be for completely
uncorrelated replicas. Moreover, we assume that the overlap between coupled replicas is the same for any pair
of replicas $a\neq b$ belonging to the same block. This {\it ansatz}, which is the simplest possible, is
called {\it 1-step replica symmetry breaking} ({\sc 1rsb}) for reasons that will be made clear below. 
The {\sc 1rsb} {\it ansatz} for $Q$ reads, for $n=2$ and $m=3$:
\begin{equation}
\label{Q1rsb}
Q=\left(
\begin{array}{cc}
\left(
\begin{array}{ccc}
1 & q & q \\
q & 1 & q \\
q & q & 1 \\
\end{array}
\right) & 0 \\
0 & 
\left(
\begin{array}{ccc}
1 & q & q \\
q & 1 & q \\
q & q & 1 \\
\end{array}
\right) \\
\end{array}
\right) \ .
\end{equation}
Note that here there is no spontaneous symmetry breaking: replica symmetry is broken explicitly by the coupling term,
and Eq.~(\ref{Q1rsb}) is the simplest {\it ansatz} that is compatible with the symmetries of the action $X(Q)$.

We can use the relation
\begin{equation}
\det  \left(
\begin{array}{ccc}
1 & q & q \\
q & 1 & q \\
q & q & 1 \\
\end{array}
\right) = (1-q)^{m-1} [ 1+(m-1)q ] \ ,
\end{equation}
to get
\beq
\det Q = \big\{ (1-q)^{m-1} [ 1+(m-1)q ] \big\}^n \ .
\eeq
Substituting this ansatz in (\ref{XQeps}) we ge
\beq
X(Q) = -\b n m \phi_\mathrm{1RSB}(m,q,T) + \b \ee n m (m-1) q
\eeq
where the {\sc 1rsb} free energy is
\beq\label{phi1RSBpspin}
\phi_\mathrm{1RSB}(m,q,T) = -\frac{1}{2\b}\left\{ \frac{\b^2}{2}\big[1+(m-1)q^p\big]
+\frac{m-1}{m} \log(1-q) +\frac{1}{m}\log\big[1+(m-1)q\big] \right\} \ .
\eeq
Note that this expression, which has been derived for integer $m$ and $n$, is also perfectly
well defined for real $m$ and $n$, so that now we can perform the continuation to $n\to 0$
and real $m$.
Finally
\beq\label{repPhi}
\Phi(m,T) =  -\frac{T}{N}
\lim_{n\to 0} \partial_n\exp \big[ -\b n m N \phi_\mathrm{1RSB}(m,q^*,T) + N \b \ee n m (m-1) q^*   \big] = 
m \phi_\mathrm{1RSB}(m, q^*,T) - \ee m (m-1) q^*    \ ,
\eeq
and $q^*$ is the stationary point of this expression.

An important remark is that, due to the simple structure of the saddle-point matrix $Q$,
the result for $\Phi(m,T)$
is equivalent to the one that would be obtained by a direct computation of 
$-\frac{T}{N} \log \overline{Z_m}$ without introducing the $n$ additional replicas.
This is due to the fact that replicas in different blocks are completely uncorrelated, and
therefore $X(Q)$ is simply proportional to $n$. This fact is very specific to the $p$-spin
model and is false in most other
cases. In these cases the calculation might
be more complicated, but the general strategy outlined above remains valid. We will see
an example later on, in {\bf $\Rightarrow$ Ex.\ref{sec:optim}.\ref{ex:II.1}} (but it's too early to
do the exercise at this point). 

So, finally, we can discuss the role of the coupling $\ee$. 
At low enough temperatures, three solutions of the stationary equation for $q$ exist: two of them are 
minima (or maxima depending on $m$) and are separated by a minimum (maximum).
Clearly, in this situation a large enough $\ee$ will always favor the large $q$ solution. However, we do not want
to perturb the TAP states, or in other words we want replicas to be uncorrelated within a state, and for this reason we must send $\ee\to 0$.
While doing this, however, we want to maintain the high correlation between replicas. Hence, the prescription is to set $\ee=0$ but always
select the large $q$ solution of the stationary condition, if it exists (see~\cite{Me99} for a more detailed discussion).
Otherwise, if for $\ee=0$ only the solution $q=0$ exists, we get a trivial result: $\Phi(m,T) = -\frac{\b m}4 =  m f_{\rm para}$ and using
Eq.~(\ref{mcomplexity}) we find that the complexity vanishes.

\subsubsection{The phase diagram of the spherical $p$-spin model}

Let us summarize the previous discussion.
We have assumed that replicas in different blocks are uncorrelated and have zero overlap, while replicas in
the same block are correlated and so we assigned them an overlap $q$. This correlation is due to the external
coupling $\ee$, whose role is only to select the solution for $q$ that has the highest overlap.
Indeed, the free energy (\ref{repPhi}) at $\ee=0$ and as a function of $q$ always has a stationary point at $q=0$, but
for low enough temperatures a second stationary point appears at $q^* \neq 0$, and the coupling is only
introduced to select this solution.
This approach is very similar to what is usually done in first order phase transitions
by selecting one state or the other using an infinitesimal external field (\eg in a ferromagnet the
states with positive or negative magnetization can be selected by an infinitesimal positive or negative magnetic field).
Therefore, if we are interested in the partition function of $m$ replicas {\it in the same state},
as we did in the previous section in order to compute the complexity, we should {\it always}
take the solution with $q \neq 0$, if it exists. 

\begin{figure}[t]
\centering
\includegraphics[width=10cm]{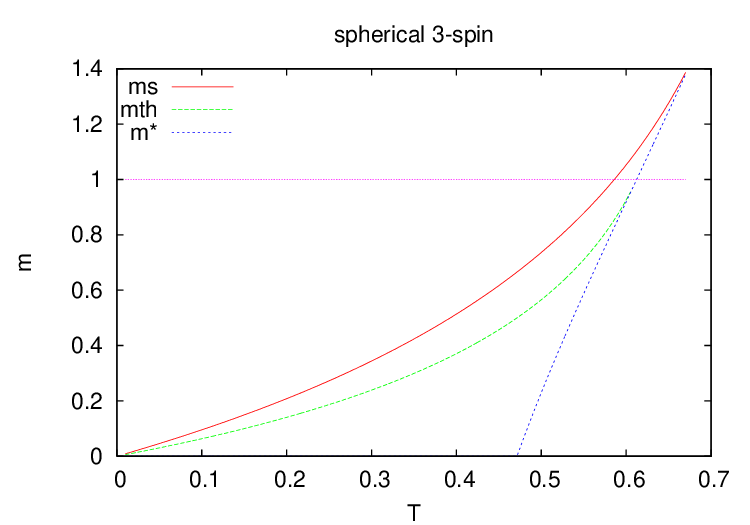}
\caption{Phase diagram of the $3$-spin spherical model in the $(m,T)$ plane.
}
\label{fig3:dia}
\end{figure}

The final result is that
\beq
\Phi(m,T) = m \phi_\mathrm{1RSB}(m, q^*(m,T),T) \ .
\eeq
From this, we can draw a ``phase diagram'' in the $(m,T)$ plane, which is reported in
Fig.~\ref{fig3:dia} for the spherical 3-spin model ({\bf $\Rightarrow$ Ex.\ref{sec:fully}.\ref{ex:1.3}, \ref{sec:fully}.\ref{ex:1.4} and \ref{sec:fully}.\ref{ex:1.5}}). 
It results from a series of considerations:

\begin{itemize}

\item
First of all we must identify the region where a solution with $q \neq 0$
is found.
This region is delimited by the line $m^*(T)$, 
For $m<m^*(T)$, we have $q^*(m,T)=0$, $\Phi(m,T)$ is trivial, and the complexity is zero.
For $m>m^*(T)$,
a non-trivial solution $q^*(m,T)$ is found: in this region
we can compute, using Eq.~(\ref{mcomplexity}), the complexity and free energy 
as a function of $m$:
\beq\label{SimT}
\begin{split}
&\Si(m,T) = m^2 \partial_m \big[ m^{-1}\b \Phi(m,T) \big] = 
m^2 \partial_m [ \b\phi_\mathrm{1RSB}(m,q^*,T) ] \ , \\
&f^*(m,T) = \partial_m  \Phi(m,T) = \partial_m m \phi_\mathrm{1RSB}(m,q^*,T) \ .
\end{split}\eeq
These quantities are therefore defined {\it only} above the line $m^*(T)$. We can observe that the non-trivial solution $q^*(m,T)$
disappears at $m^*(T)$ by merging with another solution (a standard ``bifurcation" of solutions). Because of this, a zero mode
(the so-called ``longitudinal'' mode)
in the stability matrix of the {\sc 1rsb} solution appears at $m^*(T)$ ({\bf $\Rightarrow$ Ex.\ref{sec:fully}.\ref{ex:1.5}}).

\item
Next, we analyze the behavior of $f^*(m,T)$ and $\Si(m,T)$.
It turns out that starting from $m=\max\{0,m^*(T)\}$ and increasing $m$, both $f^*(m,T)$ and $\Si(m,T)$ 
first increase up to a maximum and then decrease upon further increasing $m$. This non-monotonicity leads to two branches of the
parametric curve $\Si(f)$, of which only one is physical: the one that corresponds to {\it decreasing}
$f$ and $\Si$ with $m$. Indeed, from Fig.~\ref{fig1:Scqualit} it should be clear that both $f^*$ and $\Si$ must decrease
when increasing $m$.
We then define a line $m=m_{\rm th}(T)$ where $f^*(m,T)$ is maximum, which corresponds to the threshold values
$f_\mathrm{th}(T)$. The physical region is the one for $m\geq m_{\rm th}(T)$.
One can show in fact that for $m<m_{\rm th}(T)$, the {\sc 1rsb} solution is {\it unstable}: one of the eigenvalues of the stability matrix 
(the so-called ``replicon'' mode)
becomes
negative suggesting that a more complex matrix $Q_{ab}$ should be considered ({\bf $\Rightarrow$ Ex.\ref{sec:fully}.\ref{ex:1.5}}). 
However, one can show that no consistent solution can be constructed for $m<m_{\rm th}(T)$~\cite{Ri13}.

\item
Finally, we observe that
at a fixed $T$, the complexity is finite for $m=m_{\rm th}(T)$
and decreases on increasing $m$ above $m_{\rm th}(T)$, until it vanishes on a second line $m_s(T)$, that is also reported
in Figure~\ref{fig3:dia}. Above this line, the complexity becomes
negative, indicating that states do not exist anymore. States are therefore found for $m_s(T)>m>m_{\rm th}(T)$, with $m_{\rm th}(T)$ corresponding to $f_\mathrm{th}(T)$ and
$m_s(T)$ corresponding to $f_\mathrm{min}(T)$.
\end{itemize}
Note that the two lines $m_{\rm th}(T)$ and $m^*(T)$ merge at $m=1$ and $T=T_d$. This is due to the fact that the longitudinal mode and the replicon mode coincide at $m=1$.
For $m>1$, there is no line $m_{\rm th}$ because the {\sc 1rsb} is stable in the whole
interval $m_s(T) \geq m \geq m^*(T)$ (in other words, the longitudinal mode becomes unstable before the replicon mode).
At yet higher temperatures the lines $m_s(T)$ and $m^*(T)$ touch: this is the temperature $T_\mathrm{TAP}$ above which only the paramagnetic state survives, see Figure~\ref{fig1:TAP}.

From the previous discussion, in particular equations (\ref{Zm1}), (\ref{saddleSi}) and (\ref{Zm}), (\ref{saddleSim}),
 $m=1$ obviously corresponds to the equilibrium partition function of a single copy of the system.
When the line $m_{\rm th}(T)$ crosses $m=1$, the saddle point in (\ref{Zm1}) is exactly equal to $f_\mathrm{th}$, which is the point
where the paramagnet breaks in many states. This crossing defines the temperature $T_d$. Similarly, the point where the
line $m_s(T)$ crosses $m=1$ corresponds to the point where the saddle point is equal to $f_\mathrm{min}$, 
the complexity vanishes and a few states dominate, \ie the
phase transition $T_K$.

For $T<T_K$, the saddle point of (\ref{Zm1}) is always $f_\mathrm{min}$, which can be computed by following the line
$m_s(T)$. The analytic continuation above $m_s(T)$ is not correct: the complexity becomes negative above $m_s(T)$.
In other words, the system of $m$ replicas undergoes a glass transition on the line $m_s(T)$. 
However, we do not need to wonder about what happens above $m_s(T)$, because the discussion of Section~\ref{sec:realreplica}
tells us that the value of the free energy $\Phi(m,T)$ on the line $m_s(T)$, which is $\Phi(m,T) = m f_{\mathrm{min}}(T)$,
is the one that persists up to the $m=1$ line. Note that $m<1$ and, recalling that $\Si=0$ along $m_s$, from Eq.~(\ref{mcomplexity}) one obtains:
\beq\label{fglass}
f_\mathrm{min}(T) = \left. \frac{\Phi(m,T)}{ m} \right|_{m=m_s(T)} = \phi_\mathrm{1RSB}(m_s(T),q^*(T),T) \ .
\eeq
This last result is very important. Indeed, for $T>T_K$ the free energy of the system is equal to the free
energy of the paramagnet, corresponding simply to $\phi_\mathrm{1RSB}(m=1) = -\b/4$. Below $T_K$, instead, the free
energy of the system is given, comparing (\ref{fglass}) and (\ref{SimT}), by extremizing $\phi_\mathrm{1RSB}$
with respect to {\it both} $m$ and $q$. Actually, the extremum is a {\it maximum}, as can be seen by an explicit
computation. Remarkably, the fact that the equilibrium spin glass free energy is the maximum of $\phi_\mathrm{1RSB}$ with respect to
$m$ and $q$ is the usual prescription of the replica method, and it can be rigorously proven to be correct~\cite{FT06}.

In summary, the replica method allowed us to fully characterize the thermodynamics of the spherical $p$-spin
model, by computing
\begin{itemize}
\item the free energy of the paramagnetic phase
\item the free energy of the glass phase
\item the distribution $\Si(f)$ of all the metastable states as a function of $f$ and $T$
\end{itemize}
The main assumption we made in the derivation is that {\it the states are all equivalent}, with a
self-overlap $q$ and zero mutual overlap, \ie they are randomly distributed in phase space. 
These properties are expressed
by the {\sc 1rsb} structure of the overlap matrix $Q_{ab}$ where all nonzero entries, corresponding to
the replicas in the same state, have the same value $q$.
This behavior
is exact for the spherical $p$-spin model, but it is somewhat exceptional. In particular
it is not true for the SK model, where the states are organized in a very complicated structure.

As a final remark, it is important to stress that the vanishing of the replicon mode on the line $m_{\rm th}$ has an extremely
important physical consequence: that the threshold states at $f_{\rm th}$ have a zero mode of their TAP stability matrix and 
they are therefore {\it marginal}. This is extremely important because one can show that this marginality property is intimately
connected to many important properties of the out-of-equilibrium {\it aging} dynamics~\cite{CK93,Cu02,FLPR12,Ri13}. For
reasons of space, we cannot discuss these properties here and we refer the reader to~\cite{CK93,Cu02,FLPR12,Ri13} and references
therein.

\subsubsection{Spontaneous replica symmetry breaking: the order parameter}

To conclude this section, we will discuss now the concept of {\it spontaneous replica symmetry breaking}.
Suppose that, ignoring completely the above discussion, we tried to compute directly the free energy
of {\it a single copy} of the system. Then, using the replica method, we would write
\beq\label{fdiuno}
f = -\overline{\frac{T}{N} \log Z} = -\frac{T}{N} \lim_{\nu \to 0} \partial_\nu \overline{Z^\nu} \ .
\eeq
Formally, this computation is completely identical to the one of the previous section, Eq.~(\ref{Zrepmn}),
with $\nu = mn$. This time, however, there is no external coupling that suggests us the structure (\ref{Q1rsb})
of the overlap matrix. The simplest guess would be simply to set $q=0$, \ie to assume
that $Q_{ab}$ is a diagonal matrix and that all the replicas are uncorrelated. Going through the rest of the computation would then give us
the paramagnetic solution, which is wrong below $T_K$, and would miss the interesting structuring of configuration space below $T_d$.
As guessed by Parisi \cite{MPV87}, the correct solution is to assume the structure (\ref{Q1rsb}) also
in the computation of (\ref{fdiuno}), and consider $m$ and $q$ as variational parameters to be optimized.
This solution is exactly what we obtained in Eq.~(\ref{fglass}), which shows that these two conceptually
different strategies lead to the same result for the free energy of the glass (which is the correct one~\cite{FT06}).
In this case, replica symmetry breaking is not imposed by an external field, but appears as a spontaneous symmetry
breaking, which this was historically how it was introduced in order to find the correct solution.

The above discussion shows that $q$ is the {\it order parameter} of the transition: it is zero in the paramagnetic
phase, and it jumps to a nonzero value in the glass phase signaling the spontaneous breaking of replica
symmetry. Alternatively, we can introduce an external field coupled to this order parameter in order to compute
the properties of the low-temperature phase, analogously to a magnetic field for a ferromagnet.
The transition is therefore first order from the
point of view of the order parameter, but recall that it is second order from the thermodynamical point of view.

\subsection{The SK model: full replica symmetry breaking}

We can now go back to the SK model. In this case, the situation is much more complicated.
The study of the TAP equations for the SK model was initiated 
in the original publication~\cite{TAP77} (the reprints of this and other relevant papers can be found in \cite{MPV87}).
The TAP equations for the SK model have many solutions, as is the case for the spherical $p$-spin model,
but here
the computation of their complexity is much more involved.
It was started in \cite{BM80}, but a recent revival of interest led
to a more complete understanding of the problem. It is reviewed in \cite{Pa05} and
references therein.

There are two main reasons for this difficulty. First, the equilibrium states in the spin
glass phase (\ie the lowest energy states) {\it are not independently distributed},
unlike for the spherical $p$-spin model in which the
different states have zero overlap. 
In the SK model, the mutual overlaps between equilibrium states are organized in a complicated pattern~\cite{MPV87}.
Second, the metastable states (whose free energy is larger than that of the equilibrium states) are not 
well-defined minima, again unlike for the spherical $p$-spin model. In the SK model, for finite $N$
they come in pairs, one being a minimum and the other a maximum. The two coalesce in the $N\to\io$
limit, forming a {\it saddle} point, \ie a state with one zero mode.

In the SK model the existence of these marginally stable states is not important, as far as the
equilibrium properties are concerned, because they do not appear in the computation of the partition
function. The transition is from the high-temperature paramagnetic state to the low-temperature
{\it equilibrium} states, and is second-order both thermodynamically and from the point of view of the order parameter.
Indeed, the local magnetizations $m_i$ are small close to $T_c$~\cite{TAP77} and so are the 
overlaps~\cite{MPV87}.

Here we will only discuss the definition of the order parameter for the SK model and its
solution using the replica method. For more details on the TAP approach 
see~\cite{TAP77,BM80,MPV87,Pa05}.

\subsubsection{The overlap distribution}

As we already discussed, in presence of many pure states the Gibbs measure decomposes
as $P[S] = \sum_\a w_\a P_\a[S]$, with $\sum_\a w_\a=1$. We also said that each state
is specified by its local magnetizations $m_i^\a = \la S_i \ra_\a$. We can define the
{\it overlap} between two states
\beq
q_{\a \b} = \frac1N\sum_i m_i^\a m_i^\b = \frac1N \sum_i \la S_i \ra_\a\la S_i \ra_\b \ .
\eeq

In the case of the spherical $p$-spin model, we found that the natural order parameter was the
overlap of two replicas defined in (\ref{Qab}). The infinitesimal coupling forces the
two replicas in the same state, so we have
\beq
q = \la Q_{ab} \ra =  \sum_\a w_\a \frac1N \sum_i \la S^a_i \ra_\a \la S^b_i \ra_\a =
\sum_\a w_\a \frac1N \sum_i m_i^\a m_i^\a = \sum_\a w_\a q_{\a\a} \ ,
\eeq
and the (thermal average of the) overlap is just the average self-overlap of a state $\a$.
Recall that for the spherical $p$-spin model we assumed that $q_{\a\a}\equiv q$, and that
different states were uncorrelated,
\ie that $q_{\a\b}=0$ for $\a \neq \b$.

In a more complicated situation we might be interested in the probability distribution of
the overlap:
\beq
P(q) = \sum_{\a\b} w_\a w_\b \d(q - q_{\a\b}) \ ;
\eeq
in general $P(q)$ will depend on the couplings $J$ and we will then consider its average
over the disorder, $\overline{P(q)}$.
We wish now to work out a connection between this quantity and the matrix $Q_{ab}$ that
appears in the replicated free energy (\ref{reppart}).
Let us then consider the following quantity,
\beq
q^{(1)}= \frac{1}{N}\sum_i \overline{\langle S_i\rangle^2} \ ,
\eeq
where the average is over the Gibbs measure.
By using the decomposition in pure states, we can rewrite $q^{(1)}$ as
\beq
q^{(1)} = 
\frac{1}{N}\sum_i\sum_{\alpha\beta} \overline{w_\alpha w_\beta\  \langle S_i\rangle_\alpha \langle S_i\rangle_\beta} =
\sum_{\alpha\beta} \overline{w_\alpha w_\beta \ q_{\alpha\beta }} =
\int dq \sum_{\alpha\beta} \overline{w_\alpha w_\beta \ \delta(q-q_{\alpha\beta})}\ q = 
\int dq \; \overline{P(q)}\; q
\eeq
Therefore $q^{(1)}$ is the first moment of the overlap distribution, averaged over the disorder. By using the clustering
property, we can easily find a generalization of this formula \cite{MPV87},
\beq\begin{split}
q^{(k)} &= \frac{1}{N^k}\sum_{i_1\dots i_k} \overline{\langle S_{i_1}\dots  S_{i_k}\rangle^2} = 
\frac1{N^k} \sum_{i_1\dots i_k}\sum_{\a\b} w_\a w_\b 
\overline{\langle S_{i_1}\dots  S_{i_k}\rangle_\a\langle S_{i_1}\dots  S_{i_k}\rangle_\b }
\\ &=
\frac1{N^k} \sum_{i_1\dots i_k}\sum_{\a\b} w_\a w_\b 
\overline{\langle S_{i_1}\rangle_\a\dots\langle  S_{i_k}\rangle_\a\langle S_{i_1}\rangle_\b
\dots \langle S_{i_k}\rangle_\b }=
\int dq\;  \overline{P(q)}
\; q^k \label{picco}
\end{split}\eeq
The important fact is that we can also compute these quantities using the replica trick. In particular,
\beq
q^{(1)}= \frac{1}{N}\sum_i \overline{\langle S_i\rangle^2} = 
\overline{\frac1{Z^2} \sum_{ S^1, S^2} e^{-\b (H[S^1]+H[S^2])} 
\frac{1}{N}\sum_i  S_i^1 S_i^2}
= \lim_{n\to 0}\overline{
\sum_{ S^a } \ \frac{1}{N} \sum_i  S_i^1  S_i^2 \ e^{-\beta\sum_a H( S^a)}
} \ ,
\eeq
where the last equation is obtained writing $Z^{-2} = \lim_{n\to 0} Z^{n-2}$.
If we now go on with the calculation along the lines of the previous paragraphs, introducing the 
overlap matrix $Q_{ab}$, we get,
\beq
q^{(1)}= \int DQ_{ab} \;\; e^{-NX(Q_{ab})}\, Q_{12} = Q_{12}^{\rm{SP}} e^{-NX(Q^{\rm SP}_{ab})} =  Q_{12}^{\rm{SP}}
\label{pluto}
\eeq
where $Q_{ab}^{\rm{SP}}$ is the saddle point value of the overlap matrix (from now on we will drop the 
suffix SP), and where we have exploited the fact that $S(Q_{ab})$ is of order $n$, and therefore does not contribute when 
$n\to 0$. Of course, there is something wrong about this formula, because replicas 1 and 2 cannot be different from the
others. If we decided to call them 4 and 7, we would get a different result whenever $Q_{ab}$ is not replica symmetric.
To better understand this point we note that if the saddle point 
overlap matrix is not symmetric, then there must be other saddle point solutions with the same free energy, 
but corresponding to matrices obtained from $Q_{ab}$ by a permutation of lines and columns \cite{MPV87}. 
This result is general: when a saddle point
breaks a symmetry corresponding to a given transformation, all the points obtained by applying the transformation
to that particular saddle point, are equally valid. We must therefore average over all these
saddle points, which is equivalent to symmetrizing the equation (\ref{pluto}):
\beq
q^{(1)} = \lim_{n\to 0} \frac{2}{n(n-1)} \sum_{a>b} Q_{ab} 
\label{sunto}
\eeq
This result is already telling us that there is a connection between
$q^{(1)}$ and the matrix of the overlap among replicas $Q_{ab}$. To go further, we can generalize (\ref{sunto}) 
to get
\beq
q^{(k)} = \lim_{n\to 0} \frac{2}{n(n-1)} \sum_{a>b} Q_{ab}^k 
\label{sunto2}
\eeq
Comparing with equation (\ref{picco}) gives that for a generic function $f(q)$
\beq
\int dq\; f(q)\;\overline{P(q)} = \lim_{n\to 0} \frac{2}{n(n-1)} \sum_{a>b}f(Q_{ab})\ ,
\eeq
which, in particular, for $f(q)=\delta(q-q')$ finally provides the crucial equation connecting physics to
replicas,
\beq
\overline{P(q)} = \lim_{n\to 0} \frac{2}{n(n-1)} \sum_{a>b} \delta(q-Q_{ab}) \ .
\label{funda}
\eeq
This equation shows that the average probability that two pure states of the system have overlap
$q$ is equal to the  fraction of elements of the overlap matrix $Q_{ab}$ equal to $q$. In 
other words, {\it the elements of the overlap matrix (in the saddle point) are the physical values 
of the overlap  among pure states, and the number of elements of $Q_{ab}$ equal to $q$ is related to the 
probability of $q$}.

Before turning to a more precise computation for the SK model, it is useful to discuss some
general properties of the overlaps. First of all, it is reasonable (and correct \cite{MPV87})
to assume that all
the states have the same self overlap, $q_{\a\a} \equiv q_\mathrm{EA}$, as in the spherical $p$-spin model.
Then, for any two states $\a$ and $\b$:
\beq
0 \leq \frac1N \sum_i (m_i^\a - m_i^\b)^2 = q_{\a\a}+q_{\b\b} - 2 q_{\a\b} = 2(q_\mathrm{EA}-q_{\a\b}) 
\hskip1cm
\Rightarrow
\hskip1cm
q_\mathrm{EA} = \max \{ q_{\a\b} \} \ .
\eeq

Additionally, it is convenient to remove the trivial symmetry $S \to -S$ that is present in the SK
model and is reflected in $P(q) = P(-q)$. This can be done
 by adding an infinitesimal magnetic field that will favor one of the two states $\a$ and $-\a$
related by the symmetry. Once it is done, the overlaps are all positive and one obtains a
distribution $P_+(q)$, such that $P(q) = (P_+(q)+P_+(-q))/2$. In the following we will drop
the suffix $+$ and consider that the overlaps are all positive.

Finally, it is useful to define the function
\beq
x(q) = \int_0^q P(q') dq' \in [0,1] \ , \hskip2cm \frac{dx}{dq}=P(q) \ .
\eeq
As $P(q)$ is positive, $x(q)$ is a monotonically increasing function and so we can define its inverse $q(x)$.
In particular we can write
\beq
q^{(1)} = \int_0^1 q P(q) dq = \int_0^1 q \frac{dx}{dq} dq = \int_0^1 q(x) dx \ .
\eeq

\subsubsection{The Parisi solution of the SK model}

In the case of the SK model, one can again introduce replicas to average over
the disorder, with again the result
\beq
\overline{Z^n} \sim \int dQ_{ab} e^{N X(Q_{ab})} \ .
\eeq
The explicit form of $X(Q)$ can be found, for example, in \cite{MPV87}, but it is not crucial for the rest of this discussion.
The replica symmetric solution corresponds to $Q_{ab}=\d_{ab} + q(1-\d_{ab})$
(in this case, we also allow for a finite overlap between replicas in different
states). Remarkably, this solution predicts that $q=0$ for $T>1$, while $q\neq 0$
for $T<1$, \ie one finds a phase transition at $T=T_c=1$.
The entropy, however, becomes negative at low temperatures, and, as the SK model is
formulated for discrete spins, this solution is clearly incorrect.

If one uses a {\sc 1rsb} ansatz, as in (\ref{Q1rsb}), the situation is improved,
in the sense that the entropy becomes negative at a much lower temperature
and it is negative but small at $T=0$. It is better to 
change notation, $q \to q_1$ in (\ref{Q1rsb}) and to replace the zeros by $q_0$ to
be more general.
Then, within the {\sc 1rsb} ansatz, from (\ref{funda}) one has
\beq\label{Pq1rsb}
\overline{P(q)} = \lim_{n\to 0} \frac{1}{n-1} \left[ (n-m) \d(q-q_0) + (m-1) \d(q-q_1) \right]
=  m \d(q-q_0) + (1-m) \d(q-q_1) \ ,
\eeq
meaning that two replicas have a probability $m$ of being in the different states and 
a probability $1-m$ of being in the same state. Note that it follows from this interpretation
that $m\leq 1$, as 
we found for the $p$-spin spherical model.

Parisi then introduced another
level of replica symmetry breaking, by assuming that replicas are split into
$m_1$ blocks, and that inside each block they are further split into $m_2 < m_1$ blocks.
For $n=8$, $m_1=4$ and $m_2 = 2$ the matrix $Q$ would then read
\begin{equation}
Q=\left(
\begin{array}{cc}
\begin{array}{cccc}
1 & q_2 & q_1 & q_1 \\
q_2 & 1 & q_1 & q_1 \\
q_1 & q_1 & 1 & q_2 \\
q_1 & q_1 & q_2 & 1 \\
\end{array}
 & 
\begin{array}{cccc}
q_0 & q_0 & q_0 & q_0 \\
q_0 & q_0 & q_0 & q_0 \\
q_0 & q_0 & q_0 & q_0 \\
q_0 & q_0 & q_0 & q_0 \\
\end{array}
 \\
\begin{array}{cccc}
q_0 & q_0 & q_0 & q_0 \\
q_0 & q_0 & q_0 & q_0 \\
q_0 & q_0 & q_0 & q_0 \\
q_0 & q_0 & q_0 & q_0 \\
\end{array} & 
\begin{array}{cccc}
1 & q_2 & q_1 & q_1 \\
q_2 & 1 & q_1 & q_1 \\
q_1 & q_1 & 1 & q_2 \\
q_1 & q_1 & q_2 & 1 \\
\end{array}
 \\
\end{array}
\right)\ .
\end{equation}
The {\sc 2rsb} solution has a better entropy, that becomes negative at even lower
temperatures and is now very small at $T=0$. The $P(q)$ reads
\beq\begin{split}
\overline{P(q)} &=\lim_{n\to 0} \frac1{n-1} 
\left[ (n-m_1) \d(q-q_0) + (m_1-m_2) \d(q-q_1) + (m_2 -1) \d(q-q_2) \right] \\
&= m_1 \d(q-q_0) + (m_2 - m_1)  \d(q-q_1) + (1 -m_2 ) \d(q-q_2) \ .
\end{split}\eeq
The positivity of $P(q)$ requires $0 < m_1 < m_2 < 1$.

Iterating this procedure produces the correct solution. One ends up with a sequence
of numbers $1 < m_K < m_{K-1} < \cdots < m_2 < m_1 < n$, each one corresponding to
blocks with overlap $q_0 < \cdots < q_K$, and $K\to\io$. 
The equality is reversed in the limit $n\to 0$, 
$m_0 = 0 < m_1 < m_2 < \cdots < m_K < 1 = m_{K+1}$,
and 
\beq
\overline{P(q)} = \sum_{i=0}^K (m_{i+1} - m_i) \d(q-q_i)
\eeq
becomes a continuous functions with support in $[0,\max_i q_i]$.
The function $x(q)$ is a piecewise constant function. In the limit $K\to \io$
it becomes an arbitrary function that must satisfy only the constraint that
$x \in [0,1]$, $q \in [0,1]$, and $\frac{dx}{dq} \geq 0$. The free energy becomes
a functional of $x(q)$, and therefore $x(q)$ (or $P(q)$) is the order parameter of
the transition for the SK model.

The Parisi solution has the following physical interpretation. Equilibrium states
have self-overlap $q_K = q_\mathrm{EA}$. They
are arranged in clusters such that states inside a cluster have mutual 
overlap $q_{K-1}$, but such clusters are arranged in other (super)clusters, 
and states belonging to the same (super)cluster have mutual overlap $q_{K-2}$.
Superclusters are arranged in supersuperclusters, etc.

The complicated mathematical structure of the Parisi solution, once correctly
interpreted, led to a number of non-trivial predictions, that we cannot review
here, but are discussed \eg in \cite{MPV87,BY86,DG06}. At present, it has been
proven that the free energy of the Parisi solution is correct, \ie it is equal
to the true free energy of the SK model below $T_c$ \cite{Ta03}. This proof alone
took twenty years of efforts, and most of the interesting properties of the solution,
even if confirmed by numerical simulations, have not yet been rigorously proven.

\subsection{Susceptibilities}

It is interesting to discuss at this point the behavior of the magnetic susceptibilities that
characterize the phase transitions we have investigated so far.
We will first review the case of the ferromagnet, and then turn to the spin
glass. We will discuss explicitly the case of mean field models, but the discussion applies
also to finite dimensional models, with minor modifications.

\subsubsection{The ferromagnet}

\begin{figure}
\centering
\includegraphics[width=8cm]{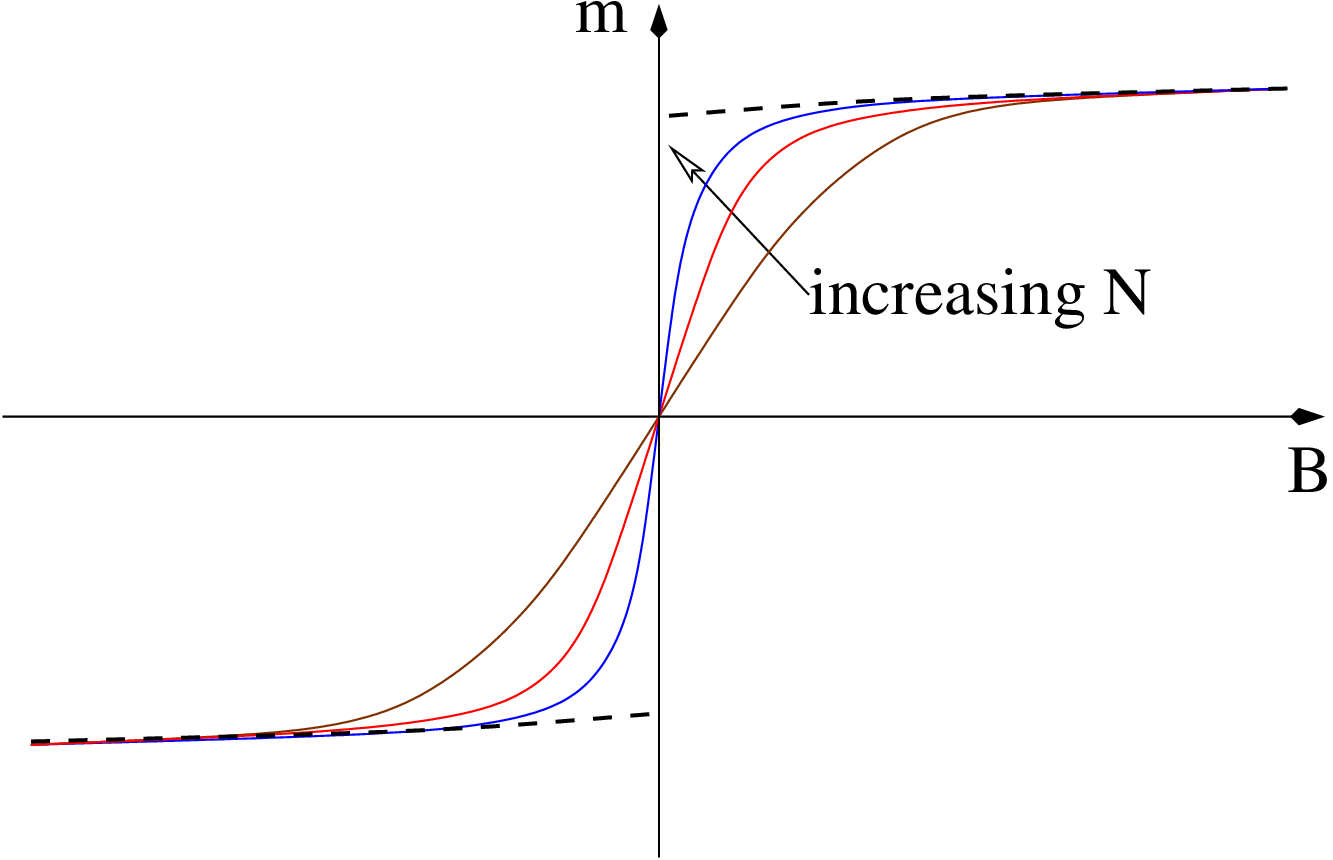}
\caption{
Magnetization $m$ as a function of magnetic field ${\cal B}$ for a ferromagnetic system at $T<T_c$,
for finite $N$ and for $N\to \io$.
}
\label{fig:mh}
\end{figure}

In a ferromagnet, the order parameter is the magnetization $m(T,\BB;N)=\frac1N \sum_i
\la S_i \ra$, or to be more precise
\beq\label{mferro}
m^*(T) = \lim_{\BB\to 0} \lim_{N\to\io}  m(T,\BB;N)  \ .
\eeq
The magnetization below $T_c$ is given by $\pm m^*$ if $\BB \to 0^{\pm}$ {\it
  after} $N \to \io$. The behavior of $m(T,\BB;N)$ is sketched in Fig.~\ref{fig:mh}.
  At finite $N$, the magnetization is an analytic function of $\BB$; but in the limit 
  $N\to \io$, it becomes steep around $\BB=0$ where the singularity develops.
  The susceptibility 
  \beq
  \chi(T,\BB; N) = \frac{dm(T,\BB;N)}{d\BB}
  \eeq
  calculated at $\BB=0$ is the slope of the curve at $\BB=0$ and diverges when $N\to \io$ at all $T < T_c$.
  This quantity is not, however, the thermodynamic magnetic susceptibility, which is instead obtained by taking first the limit
  $N\to\io$, and then $\BB\to 0$, and which is finite as one can check from Fig.~\ref{fig:mh}. 
The quantity $\chi(T,\BB$$=$$0;N$$\to$$\io)$ is the susceptibility in the full Gibbs measure, which
is infinite below $T_c$ because the Gibbs measure is unstable towards the decomposition
in the two pure states with positive and negative magnetization.  Note that because $\chi(T,\BB$$=$$0;N)$ is analytic
at finite $N$ as a function of $T$, and diverges for all $T<T_c$, it follows that $\chi(T,\BB$$=$$0;N$$\to$$\io)$ 
must diverge for $T \to T_c^+$.

We can also compute the susceptibility at finite $N$ in the Gibbs
measure (\ie in absence of external field) using the fluctuation dissipation 
relation
\beq\label{chiFLUC}
\chi(T,\BB=0;N) = \frac{d m}{d \BB} = \frac{\b}N \sum_{ij} \la S_i S_j \ra_c\ ,
\eeq
where $\la S_i S_j \ra_c = \la S_i S_j \ra - \la S_i \ra
\la S_j \ra$ is the connected correlation function.
For large enough $N$, we may think that the Gibbs measure is split between
two states $\a = \pm$, each with weight $w_\a = 1/2$, so we have $\la S_i \ra=0$.
For the fully connected model, using the
clustering property inside each state we get for $i \neq j$:
\beq\label{chiCLUS0}
\la S_i S_j \ra_c = \la S_i S_j \ra = \frac12 [\la S_i \ra_+ \la S_j \ra_+ + \la S_i \ra_- \la S_j \ra_- ] = \frac12 [(m^*)^2 + (-m^*)^2]= (m^*)^2  \ ,
\eeq
and therefore
\beq\label{chiCLUS}
\chi(T,B=0;N) = \b (N-1) [m^*(T)]^2 + \b \ ,
\eeq
which diverges for $N\to\io$ in the low temperature phase.
Note that this result is wrong, because for $T > T_c$ we obtain $\chi(T,B$$=$$0;N$$\to$$\io) = \b$, which does not diverge for $T\to T_c^+$.

Indeed, from the mean field equation $m = \tanh[\b (\BB + m)]$ (which holds only for $N\to\io$) we get, taking the derivative 
of the equation with respect to $\BB$,
\beq
\chi = (1-m^2) \b (1 + \chi)  \hskip1cm
\Rightarrow
\hskip1cm
\chi(T,\BB; N\to\io) = \frac{\b [1 - (m^*(T))^2]}{1 - \b [1-(m^*(T))^2]} \ ,
\eeq
and therefore for $T>T_c=1$ we get $\chi(T,\BB; N\to\io) = \b/(1-\b)$, which diverges for $T\to 1^+$.
In order to get this result from the fluctuation-dissipation relation we need to take into account that the clustering properties only holds
for $N\to\io$. At finite $N$ there are corrections of order $1/N$ to $\la S_i S_j \ra_c$. Because there are $N^2$ terms and a factor of $1/N$
in Eq.~(\ref{chiFLUC}), these $1/N$ corrections affect the finite part of $\chi$. Still, the calculation leading to (\ref{chiCLUS}) is correct
for the divergent term\footnote{
In a short range system, the discussion is very similar. Clustering holds only
for $|i-j| \to \io$, but if the connected correlation is finite in this limit,
the sum is dominated by the large values of $|i-j|$ and the result is the same
as in~(\ref{chiCLUS}).
}.

In summary, the calculation of the leading term of 
$\chi$ in the Gibbs measure and at finite (but large) $N$ can be done, using only the decomposition in pure states and the clustering property. This calculation
shows that $\chi(T;\BB$$=$$0; N)$ is divergent for $N\to \io$ in the whole low temperature phase, which is to be expected from
Fig.~\ref{fig:mh}. But because $\chi$ is divergent for $T<T_c$, it must also diverge for $T \to T_c^+$.
In this way we identify the susceptibility that diverges at the phase transition.

\subsubsection{Spin glasses: linear susceptibilities}

For a spin glass in absence of external field and for a
symmetric distribution of $J$, $P(J_{ij})=P(-J_{ij})$, one has
$\overline{\la S_i S_j \ra_c} =0$ for $i\neq j$. In fact, for $i \neq j$,
\beq
\overline{\la S_i S_j \ra} = \overline{\sum_\a w_\a m_i^\a m_j^\a} \ .
\eeq
Due to the symmetry of the couplings, the probability (over the states $\a$ and the disorder)
that the magnetizations
$m_i^\a$ and $m_j^\a$ have the same or opposite signs are the same, and
the average vanishes.
The term $\overline{\la S_i \ra \la S_j \ra}$ vanishes because
$\la S_i \ra$ and $\la S_j \ra$ have random signs.
The total magnetic susceptibility
is then given by
\beq\label{chieq}
\chi = \b \left(1 - \overline{ \frac1N \sum_i \la S_i \ra^2 }\right) = \b (1 - q^{(1)} ) \ .
\eeq
This susceptibility is that one would measure if the system is prepared at
equilibrium, then a small magnetic field is applied and the new equilibrium state
is reached. A small magnetic field shuffles the free energy of the states, and therefore
in general the new equilibrium states in presence of a field will be very different
from the old ones. As a result, in order to observe the susceptibility (\ref{chieq})
one has to wait a very long time for the system to reach the new
equilibrium states.

One might also be interested  in considering the linear susceptibility {\it of
a single equilibrium state} $\a$. This susceptibility is that one would observe if the
system does not have time to escape its original state after the magnetic field
is applied. Using the fluctuation-dissipation inside the state $\a$ we get, under
the assumption $q_{\a\a} \equiv q_\mathrm{EA}$,
\beq\label{chiLR}
\chi_\a = \b \left(1 - \frac1N \sum_i \la S_i \ra_\a^2 \right) \ ,
\hskip2cm
\chi_\mathrm{LR} = \overline{ \sum_\a w_\a \chi_\a } = \b (1- q_\mathrm{EA}) \ .
\eeq
The name $\chi_\mathrm{LR}$ comes from the fact that this is the susceptibility associated to linear response inside a state.
If $P(q)$ is not trivial, \ie it is not a delta function, $q^{(1)} < q_\mathrm{EA}$ and the two
susceptibilities are different, in particular $\chi_\mathrm{LR} < \chi$. This result is very reasonable.
Once we switch on a small magnetic field $\BB$, the system first responds by acquiring
a small magnetization inside the state $\a$, $m_\mathrm{LR} \sim \chi_\mathrm{LR} \BB$. It will then escape
from state $\a$ and find a better state in presence of $\BB$, which will then give a larger
magnetization $m \sim \chi \BB > m_\mathrm{LR}$. This last property of the spin glass
phase is very important.

Note that the time $\t_N$ needed to change state diverges with $N$ if there is a true
phase transition. Then, we can define (at finite $N$) 
the dynamical magnetic susceptibility
\beq\label{chi_t}
\chi(t) = \frac{dm(t)}{d\BB} = 
\b \left(1 - \overline{ \frac1N \sum_i \la S_i(t) S_i(0) \ra }\right) \ ,
\eeq
where it is assumed that the system is in equilibrium in absence of field
and the field is switched on at $t=0$.
For $1 \ll t \ll \tau_N$, the correlation function $ \la S_i(t) S_i(0) \ra$ decorrelates
inside the state $\a$ to which the initial configuration begins, $ \la S_i(t) S_i(0) \ra \sim \sum_\a w_\a \la S_i \ra_\a \la S_i \ra_\a$. 
Hence, we have
\beq
\chi(1 \ll t \ll \tau_N) = \b \left(1 - \overline{ \frac1N \sum_i \sum_\a w_\a \la S_i \ra_\a \la S_i \ra_\a }\right) = \chi_\mathrm{LR} \ .
\eeq
On the other hand, for $t\gg \t_N$, the system decorrelates completely in the full Gibbs measure:  
$ \la S_i(t) S_i(0) \ra \sim \la S_i \ra \la S_i \ra$, and $\chi(t \gg \t_N) = \chi$, the equilibrium susceptibility.
Recalling that $\t_N$ diverges with $N$, we get
\beq\begin{split}
&\lim_{N\to\io} \lim_{t\to\io} \chi(t) = \chi \ , \\
&\lim_{t\to\io} \lim_{N\to\io} \chi(t) = \chi_\mathrm{LR} \ .
\end{split}\eeq
The behavior of $\chi(t)$ on a time scale $t$ that diverges with $N$ is more 
complicated, and its discussion is behind the scope of these notes.

In either case, {\it for spin glasses the magnetic susceptibilities are finite at the
  transition}. For a continuous transition, 
both $\chi$ and $\chi_\mathrm{LR}$ show a cusp at $T_c$, when $q_\mathrm{EA}$
becomes non-zero. For a discontinuous {\sc 1rsb} transition, $\chi_\mathrm{LR}$ jumps
  at $T_d$ where states appear. However, from (\ref{Pq1rsb}) with $q_0=0$, we have 
$q^{(1)} = (1-m) q_1 = (1-m) q_\mathrm{EA}$, so
$\chi$ is analytic across $T_d$ because $m=1$ and
$q^{(1)}=0$, and has a cusp at $T_K$ where $m \neq 1$.

\subsubsection{The static spin glass susceptibility}

The diverging susceptibility is more complicated in the spin glass case.
In order to keep the discussion more general, we will avoid 
performing explicitly the average over the disorder and consider a single 
(large enough) sample, using the fact the susceptibilities are self-averaging
quantities. In this way the following discussion can be extended
straightforwardly to glassy systems without quenched disorder.

The analog of the magnetization for a spin glass system is the self-overlap
$q_\mathrm{EA} = q_{\a\a}$, which becomes non-zero when states appear. As we discussed
in section~\ref{sec:realreplica}, to compute $q_\mathrm{EA}$ 
we should put a coupling between replicas
in order to force them to be in the same state. We consider two replicas and
choose a coupling $\d H = -\ee \sum_i S_i^1 S_i^2$. Then (\ref{mferro}) becomes
\beq
q_\mathrm{EA} = \lim_{\ee \to 0} \lim_{N\to\io} \frac1N \sum_{i} \la S_i^1 S_i^2
\ra_{\ee} \ .
\eeq
It is worth noting at this point that the quantity above is really the self-overlap of the states
and not $q^{(1)}$, which is the thermodynamic average of $q_{\a\b}$. The reason is the
following. Suppose we are in a phase where the $P(q)$ is not trivial (for instance, we consider
the SK model at $T<T_c$ or the $p$-spin model for $T<T_K$). In this case, the Gibbs measure is
dominated by the set of states that have the lowest intensive free energy. The fluctuations in the weight
are due to $1/N$ corrections to the free energy. We can write $f_\a = f_\mathrm{min} + \D f_\a/N$, and therefore
\beq
w_\a = \frac{e^{-\b N f_\a}}{\sum_\a e^{-\b N f_\a}} = \frac{e^{-\b \D f_\a}}{\sum_\a e^{-\b \D f_\a}} \ .
\eeq
It can be shown~\cite{MPV87} that these weights have finite fluctuations, and although the number of states 
with finite weights is infinite for $N\to\io$, the number of states that
are needed to cover a fraction $1-\ee$ of the total weight is finite for any $\ee$.
The partition function of two replicas with the coupling discussed above can thus be written as
\beq
Z_\ee = e^{-2 \b N f_\mathrm{min}} \sum_{\a\b} e^{-\b (\D f_\a + \D f_\b) } e^{\b \ee N q_{\a\b}} \ . 
\eeq
Clearly, because the number of relevant terms in the sum is finite, the weights are finite, and $q_\mathrm{EA} = q_{\a\a} > q_{\a \neq \b}$,
the coupling term for any $\ee>0$ makes the contribution of the terms with $\a = \b$ exponentially bigger than that of
$\a \neq \b$. The replicas are therefore in the same state and the average overlap is given by $q_\mathrm{EA}$.
On the contrary, $q^{(1)}$ is related to a full average in the Gibbs measure in absence of any coupling\footnote{By a similar argument
one can show that $P(q) = \d(q-q_\mathrm{EA})$ for any $\ee > 0$}.

The susceptibility associated to $q_\mathrm{EA}$ is therefore\footnote{The factor
  $\b$ is sometimes omitted or replaced by $\b^2$ in the literature.} (at finite $N$ and 
$\ee \to 0$, when the replicas are uncoupled):
\beq\label{chiSG}
\chi_\mathrm{SG} = \frac{dq_\mathrm{EA}}{d\ee} = \frac{\b}{N} \sum_{ij} \big[\la S_i^1 S_j^1 \ra
\la S_i^2 S_j^2 \ra - \la S_i^1 \ra\la S_j^1 \ra
\la S_i^2 \ra\la S_j^2 \ra \big] \ .
\eeq
Performing the decomposition in pure states and using the clustering property
it is easy to show that
\beq
\chi_\mathrm{SG} = \b N \left[ \sum_{\a\b} w_\a w_\b (q_{\a\b})^2 -
  \sum_{\a\b\g\d}w_\a w_\b w_\g w_\d q_{\a\b} q_{\g\d} \right] =
\b N \big[ q^{(2)} - (q^{(1)})^2 \big] \ .
\eeq
Then $\chi_\mathrm{SG}$ is divergent in the spin glass phase where the $P(q)$ is not trivial and $q^{(k)} \neq 0$.
This happens for $T<T_c$ in the SK model (and in all models with a second-order spin glass transition), 
and for $T<T_K$ in the spherical $p$-spin model (and in all models with a discontinuous transition).
In all these cases the partition function is dominated by the low free energy states, so that an infinitesimal 
coupling suffices to force two replicas to be in the same state.

\subsubsection{The dynamic spin glass susceptibility}

The static spin glass susceptibility defined above does not diverge
between $T_d$ and $T_K$ in the discontinuous case, as it should, because it is a thermodynamic quantity and
we showed that the thermodynamics has no singularity at $T_d$.
How can we obtain a susceptibility that diverges at the clustering transition?
For this, we would like to probe each state separately and not the full Gibbs measure.

Similarly to what we did in Eq.~(\ref{chi_t}), the solution is to consider a
dynamical susceptibility. We define it as follows:
\beq
\chi_\mathrm{SG}(t) = \frac{\b}{N} \sum_{ij} \big[\la S_i(t) S_j(t)
 S_i(0) S_j(0) \ra - \la S_i(t) S_i(0) \ra\la S_j(t)S_j(0) \ra ]  \ .
\eeq
Note that $\chi_\mathrm{SG}(0) = 0$. Again, the time needed to change state, $\t_N$, diverges with $N$.
For $t \gg \t_N$ the system decorrelates in the full Gibbs measure and
\beq
\chi_\mathrm{SG}(t \gg \t_N) = \frac{\b}{N} \sum_{ij} \big[ 
\la S_i S_j \ra
\la S_i S_j \ra - \la S_i \ra \la S_i \ra\la S_j \ra \la S_j \ra \big] = \chi_\mathrm{SG}  \ ,
\eeq
so that it reduces to the static spin glass susceptibility, and is finite between $T_K$ and $T_d$.

In the region $1 \ll t \ll \t_N$, the system is only able to decorrelate within one state, so the dynamical susceptibility is
\beq\begin{split}
\chi_\mathrm{SG}(1 \ll t \ll \t_N) & = \frac{\b}{N} \sum_{ij} \big[\sum_a w_\a \la S_i S_j \ra_\a
\la S_i S_j \ra_\a - \sum_\a w_\a \la S_i \ra_\a \la S_i \ra_\a
\sum_\b w_\b \la S_i \ra_\b \la S_j \ra_\b \big]  \\
&= \b N  \big[ \sum_a w_\a q_{\a\a}^2 - 
\sum_{\a\b} w_\a w_\b q_{\a\b}^2 \big] = 
\b N \big[ ( q_\mathrm{EA} )^2 - q^{(2)}  ]  = \chi_\mathrm{SG,LR} \ .
\end{split}\eeq
Now, $\chi_\mathrm{SG,LR}$ also diverges in the discontinuous {\sc 1rsb} case between
$T_d$ and $T_K$, where $q_\mathrm{EA} \neq 0$ and $q^{(2)}=0$.

In summary, we have $\chi_\mathrm{SG}(0)=0$ and
\beq\begin{split}
&\lim_{N\to\io} \lim_{t\to\io} \chi_\mathrm{SG}(t) = \chi_\mathrm{SG} \ , \\
&\lim_{t\to\io} \lim_{N\to\io} \chi_\mathrm{SG}(t) = \chi_\mathrm{LR,SG} \ .
\end{split}\eeq
Note that in the cluster phase $T_K < T < T_d$, $\chi_\mathrm{SG}(t)$ has a very
peculiar behavior\footnote{This behavior has been recently observed in structural
  glasses, for which  $\chi_\mathrm{SG}(t)$ is usually called $\chi_4(t)$ in the literature.}: 
it grows up to a very large ($\sim N$) value for $t \sim \t_N$, and
then decays back to a finite value for really large times $t \gg \t_N$. By contrast,
in the thermodynamic spin glass phase, when $T<T_K$, $\chi_\mathrm{SG}(t)$ is of order $N$ for all times $t \gg 1$.

\subsection{Exercises}

\begin{enumerate}
\item \label{ex:1.1}
{\bf Dynamics of the fully connected Ising model} - Write a program to simulate the Metropolis dynamics of the
fully connected Ising model, Eq.~(\ref{HFCIM}). The dynamics is defined as follows. At each step try to flip a random
spin, $S_i \to -S_i$. Draw a random number $x$ uniformly in $[0,1]$ and accept the move if $x < \exp[-\beta (H[S']-H[S]) ]$,
where $S'$ is the configuration with $S_i$ flipped.
(Hint: in addition to the spins $S_i$, keep in memory the total magnetization $M = \sum_i S_i$, updating it at each step. Use it
to compute the energy change.)
Using the program, simulate the time evolution of a small number of spins in the low temperature phase. Check that the
system jumps from one state to the other and try to observe the scaling of the persistence time with $N$.
\item \label{ex:1.2}
{\bf The Random Energy Model} - The Ising $p$-spin Hamiltonian (\ref{Hpspin}) for Ising spins is, for a given 
configuration $S$ (with $S_i = \pm 1$), a Gaussian random variable with zero average.
\begin{itemize} 
\item Show that the covariance $\overline{H[S] H[S']}=\frac12 N Q(S,S')^p$, where $Q(S,S')=\frac1N \sum_i S_i S'_i$ is the overlap
of $S$, $S'$.
\item Deduce that for $p\to\io$ the energy of a configuration is a Gaussian random variable with variance $N/2$ and that
the energy of different configurations are uncorrelated (this is trivial).
\end{itemize}
Therefore in the large $p$ limit the $p$-spin converges to the {\it Random Energy Model} (REM): there are $2^N$ configuration
with energies $E_i$ that are Gaussian and independent, with zero average and variance $N/2$.
\begin{itemize} 
\item Compute the average number of configurations that have energy $E$, call it $\Omega(E)$; 
compute the complexity $\Si(e)= \frac1N \log \Omega(N e)$
(the result is $\Si(e)=\log 2 - e^2$). Show that the complexity vanishes for $e < e_K$; compute~$e_K$.
\item The previous result implies that for $e>e_K$ the number of configurations is very large; conversely, for $e<e_K$ 
the probability of finding a configurations is exponentially small ($\Si<0$) and we can assume that there are no configurations.
Defining the partition function as
\beq
Z(T) = \sum_{i=1}^{2^N} e^{-E_i/T} \sim \int_{-N |e_K|}^{N e_K} dE \Omega(E) e^{-E/T} \ ,
\eeq
compute the free energy $f(T)$ via a saddle point. Show that there is a phase transition at a certain temperature $T_K$,
in particular that the second derivative of $f(T)$ (the specific heat) has a jump.
\end{itemize}
The REM has exactly the same phenomenology of the $p$-spin, except for the fact that in the REM {\it each configuration
is a state} (we did not show it but it can be done for instance by showing that, for the $p$-spin model, the overlap in a state goes to 1 when $p\to\io$). 
Therefore $T_d$ does not exist since states are stable at all temperatures, and moreover the internal entropy
$s(T)$ is identically zero. See \cite{De81,MPV87} for details.
\item \label{ex:1.3}
{\bf Zero-temperature complexity of the $p$-spin model} -
Consider the zero-temperature limit of (\ref{phi1RSBpspin}). Show that for $\b \to \io$ it has a finite
limit at fixed $\mu = \b m$ and with $q = 1 - \a T$, with $\a$ of order 1, given by
\beq
\phi_\mathrm{1RSB} = -\frac14 (p \a + \mu) + \frac1{2\mu} \log \frac{\a}{\a + \mu} \ .
\eeq
Deduce that $\a$ satisfies the equation
$\a^2 + \mu \a - 2/p = 0$,
and compute its positive solution (recall that $q \leq 1$).
Now show that
\beq\begin{split}
& e(\mu) = \lim_{\b\to\io , \b m=\mu} f(m,T) = \partial_\mu [\mu \phi_\mathrm{1RSB}] \ , \\
& \Si(\mu) = \mu^2 \partial_\mu \phi_\mathrm{1RSB} \ ,
\end{split}\eeq
and compute their explicit expressions as function of $\mu$. Check that the physical region
corresponds, for $p=3$, to $\mu \in [0.577,0.884]$, that the threshold energy is
$e_{th} = -\sqrt{4/3}\sim -1.155$ and that the ground state energy is $e_0 \sim -1.172$.
Show that in general $e_{th} = -\sqrt{2(p-1)/p}$.
\item \label{ex:1.4}
{\bf Complexity of the $3$-spin model} - 
Starting from Eq.~(\ref{phi1RSBpspin}), and using (\ref{mcomplexity}), compute $\Si(f)$ for the spherical $3$-spin
model at different temperatures, \eg $T=0.63 > T_d$, $T_d> T=0.6 > T_K$, $T=0.5 < T_K$ (it can be done by 
Mathematica\footnote{A Mathematica sheet that does the job can be downloaded from
\url{http://www.lpt.ens.fr/~zamponi}, section ``Teaching''.}
or writing a program in C/Fortran/...). 
There are two possibilities to solve the equation for $q$: {\it i)} use that it is cubic and write the explicit solution;
{\it ii)} write it as $q=f(q)$ and solve it iteratively.
\item \label{ex:1.5}
{\bf Stability of the {\sc 1rsb} solution for the $3$-spin model} - 
The aim of this exercise is to obtain the lines $m_{\rm th}$ and $m^*$ from the stability matrix of the {\sc 1rsb} solution
used in the previous exercise. Most of these results have been obtained in \cite{FLPR12} which you can consult in case
of difficulties.
We consider $m$ coupled replicas, in the limit of zero coupling.
Following the discussion of section~\ref{sec:realreplicapspin}, the free energy is given by $X(Q)$ in Eq.~\eqref{reppart}
and the {\sc 1rsb} ansatz corresponds in this language to choose $Q_{ab} = \d_{ab} + q (1-\d_{ab})$. Furthermore,
we have to consider the solution with the largest non-zero $q = q^* > 0$.
\begin{itemize}
\item
Keep in mind that the matrix $Q_{ab}$ is symmetric, $Q_{ab} = Q_{ba}$. Take the elements with $a<b$ as the independent ones.
Show that the stability matrix of the second derivatives is
\beq
M_{a<b;c<d} = \frac{d^2 X}{dQ_{a<b} dQ_{c<d}} = \frac{\b^2}2 p (p-1) Q_{ab}^{p-2} ( \d_{ac} \d_{bd} + \d_{ad}\d_{cb})
- Q^{-1}_{ac} Q^{-1}_{bd} - Q^{-1}_{ad} Q^{-1}_{cb} \ .
\eeq
\item
 Consider a small perturbation around the non-trivial {\sc 1rsb} solution. Show that in this case
\beq
Q^{-1}_{ab} =  \frac1{1-q} \left[ \d_{ab} - \frac{q}{1+(m-1)q} \right]
\eeq
and the matrix $M$ takes the form
 \beq\begin{split}
M_{a<b;c<d} &= M_1 \frac{    \d_{ac} \d_{bd} + \d_{ad}\d_{cb} }2 + M_2 \frac{ \d_{ac} + \d_{ad} + \d_{bc} + \d_{bd}}4 + M_3 \ , \\
M_1 &= \b^2 p ( p-1) q^{p-2} - \frac{2}{(1-q)^2} \ , \\
M_2 & = \frac{4}{(1-q)^2}\frac{q}{1+(m-1)q}  \ , \\
M_3 &= -2 \frac{1}{(1-q)^2} \left( \frac{q}{1+(m-1)q} \right)^2 \ .
\end{split} \eeq
\item Following \cite{BM79,CS92}, show that the matrix above has three independent eigenvalues,
 \beq\begin{split}
\l_R &= M_1 \ , \\
\l_L & = M_1 + (m-1) (M_2 + m M_3) \ , \\
\l_A & = M_1 + \frac{m-2}2 M_2 \ . 
\end{split} \eeq
\item
Show that $\l_L \propto \frac{\partial^2 \phi_{\rm 1RSB}(m,q,T)}{\partial q^2}$. From this deduce that one must have 
$\l_L=0$ on the line $m^*(T)$ where the non-trivial
solution $q^*$ disappears.
\item Fix $p=3$ and compute the three eigenvalues numerically. Check numerically that $\l_L$ vanishes on the line $m^*$. 
Next, compute the line $m_{\rm th}(T)$ corresponding to threshold states as the line where $\l_R =0$. Show that you reproduce
the results of figure~\ref{fig3:dia}.
\end{itemize}

\end{enumerate}


\clearpage

\section{Diluted models and optimization problems}
\label{sec:optim}

\subsection{Definitions}

\subsubsection{Statistical mechanics formulation of optimization problems}

The theory of computational complexity~\cite{PS98} establishes a classification
of constraint satisfaction problems (CSP) according to their difficulty in the 
worst case. For concreteness it is convenient to introduce three problems we shall 
use as running examples in the following:
\begin{itemize}
\item[$\bullet$] $k$-XORSAT. 
Find a vector $\vec{x}$ of boolean variables satisfying the linear equations 
$A \vec{x}=\vec{b} \ ({\rm mod} \ 2)$, where each row of the $0/1$ matrix $A$ 
contains exactly $k$ non-null elements, and $\vec{b}$ is a given boolean
vector.
\item[$\bullet$] $q$-coloring ($q$-COL). 
Given a graph, assign one of $q$ colors 
to each of its vertices, without giving the same color to the two extremities 
of an edge.
\item[$\bullet$] $k$-satisfiability ($k$-SAT). 
Find a solution of a boolean formula 
made of the conjunction (logical AND) of clauses, each made of the disjunction 
(logical OR) of $k$ literals (a variable or its logical negation).
\end{itemize}
Each of these problems admits several variants, for instance
\begin{itemize}
\item[$\bullet$] {\bf  Decision:}
one has to assert the existence or not of a solution,
for instance a proper coloring of a given graph.
\item[$\bullet$] {\bf Sampling:}
one has to sample the solution according to a given distribution (e.g. uniform over all the solutions), 
and estimate for instance the total number of solution.
\item[$\bullet$] {\bf Optimization:}
one has to discover optimal configurations; these are solutions, if present, or otherwise 
configurations 
minimizing the number of violated constraints: for instance colorings
minimizing the number of monochromatic edges.
\end{itemize}
The decision variant of the 
three examples stated above fall into two distinct complexity classes: 
$k$-XORSAT is in the P class, while the two others are NP-complete for 
$k,q \ge 3$.
This means that the existence of a solution of the XORSAT problem can
be decided in a time growing polynomially with the number of variables,
for any instance of the problem; one can indeed use the Gaussian elimination 
algorithm. On the contrary no fast algorithm able of solving every coloring or
satisfiability problem is known, and the existence of such a polynomial
time algorithm is considered as highly improbable.

A general statistical mechanics formulation of a CSP is the
following. One is given $N$ variabls $\s_i$, $i=1,\cdots,N$, 
that might be Ising spins (Boolean
variables) or Potts spins, real numbers, or more complicated variables.
Then, one is given a set of $M$ constraints labeled by $a=1,\cdots,M$; 
each constraint involves a certain set of variables that is denoted by
$\s_a = (\s_{i_1},\cdots,\s_{i_{K_a}})$. 
The constraint is a function $\psi_a(\s_a)$ that is $0$
when $\s_a$ does not satisfy the constraint and $1$ otherwise.
In the following we will use also the words ``clause'' or ``test''
to denote a constraint.

The CSP consists in finding a 
configuration $\s$ verifying the condition $\prod_a \psi_a(\s_a) = 1$;
the number of solutions is
\beq\label{ZTzero}
Z = \sum_{\s} \prod_a \psi_a(\s_a) \ .
\eeq

The optimization version of the same
problem is obtained by replacing the hard constraint $\psi_a$ with
a ``soft'' one, $\psi_a^{\b} = e^{-\b}$ if the constraint is not satisfied
and $\psi_a^{\b} = 1$ otherwise. This corresponds to a problem
at finite temperature $\b$ such that the energy of a violated constraint is
equal to $1$. The optimization problem is that of finding the ground state,
while the decision problem amount to decide whether a ground state with zero energy
exists. In both cases we are interested in working at zero (or very small)
temperature.

The three illustrative examples presented above admits a simple representation
in this formalism:
\begin{itemize}
\item[$\bullet$] $k$-XORSAT. The degrees of freedom of this CSP are boolean
variables that we shall represent, following the physics conventions, by
Ising spins, $S\in\{-1,+1\}$. 
Each constraint involves a subset of $k$ variables,
$S_a=(S_{i_a^1},\dots,S_{i_a^k})$, and reads 
$\psi_a(S_a) = \I( S_{i_a^1} \dots S_{i_a^k}  = J_a)$, where here and in
the following $\I(\cdot)$
denotes the indicator function of an event and $J_a \in\{-1,+1\} $ is a given
constant. This is equivalent to the definition given in 
the introduction: defining $x_i,b_a \in \{0,1 \}$ such that 
$S_i = (-1)^{x_i}$ and $J_a=(-1)^{b_a}$, the constraint imposed by $\psi_a$
reads $x_{i_a^1} + \dots + x_{i_a^k} = b_a \pmod 2$, which is nothing but 
the $a$'th row of the matrix equation $A \vec{x} = \vec{b}$. The addition
modulo 2 of Boolean variables can also be read as the binary exclusive OR 
operation, hence the name XORSAT used for this problem.

\item[$\bullet$] $q$-COL. Here $\s\in\{1,\dots,q\}$ is the set of allowed
colors on the $N$ vertices of a graph. Each edge $a$ connecting the vertices
$i$ and $j$ prevents them from being of the same color: 
$\psi_a(\s_i,\s_j) = \I(\s_i \neq \s_j)$.

\item[$\bullet$] $k$-SAT. As in the XORSAT problem one deals with Ising 
represented boolean variables, but in each clause the XOR operation between
variables is replaced by an OR between literals (i.e. a variable or its 
negation). In other words a constraint $a$ is unsatisfied only when all 
literals evaluate to false, or in Ising terms when all spins $S_a=(S_{i_a^1},\dots,S_{i_a^k})$
involved in the constraint take their wrong values, that we denote $J_a=(J_{i_a^1},\dots,J_{i_a^k})$:
 $\psi_a(S_a) = \I (S_a \neq J_a) =   1-\I ( S_a = J_a )$.

\end{itemize}

In all these cases we can define a Hamiltonian $H = \sum_{a=1}^M E_a(\s_a)$,
such that $E_a = 0$ if the constraint is satisfied and $E_a = 1$ otherwise.
Then, $\psi^\b_a = \exp(-\b E_a)$ and $\psi_a$ corresponds to $\b\to\io$.
For instance, in the case of the $q$-COL, the Hamiltonian corresponds to
a Potts antiferromagnet $H=\sum_{(ij)} \d(\s_i,\s_j)$. In the following we
will often drop the explicit dependence on $\b$ to lighten the notation.

\begin{figure}
\includegraphics[width=6cm]{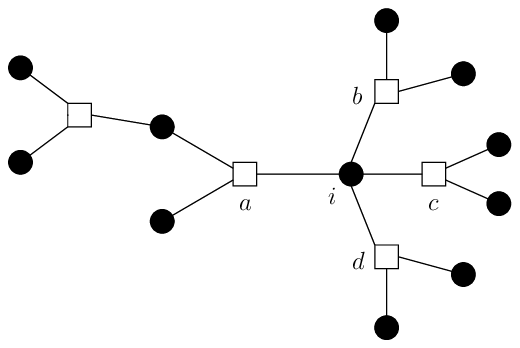}
\caption{An example of factor graph. The neighborhoods are for instance 
$\partial i = \{a,b,c,d\}$ and $\partial i \setminus a = \{b,c,d\}$}
\label{fig_factor}
\end{figure}

Factor graphs~\cite{Ja00} provide an useful representation of a CSP.
These graphs (see Fig.~\ref{fig_factor} for an example) have two kind of 
nodes. Variable nodes (filled circles on the figure) are associated to
the degrees of freedom $\s_i$, while constraint nodes
(empty squares) represent the clauses $\psi_a$. An edge between constraint
$a$ and variable $i$ is drawn whenever $\psi_a$ depends on $\s_i$. 
The neighborhood $\partial a$ of a constraint node is the set of variable 
nodes $(i_1,\cdots,i_{K_a})$ that appear in $a$. 
Conversely we will denote by $\partial i$ the set of all constraints
$(a_1,a_2,\ldots)$ in which variable $i$ is involved. We shall conventionally use the indices
$i,j,\dots$ for the variable nodes, $a,b,\dots$ for the constraints,
and denote $\setminus$ the subtraction from a set. 
The graph distance
between two variable nodes $i$ and $j$ is the number of constraint nodes
encountered on a shortest path linking $i$ and $j$ (formally infinite if the
two variables are not in the same connected component of the graph).

Note that if the constraints only involve two variables $i$ and $j$
(as for instance in the SK model, in $2$-XORSAT or in $q$-COL),
then each constraint is equivalent to a link connecting $i$ and $j$
and the factor graph reduces to a standard graph.

\subsubsection{Random optimization problems: random graphs and hypergraphs}

The common notion of computational complexity, being based on worst-case 
considerations, could overlook the possibility that ``most'' of the
instances of an NP problem are in fact easy and that the difficult cases are
very rare. Random ensembles of problems have thus been 
introduced in order to give a quantitative content to this notion of typical 
instances; a
property of a problem will be considered as typical if its probability
(with respect to the random choice of the instance) goes to one in the limit 
of large problem sizes. Of course the choice of a distribution over the instances
is arbitrary and could not reflect the properties of the instances that relevant
for a given practical application. Still, the introduction of a distribution over
the instances allows to formulate the problem in terms of the statistical mechanics
of a spin-glass-like model. We will see that this formulation provides important
insight in the properties of difficult instances of these problems.

An instance of a random CSP is defined by two objects: the underlying
factor graph and,
as in fully connected models, the set of {\it couplings} $J$ appearing in the constraints
(\eg the right hand side of an equation in XORSAT). 
Both the factor graph and the couplings can be taken as random variables to define
a probability distribution over instances.
Recall that we have $N$ variable nodes and $M$ constraint nodes. In the statistical
mechanics approach we will be
interested in the thermodynamic limit of large
instances where $N$ and $M$ both diverge with a fixed ratio 
$\alpha=M/N$.

Among many possible ensembles of graphs, two have been investigated in great
detail:
\begin{itemize}
\item {\it Random regular graphs} (or fixed connectivity)~\cite{Wo99}: each constraint involves $k$ 
distinct variables
($k$ is a free parameter for (XOR)SAT and $k=2$ for COL),
and each variable enters in {\it exactly} $c$ different constraints.
Uniform probability is given to all graphs satisfying this property.
Note that one must have $M k = N c$, \ie $c= k M/N = k \a$.
\item {\it Erd\"os-R\'enyi random graphs}~\cite{Ja00}:
For each of the $M$ clauses $a$ a $k(\geq2)$-uplet of distinct 
variable indices $(i_a^1,\dots,i_a^k)$ is chosen uniformly at random among the
$\binom{N}{k}$ possible ones.
For large $N,M$ the degree of a variable node of the factor graph converges to 
a Poisson law of average $\alpha k$.
To compare with regular graphs we shall use the notation $c=k \alpha$ for the average connectivity. 
\end{itemize}
In principle one might allow also the connectivity of the constraints to be
a random variable but we do not discuss such case here. Note that the limit $c\to\io$ with
a proper scaling of the couplings gives back the fully connected model. In this limit, the fluctuations
of $c$ in Erd\"os-R\'enyi graphs can be neglected and the two ensembles of graph are equivalent.

Random (hyper)graphs
have many interesting properties in this limit~\cite{Ja00}.
In particular, picking at random one variable node $i$ and isolating the subgraph 
induced by the variable nodes at a graph distance smaller than a given constant
$L$ yields, with a probability going to one in the thermodynamic limit, a
(random) tree. This tree can be described by a Galton-Watson branching 
process: 
the root $i$ belongs to $l$ constraints, where $l$ is a Poisson random variable
of parameter $\alpha k$ (or $l=\a k$ in the fixed connectivity case). The variable nodes
adjacent to $i$ give themselves birth to new constraints, in numbers which
are independently Poisson distributed with the same parameter. This
reproduction process is iterated on $L$ generations, until the variable nodes
at graph distance $L$ from the initial root $i$ have been generated.

The algorithm to construct Erd\"os-R\'enyi graphs is trivial, because it is given by the definition. 
Fixed connectivity
graphs can be constructed as follows: first one attach to each variable node a number
$c$ of links, thus obtaining $Nc$ links. These links have to be connected to the 
$M k = Nc$ links attached to constraint nodes. To do this, one simply numbers the links
from $1$ to $Nc$ and then pick up a random permutation of these numbers in order to
decide which of the variable links has to be attached to a given constraint link.
The resulting graph however might have variables that are connected twice or more to
the same node. In this case the permutation is discarded and a new one is picked until
a good graph is reached. In practice the probability of such event is small if $N$ is
large and $c$ not too large, so that the procedure converges quickly to a good graph.

\subsubsection{Connectivity-temperature phase diagram}

\begin{figure}
\includegraphics[height=6cm]{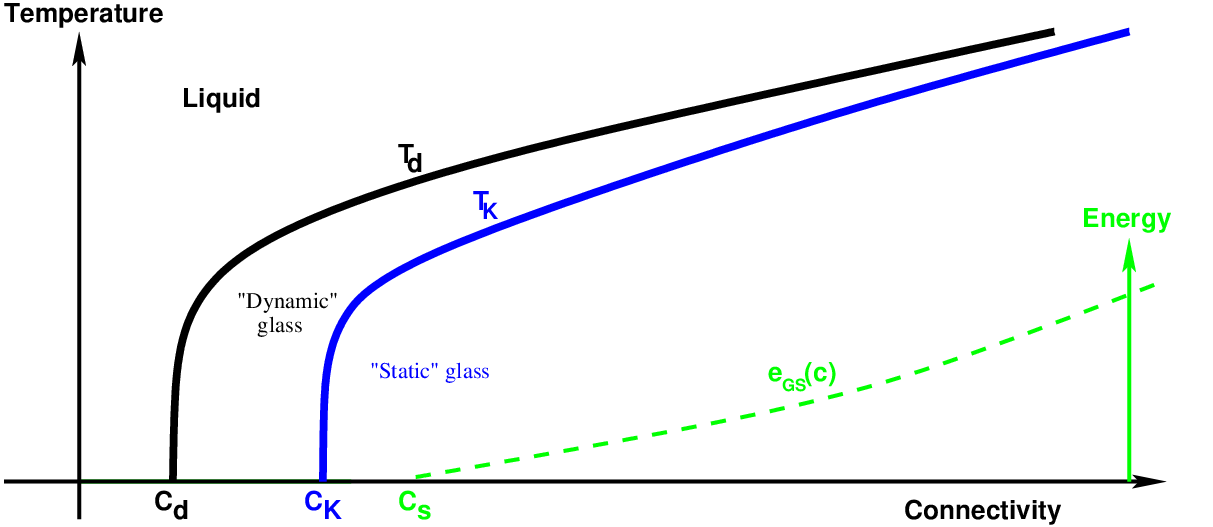}
\caption{Sketch of the phase
    diagram in the coloring problem at finite temperature (from
    \cite{KZ08}). At $T_d$, the system falls out of equilibrium
    (``dynamic'' transition). At $T_K$ the system undergoes a
    ``static'' glass transition. 
    $e_{GS}$ represents the ground state energy, which is nonzero above
 the connectivity $c_s$.}
\label{fig:Tc}
\end{figure}

In the previous sections we discussed the appearance of an exponential
number of states as a function of the temperature, therefore focusing 
on the free energy $f$ of the states.
In the case of a random CSP, the control parameters are the temperature $T$
and the connectivity of the underlying graph $c$ (or equivalently the ratio of constraints
per variable $\a = c/k$), and we are mostly interested
in the $T=0$ limit.

We anticipate that the typical phase diagram of a random CSP looks like the
one in figure~\ref{fig:Tc}. The fully connected case corresponds to the limit of large
$c$. In this case, as a function of temperature, we have shown that there is a ``clustering''
transition at $T_d$ and a spin glass transition at $T_K < T_d$. Both $T_d$ and $T_K$ depend
on $c$ and they vanish at some given values of $c$, $c_d$ and $c_K$ respectively.
Therefore we expect that for connectivities below $c_d$, there is a unique cluster of
ground states of zero energy (solutions), while for $c>c_d$ these ground states are
arranged in many clusters, each cluster being characterized by the number of solutions $\NN_{in}$ 
belonging to it. We call ``internal entropy'' of a cluster the logarithm $s = \frac1N \log \NN_{in}$,
and again we call complexity $\Si$ the logarithm of the number of clusters.

In the case of the $p$-spin model we used the modified partition
function (\ref{Zm}) in order to compute the complexity.
We now comment briefly on the way this general
formalism is applied to constraint satisfaction problems (CSP).
First we split the free energy of a state in $f= e - T s$,
and we introduce a complexity as a function of $e$ and $s$,
$\Si(e,s)= \frac1N \log \NN(e,s)$.
Then we rewrite (\ref{Zm}) as
\beq\label{Zsplit}
Z_m = \int de ds \, e^{N [\Si(e,s) - \b m (e-T s)]} \ .
\eeq
It would have been useful to introduce two parameters, one conjugated to
energy and the other to entropy; however, in general the partition function
above, that contains only $m$, is easier to compute (\eg using replicas as we already discussed).
Therefore, since we have only one parameter $m$, we cannot
reconstruct the full $\Si(e,s)$ via a double Legendre transform.
The point we want to discuss now is that depending on how we 
take the $T\to 0$ limit, we can obtain information on the entropy
or the energy of the states.

For a satisfiable instance of a CSP,
the law defined in (\ref{ZTzero}) is the
uniform distribution over the solutions of the CSP, and the partition
function counts the number of such solutions.
If we take the limit $\b\to\io$ at fixed $m$ of (\ref{Zsplit}), the term
$\exp[ -N\b m e]$ restricts the integral to $e=0$ and we get
\beq
Z_m = \int ds \, e^{N [\Si_s(s) + m s]} \ ,
\eeq
where $\Si_s(s)=\Si(e=0,s)$.
We can define a ``free entropy''
\beq\label{fentropy}
\SS(m) = \frac1N \log Z_m = \max_s [\Si_s(s) + m s] \ ,
\eeq
from which we can compute $\Si_s(s)$ by a Legendre transform.
This ``entropic''
method allows to obtain information on the distribution of the entropies
of the zero energy states, when they exist.
This approach is however ill-defined for unsatisfiable instances. In this
case Eq.~(\ref{Zsplit}) gives
\begin{equation}
\SS(m) = \max_{s,e} [ \Sigma(s,e) + m (s-\beta e)] \ .
\end{equation}
If one takes now the limit $\beta \to \infty$ in the region where $e>0$, 
the complexity term becomes subdominant and we do not get much
information.

In order to obtain a meaning result, one has to
take the limit $m\to 0$ and $\b\to\io$ simultaneously,
in such a way that the product $\beta m$,
usually denoted $y$, remains finite. One therefore obtains
\begin{equation}
\SS_{\rm e}(y) = \max_e [\Sigma_e(e)- y e] \ , \;\;\;\;\;
\Sigma_e(e) \equiv
\max_s \Sigma(s,e)\, .
\end{equation} 
This ``energetic'' cavity approach allows to obtain the distribution 
$\Si_e(e)$ of the energies of the states (irrespective of their entropy) and
allows in particular to compute the ground state energy
of the problem~\cite{MeZe,MeMeZe}.

We expect (and it can be verified by the explicit solution of these models)
that the static value $m_s(T,c)$, that results from optimization of the free energy,
will be 1 in the ``liquid'' (or ``paramagnetic'') high temperature
phase. Then $m_s(T=0,c)=1$ on the $T=0$ line below $c_d$. On increasing 
$c$ above $c_d$, $m_s(T=0,c)$ will become smaller than one and decrease,
still remaining finite. Only at $c= c_s$, $m_s(T=0,c_s)=0$, signaling the transition
to the unsatisfiable phase. Above $c_s$, $m_s(T,c)$ vanishes linearly in $T$
for $T\to 0$, defining the corresponding $y$. For this reason, the entropic cavity
method is mostly appropriate below $c_s$, while the energetic cavity method
is mostly appropriate above $c_s$. 

Both methods can be used to detect $c_s$.
Indeed, $c_s$ is the point where solutions at $e=0$ disappear, hence
$\max_s \Si_s(s) = \max_{s} \Si(e=0,s)$ vanishes at $c_s$. In the entropic
method, $\Si_s(s)$ is computed directly and its maximum corresponds to $m=0$.
In the energetic method, in the satisfiable region, we can take a second limit
$y \to \infty$ (after $\beta \to \infty$ at fixed $y = \b m$) to concentrate on the
states with $e=0$. The complexity computed in this limit is
$\Sigma_e(0) = \max_s \Sigma(s,e=0)$, so that it gives back the maximum of the entropic
complexity. In other words the procedure $y \to \infty$ after $\beta
\to \infty$ is equivalent to perform the entropic computation with a Parisi
parameter $m=0$, \ie to weight all the pure states in the same way,
irrespectively of their sizes, which is the correct way to
determine the satisfiability threshold $c_{s}$.
The determination of the clustering transition is more subtle. A calculation of
$c_{d}$ using the energetic method was first performed in~\cite{MeZe,MeMeZe}; this 
corresponds to the appearance of a solution of the 1RSB equations with $m=0$.
Later it has been shown that the calculation of $c_d$ at $m=1$ can be performed using the
entropic cavity method~\cite{KrMoRiSeZd}; this is the equilibrium clustering threshold, which
can be related to a dynamical transition at zero temperature as a function of $c$.

\subsection{XORSAT: clustering and SAT/UNSAT transition}
\label{sec:XORSAT}

Before discussing the cavity method, we would like to analyze in some
detail the XORSAT problem. We will focus on the XORSAT problem defined
on Erd\"os-R\'enyi random graphs, and with couplings $J_a = \pm 1$ 
(equivalently $b_a = 0,1$) with
probability $1/2$. We will refer to this distribution as ``random XORSAT" in this
section.
Similarly to the fully connected $p$-spin model,
random XORSAT can be fully analyzed at zero temperature and the phase diagram
can be understood in detail. These results have been originally derived
in \cite{xor_1,xor_2}.
A very nice review on phase transition in optimization
problems, from which this section is reprinted, is \cite{Mo07}.

\subsubsection{Bounds from the first and second moments methods}

Let ${\cal N}$ be a random variable taking values on the positive integers (it will be the number
of solution of an instance drawn from the assigned probability distribution), and
call $p_{\cal N}$ its probability. We denote by $\langle{\cal N}\rangle$ and 
$\langle {\cal N} ^2\rangle$ the first and second moments of 
${\cal N}$ (assumed to be finite), and write 
\begin{equation}
P_{SAT}=p({\cal N}\ge 1) =\sum _{{\cal N}=1,2,3, \ldots} p_{\cal N} = 1 -p_0
\end{equation}
the probability that ${\cal N}$ is not equal to zero.
Our aim is to show the inequalities
\begin{equation} \label{ineq}
\frac{\overline{\cal N}^2}{\overline{{\cal N}^2}} \le  
p({\cal N}\ge 1) \le {\overline{\cal N}} \ .
\end{equation}
The right inequality, called ``first moment inequality'', is straightforward:
\begin{equation}
{\overline{\cal N}} = \sum _{\cal N} {\cal N}\; p_{\cal N} = 
\sum _{{\cal N}\ge 1} {\cal N} \; p_{\cal N} \ge
\sum _{{\cal N}\ge 1}  p_{\cal N} =  p({\cal N}\ge 1) .
\end{equation}
Consider now the linear space made of vectors 
${\bf v}=(v_0,v_1,v_2,\ldots\}$ whose components are labelled by positive 
integers, with the scalar product
\begin{equation}
{\bf v} \cdot {\bf v}' = \sum _{\cal N} p_{\cal N}\; v_{\cal N}\; 
v_{\cal N}' \ .
\end{equation}
Choose now $v_{\cal N}={\cal N}$, and $v'_0=0,v'_{\cal N}=1$ for 
${\cal N}\ge 1$. Then
\begin{equation}
{\bf v} \cdot {\bf v} = \overline{{\cal N}^2} \ , \ 
{\bf v} \cdot {\bf v}' = \overline{\cal N} \ , \
{\bf v}' \cdot {\bf v}' = p({\cal N}\ge 1) \ .
\end{equation}
The left inequality in (\ref{ineq}) is simply the Cauchy-Schwarz inequality 
for ${\bf v},{\bf v}'$: $({\bf v} \cdot {\bf v}')^2 \le  ({\bf v} \cdot {\bf v} ) \times ({\bf v}' \cdot {\bf v} ')$. 
If $\NN$ represents the number of solution of a given XORSAT instance, we can use
these bounds to obtain bounds on $P_{SAT} = p(\NN \geq 1)$.

\begin{figure}[t]
\includegraphics*[width=8.truecm,angle=0]{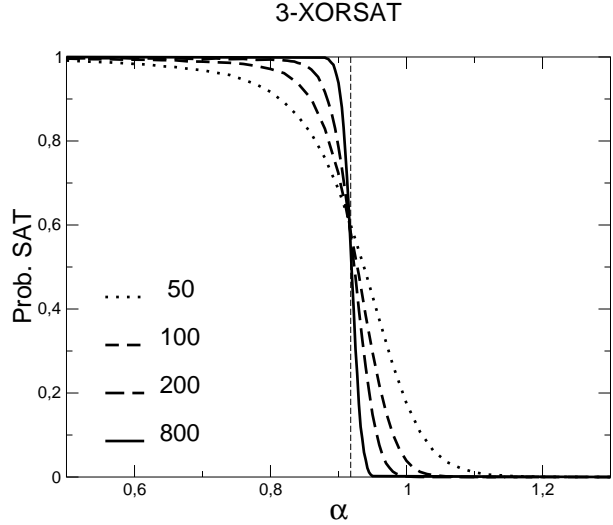}
\caption{Probability that a random 3-XORSAT formula is satisfiable as
a function of the ratio $\alpha$ of equations per variable, and for
various sizes $N$. The dotted line locates the threshold $\alpha_s
\simeq 0.918$.} 
\label{proba3-fig}
\end{figure}

Figure \ref{proba3-fig} shows the probability that a random 3-XORSAT
formula is satisfiable as a function of $\alpha$ for increasing sizes
$N$. It appears that formulas with ratio $\alpha < \alpha _c \simeq
0.918$ are very likely to be satisfiable in the large $N$ limit,
while formulas with ratios beyond this critical value are almost
surely unsatisfiable. 
Use of the first and second moment inequalities (\ref{ineq}) for
the number ${\cal N}$ 
of solutions provides us with upper and lower bounds to the
Sat/Unsat ratio $\alpha_s$. 

To calculate the first moment of ${\cal N}$ remark that an equation
is satisfied by one half of the configurations. When we average over the possible
choices of the second member of the equation, we have that two equations
are satisfied simultaneously by $1/4$ of the configurations, and $M$ equations
are satisfied simultaneously by $1/2^M$ of the configurations.
The average number
of solutions is thus $2^N/2^M$, from which we get
\begin{equation} \label{lowerom2}
P_{SAT} \leq \overline{\NN} = 2^{N(1-\a)} \ .
\end{equation}
The first moment vanishes for
ratios larger than unity, showing that
\begin{equation}
\alpha _c \le  1 \ .
\end{equation}
This upper bound is definitely larger than the true threshold from the 
numerical findings of Figure~\ref{proba3-fig}. 
Close to $\a=1$ formulas are unsatisfiable with probability one (when $N\to\infty$),
yet the average number of solutions is exponentially large! The reason
is that the average result is spoiled by rare, satisfiable
formulas with many solutions.

As for the lower bound we need to calculate the 
second moment $\overline{{\cal N}^2}$ of ${\cal N}$. 
We use here the representation in terms of bits and we denote
by $X = \{x_1,\cdots, x_N\}$ the configuration of the $N$ bits.
For a given instance, $\NN = \sum_X \I(X) = \sum_X \prod_a \I_a(X)$, where
$\I(X)$ is the indicator function of the event that $X$ is a solution to that instance,
and $\I_a(X)$ is the indicator of $X$ being a solution to clause $a$.
As equations are 
independently drawn 
\begin{equation} \label{n2}
\overline{ {\cal N} ^2} = \overline{ \sum_{X,Y} \I(X) \I(Y)} =
\sum_{X,Y}\overline{\prod_a \I_a(X) \I_a(Y)} =
\sum_{X,Y}\prod_a \overline{\I_a(X) \I_a(Y)} =
\sum _{X,Y} p(X,Y) ^M \, 
\end{equation} 
where the sum is carried out over the pairs $X,Y$ of configurations
of the $N$ variables, and $p(X,Y)$ is the
probability that both $X$ and $Y$ satisfies the same randomly
drawn equation. The latter can be easily expressed in terms of 
the Hamming distance (per variable) $d$ between
$X$ and $Y$, defined as the fraction of variables having opposite
values in $X$ and $Y$. The general
expression for $k$-XORSAT is\footnote{The
equation is satisfied by $X$ with probability $1/2$. Given that $X$ is a solution,
it is satisfied also by $Y$ if the number of variables entering the equation and taking opposite
values in $Y$ as in $X$ is even.
By definition of $d$ the probability (over its index $i$) that a variable
takes different value in $X$ and $Y$ is $d$. To construct the equation
one has to extract $k$ independent indexes $i$, then the probability
that the extracted variables are equal in $X$ and $Y$ is $\binom{k}{0} d^0 (1-d)^{k-0}$.
Similarly the probability to have two different variables is
$\binom{k}{2} d^2 (1-d)^{k-2}$, and so on. The sum of all even numbers
gives expression (\ref{qk}) for $p(d)$. See \cite{Mo07} for details.}
\begin{equation} \label{qk}
p(d)=\frac 14( 1 + (1-2d)^k) \ .
\end{equation}
Going back to (\ref{n2}) we can sum over $Y$ at fixed $X$, that is,
over the distances $d$ taking multiple values of $\frac 1N$ with the 
appropriate binomial multiplicity, and then sum over $X$ with the result
\begin{equation} \label{n2p}
\overline{{\cal N} ^2} = 2^{N} \sum _{d} {N \choose N\,d} \; p(d)^M
= \exp( N\, \max_{d \in [0;1]}A(d,\alpha) )
\end{equation} 
in the large $N$ limit, where
\begin{equation} \label{upperom2}
A(d,\alpha) =   \log 2 
  -d \log d -(1-d) \log (1-d) +\alpha \ln  p(d)  \ .
\end{equation}
The absolute maximum of the
function $A(d,\alpha)$ is located in
$d^* = \frac 12$ when $\alpha < \alpha _2 \simeq 0.889$, and
$d^* < \frac 12$ when $\alpha > \alpha_2$. In the latter case 
$\overline{{\cal N}^2}$ is exponentially larger than $\overline{
{\cal N}}^2$, and the second moment inequality
(\ref{ineq}) does not give any information about $P_{SAT}$. In
the former case $\overline{{\cal N}^2}$ and $\overline{
{\cal N}}^2$ are equivalent to exponential-in-$N$
order. It is shown in \cite{Mo07}, by computing the saddle point corrections to (\ref{n2p}),
that their ratio actually tends to 
one as $N\to\infty$. We conclude that formulas with
ratios of equations per variable 
less than $\alpha_2$ are satisfiable with high probability in
the infinite size limit, or, equivalently,
\begin{equation}
\alpha _c \ge \alpha _2 \simeq 0.889 \ .
\end{equation}
Unfortunately the lower and upper bounds do not match and the precise
value of the threshold remains unknown at this stage. We explain in
the next section how a simple preprocessing of the formula, before the
application of the first and second moment inequalities, can close the
gap, and shed light on the structure of the space of solutions.

\subsubsection{The leaf-removal algorithm}

We now sketch how clustering in the random 3-XORSAT problem may be analyzed rigorously.
The techniques used are borrowed from the analysis of algorithms,
and probability theory.

Let ${\cal S}$ be a randomly drawn 3-XORSAT system.
A crucial remark is that removal of an equation containing a
single-occurrence variable preserves the satisfiability, or
unsatisfiability of ${\cal S}$. Consider for instance $x_1 + x_2 + x_3 = b$.
Clearly if $x_1$ appear only in this equation, for any value of $x_2$ and
$x_3$ the equation can be satisfied by choosing the appropriate value of $x_1$.
Therefore if the system of the remaining solution is satisfiable, the same
will be for the full system.

This procedure can be iterated to
further simplify ${\cal S}$ until no single-occurrence variable is left. The
output of the procedure is ${\cal S}'$, the largest subsystem
of ${\cal S}$ where variables appear at least twice\footnote{From this 
definition, it is clear that ${\cal S}'$ is unique and independent of the
order in which equations with single-occurrence variable are removed.}. 
In the following, we analyze:
\begin{itemize}
\item the procedure which allows us to extract system ${\cal S}'$ from
 ${\cal S}$;
\item the statistical properties of ${\cal S}'$, and their consequences
in terms of satisfiability for ${\cal S}$;
\item how the solutions of the original system ${\cal S}$ can be reconstructed 
from the solutions of ${\cal S}'$, and how clustering emerges from this
reconstruction process. 
\end{itemize}

Intuition on single-occurrence variable removal is made easier once we
introduce the graphical representation of Boolean systems in terms of 
factor graphs discussed above. 

Removal of a single-occurrence variable and of its attached equation
from the system ${\cal S}$ is equivalent to removal of a leaf variable 
and its attached constraint from the factor graph. 
Removal of a leaf may ``uncover" vertices and produce new leaves. The process
may therefore be iterated well after all leaves present in the
original hypergraph have been removed. How can we quantitatively
track the reduction of the hypergraph as removal goes on? Let us call
step of the procedure the action of choosing a leaf and removing it
together with its constraint, and $\ell$--vertex a vertex of connectivity $\ell$,
\ie which appears in $\ell$ distinct constraints (with $\ell \ge 0$). 
Removed leaves will be considered as $0$-vertices in order to conserve the total
number of vertices.
The number of
$\ell$--vertices after $T$ steps is a stochastic variable, depending
upon the system ${\cal S}$ and the sequence of leaves removed by the
procedure, denoted by $N_\ell (T)$. Obviously, for all $T$,
\begin{equation} \label{xorrigorconst}
\sum _{\ell \ge 0} N_\ell (T) = N \quad \hbox{\rm and} \quad
\sum _{\ell \ge 0} \ell \;N_\ell (T) = 3\,(M-T) \quad , 
\end{equation}
as a result of the conservation of the total number of vertices and 
constraints respectively. Removal goes on as long as $N_1(T)\ge 1$.
Denote ${\sf N} (T)$, and call population the set $\{ N_\ell(T) , \ell
\ge 0\}$ of all numbers of $\ell$--vertices.  Knowledge of the
population is generally not sufficient to unambiguously determine the
hypergraph produced by $T$ steps of the removal procedure.  Indeed,
many hypergraphs have the same population ${\sf N} (T)$. An essential
observation is, however, that the output of $T$ steps of the removal
procedure is equally distributed among the set of all hypergraphs
having population ${\sf N} (T)$.  In other words, a complete
statistical information about hypergraphs is obtained from the
knowledge of the population. 

Let us now see how this population ${\sf N} (T)$ is modified
during step $T\to T+1$. The variations of the $N_\ell$
are stochastic variables due to the randomness in the system ${\cal S}$ 
and the choice of the leaf to be removed, with conditional expectations
with respect to ${\sf N} (T)$ given by
\begin{equation}
\label{flow}
{\mathbf{E}}\big[ N_\ell(T+1) - N_\ell (T) |{\sf N}(T)\big]
=  -\d_{\ell 1} + \d_{\ell 0} +  
2 \,p_{\ell+1}(T) -2 \,p_{\ell}(T) \quad .
\end{equation}
When a constraint is
removed, a 1-vertex disappears ($-\d_{\ell 1}$ term in (\ref{flow})) to
become a 0-vertex ($\d_{\ell 0}$). This constraint is attached to
two other vertices. The numbers $\ell, \ell '$ of occurrences of 
each of these two vertices are distributed with
probability
\begin{equation}\label{probpxor}
p_\ell (T) = \frac{\ell\,N_\ell(T)}{3\, (M-T)} \quad , 
\end{equation}
and are diminished by one once the constraint is taken
away\footnote{The probability $p_1$ of picking a variable, say, $x_1$, 
is equal to the number $\ell _1$ of its occurrences divided by the
total number of occurrences of variables in the system, $3(M-T)$. The probability
of picking any variable with occurrence $\ell$ is $p_1$ multiplied by
the number $N_\ell(T)$ of such variables, hence (\ref{probpxor}). 
Equation (\ref{xorrigorconst}) ensures that $p_\ell$ is
properly normalized.}. From
(\ref{flow}), the expectation values of $N_\ell$ vary by a quantity
of the order of unity at each time step, and the expected
densities of $\ell$-vertices, defined through
\begin{equation}
 n _\ell (T) = \frac{ {\mathbf{E}}\big[ N_\ell(T) \big]}{N} \quad ,
\end{equation}
undergo changes of the order of $1/N$ only. We are naturally led to
conclude that densities are function of a much longer ``time scale",
$t=T/N$. In other words, given $t\in [0,1]$, $n_{\ell} ([tN])$ and $n_{\ell}
([tN]+ o(N))$ are equal to within $o(1)$ terms as $N$ tends to
infinity, and we denote by $n_\ell (t)$ their common limit.  
Assuming that the densities $n_\ell (t)$
are differentiable functions of this time, we obtain 
from evolution equation (\ref{flow}) a set of coupled first order
differential equations\footnote{The left hand side of (\ref{flow}) reads
$N[n_\ell(t+1/N)-n_\ell (t)] = dn_\ell/dt + O(1/N)$ through a Taylor
expansion.}, 
\begin{equation}
\label{diffxor}
\frac {d n_\ell }{dt} (t)= -\d_{\ell 1} + \d_{\ell 0} + 
\frac{2}{3(\a-t)} \big[ (\ell+1)\, n_{\ell+1}(t)-\ell\,
n_{\ell}(t) \big]  \ .
\end{equation}
Resolution of these equations require to assign the initial conditions.
At time $t=0$, that is, prior to any removal of constraint, 
densities are Poisson distributed with parameter $3\,\a$,
\begin{equation} \label{initcondxor}
n_\ell(0)=e^{-3\,\a}\,\frac{(3\,\a)^{\ell}}{\ell!} \quad, \forall
\; \ell \ ,
\end{equation}
as we discussed above.

\begin{figure}
\begin{center}
\includegraphics[height=8cm,angle=-90]{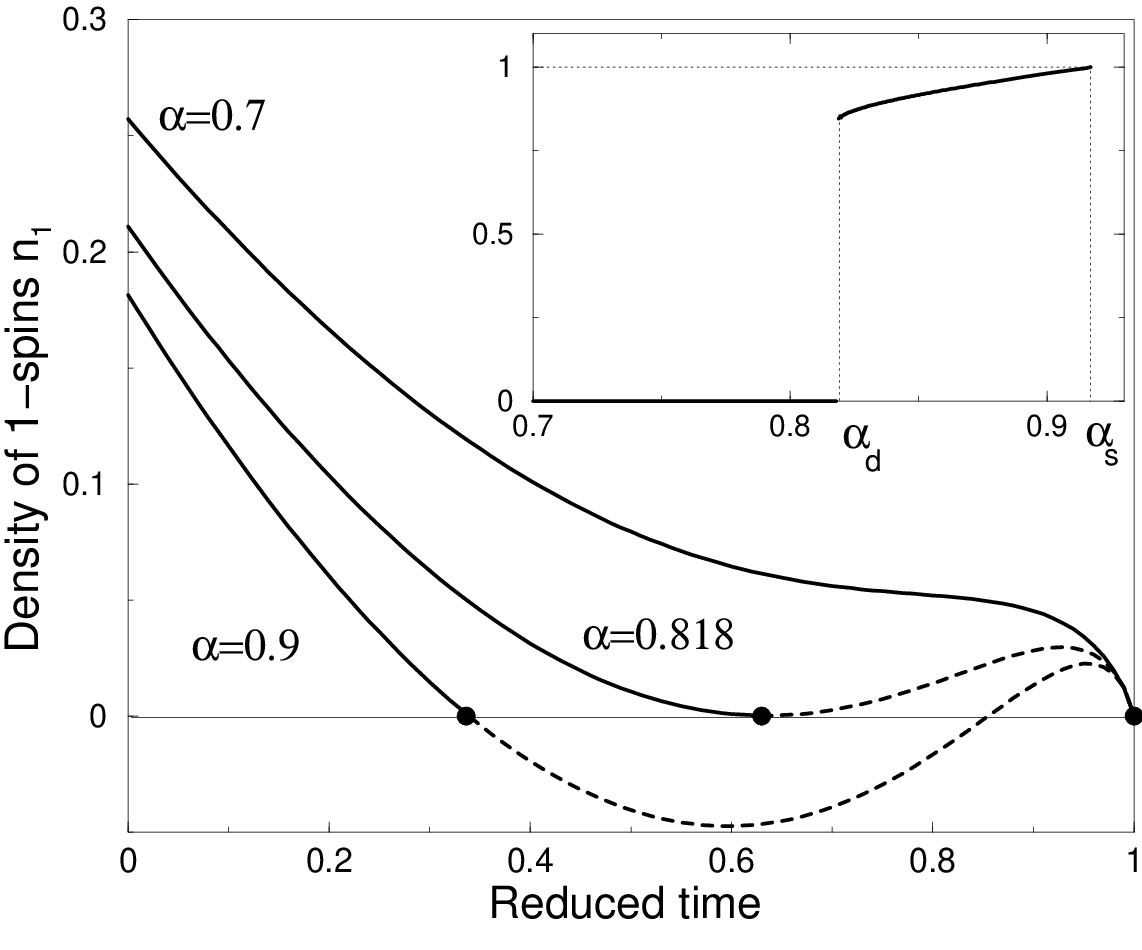}
\end{center}
\caption{(From~\cite{Mo07})
Evolution of the density of 1-vertices, $n_1(t)$, under the
operation of the leaf removal procedure as a function of the reduced
time $t/\a$. For $\a<\a_d \simeq 0.818$, $n_1(t)$ remains
positive until all the constraints are eliminated at $t^*=\a$. For
$\a>\a_d$ the procedure stops at the time $t^*$ for which
$n_1$ vanishes (black dots), and becomes negative 
for $t>t^*$ (dashed part of the curves). 
Notice that $t^*$ discontinuously jumps down at $\a=\a_d$.
In the inset, $\a'$ defined in Eq.~(\ref{reducedratioxor}) is plotted as a function of $\a$.
}
\label{algo}
\end{figure}

Solutions of equations (\ref{diffxor}) with initial
conditions (\ref{initcondxor}) read  
\begin{eqnarray}
\label{sol}
n_0 (t)&=& e^{-3\,\a\,b(t)^2} + 3\,\a\, b(t)^2\;\big( 1-b(t) \big) \ ,
\nonumber \\
n_1 (t)&=& 3\,\a\, b(t)^2\;\left( e^{-3\,\a\,b(t)^2}+b(t)-1 \right) \ ,
\nonumber \\
n_\ell(t)&=&e^{-3\,\a\, b(t)^2}\,\frac{\big(3\,\a\, b(t)^2\big)
^{\ell}}{\ell!} \ , \quad \forall\; \ell \ge 2 \ ,
\end{eqnarray}
where 
\begin{equation} \label{defxorb}
b(t) \equiv \bigg(1- \frac t\a \bigg)^{1/3} \quad .
\end{equation} 
These equations are valid
as long as $n_1$ is positive, since the procedure stops when no
leaf is left. The density $n_1(t)$
of 1--vertices is showed on Fig.~\ref{algo} for various initial 
ratios $\a$ of equations per variable. The shape of the curve reflects
the competition between two opposite effects: the annihilation of
1--vertices due to leaf removal and their creation as a result of 
reduction of $2$-vertices to $1$-vertices.
At small ratios e.g. $\a=0.7$, the former mechanism dominates and $n_1$
is a decreasing function of time. For slightly larger ratios, 
the density $n_1$ does not monotonously decrease with time
any longer, but still vanishes at time $t^*=\a$ {\em i.e.} when no constraint
is left.

According to the solution (\ref{sol}) of the equation of motion, 
this statement holds for ratios $\a$ for which the equation
\begin{equation}
b = 1 - e ^{- 3 \,\a \, b^2} \quad ,
\label{backbone}
\end{equation}
has no strictly positive root. 
When $\a$ is larger than
\begin{equation} \label{xorcd}
\a_d = \min _{0<b<1} \bigg[ - \frac{\ln (1-b)}{3\, b^2} 
\bigg] \simeq 0.818469 ... \quad ,
\end{equation} 
there exist
non zero real solutions to (\ref{backbone}), and we denote by $b^*$
the largest one. 
As shown in Fig. \ref{algo}, the density of 1--vertices
decreases and vanishes at time $t^*<\a$ where $b(t^*)$, given in Eq.~(\ref{defxorb}),
reaches $b^*$. The output of the procedure is a non-empty subset
${\cal S}'$ of ${\cal S}$ with $M'=N\,(\a-t^*)$ equations, such that 
any of the 
\begin{equation}\label{defnpxor}
N' = N\;  \sum _{\ell \ge 2} n_\ell (t^*)
\end{equation}
variables present in ${\cal S}'$ appears at least twice.
The ratio of equations per variable in ${\cal S}'$ is given by
\begin{equation}\label{reducedratioxor}
\a' =\frac{M'}{N'}= \frac{ \a \;b^{*\,2}}{1-3\,\a\, b^*\, (1-b^*)} \quad ,
\end{equation}
and is shown in the inset of Fig. \ref{algo}.

Let us briefly sum up the outcome of the above analysis. There
exists a critical ratio $\a_d$ (\ref{xorcd}) for the original system ${\cal S}$ 
such that
\begin{itemize}
\item if $\a<\a_d$, the leaf removal procedure succeeds in eliminating
all variables and equations, thus ${\cal S}'$ is empty;
\item if $\a>\a_d$, the output of the procedure is a non-empty subset
${\cal S}'$, with ratio of equation per variables equal to $\a'$ 
(\ref{reducedratioxor}).
\end{itemize}
The reader could feel concerned about fluctuations around those average
results. Equation of motion (\ref{diffxor}) is indeed true for the average
density of vertices only. Fortunately, one can show that, for large
sizes $N$, the numbers of $\ell$--vertices are highly concentrated around their
average values,
\begin{equation}
N_\ell (T) = N \; n_\ell \bigg(\frac TN \bigg)
+o (N) \quad ,
\end{equation}
whatever the initial instance ${\cal S}$ and the sequence of choices done
by the removal procedure.
Therefore, the evolution of the population
cannot deviate from the average behaviour when $N\to\infty$~\cite{Mo07}.

\subsubsection{Clustering and SAT/UNSAT transitions.}

As before, we consider a system ${\cal S}$ with ratio $\a$ and apply
the leaf removal procedure. The output is the subsystem ${\cal S}'$.
Can we reconstruct the solutions of ${\cal S}$ from the ones of
${\cal S}'$? 
The answer is positive, and the reconstruction process
permits us to characterize the structure of the set of solutions in 
an accurate way.

Assume that we have, in the course of the removal procedure, stored all
removed equations on top of each other in a stack. Let $X'$ be a solution
of ${\cal S}'$.
The length of $X'$,
that is, the number of variables it includes is $N'$ defined
in (\ref{defnpxor}). Through a relabelling of variables we can always
assume that $X'$ specifies the values of the first $N'$ variables $x_i$. 
We are now going to reinsert in ${\cal S}'$ the equations that were 
removed from ${\cal S}$ following a Last--In--First--Out unstacking order.

Consider the first reinserted  {\em i.e.} on top of stack equation,
\begin{equation} \label{lastxor}
x_i + x_j + x_k = v \quad ,
\end{equation}
where $v=0,1$ is the second member, and $(i,j,k)$ a triplet of
distinct integers comprised between 1 and $N$. Call $\nu _1$ the number
of these integers strictly larger than $N'$. That this
equation was eliminated by the leaf removal procedure and was the last
one to be so tells us that $\nu_1 \ge 1$: at least one of the three
variables in (\ref{lastxor}) is not assigned by the particular
solution $X'$.  If $\nu$ is precisely equal to unity, we have,
say, $i,j\le N'$ and $k\ge N'+1$. Then the values of $x_i,x_j$ are known
for $X'$, and there is a unique way to assign $x_k$ to satisfy 
(\ref{lastxor}). If $\nu_1=2$, we know, say, $x_i$ from $X'$, and
have two possible combinations of $x_j,x_k$ fulfilling (\ref{lastxor}).
Therefore we can construct two distinct solutions of the system
${\cal S}'_1={\cal S}'+$ 
reinserted equation. The reasoning is straightforwardly
extended to the last $\nu_1=3$ case, with the general result that 
$2^{\nu_1-1}$ distinct solutions can be reconstructed from $X'$.
Then, after reinsertion of the second-to-top equation, we reconstruct
$2^{\nu_2-1}$ solutions from any of the solutions of ${\cal S}'_1$
where $\nu_2$ is the number of new variables {\em i.e.} present in
the reinserted equation and not fixed by $S_1'$.
Iterating this procedures up to the reinsertion of all $M-M'$ equations
permits us to obtain (labeling $R$ the reinsertion ``time'')
\begin{equation} \label{defxorclu}
{\cal N} _{rec} = \prod _{R=1} ^{M-M'} 2^{\nu _R-1} 
\end{equation}
solutions of ${\cal S}_{M-M'}'={\cal S}'+$ all the reinserted equations. 
Notice that solutions of ${\cal S}_{M-M'}'$ do not
quite coincide with the ones of the original system ${\cal S}$.
Variables that were absent from equations in ${\cal S}$ (\ie that had zero connectivity)
are still 
undetermined from ${\cal S}_{M}'$, and can be freely chosen, giving
an extra multiplicative factor to the total number of solutions 
of ${\cal S}$ that can be reconstructed from $X'$,
\begin{equation} \label{defxorclu1}
{\cal N} _{in} = 2^{N_0} \;{\cal N} _{rec} 
= 2^{N_0} \prod _{R=1} ^{M-M'} 2^{\nu _R-1} \quad ,
\end{equation}
where $N_0$ is the number of $0$--occurrence variables.
This set
of solutions and $X'$ are respectively called {\it cluster of
solutions} and {\it seed of the cluster}.
Taking the logarithm of 
(\ref{defxorclu}) and dividing by $N$, we obtain the following expression
for the entropy of the reconstructed cluster from seed $X'$,
\begin{equation} \label{defxorclu2}
\frac{s}{\log 2} = \frac1N \log_2 {\cal N}_{in} =  \frac {N_0}N +
\frac 1N \sum _{R=1} ^{M-M'} (\nu _R-1) \quad .
\end{equation}
The average value of this entropy can be calculated from
the knowledge of the average values of the $\nu_R$s. 
Note that time $R$ of the reconstruction process corresponds to time
$T = 1 + M- M'-R$ of the leaf removal process.
As equations were
carefully introduced in opposite order to their removal, the 
expectation value of $\nu_R-1$ (conditioned to population
${\sf N}(T)$, at time $T = 1+M-M'-R $) can be estimated from the analysis of the previous
section and is equal to $2\, p_1(1+M-M'-R)$ with $p_1$ given in
(\ref{probpxor}). Changing variable from $R$ back to $T$ and taking the limit $N\to\io$ we end 
up with
\begin{equation} \label{sentxor}
\frac{s}{\log 2}= e^{-3\,\a} + \int_0^{t^*}\, dt\, \frac{2\,n_1(t)}{3\, (\a-t)} 
\ ,
\end{equation}
where the first term comes from the contribution $n_0(0)$ 
(\ref{initcondxor}) of absent variables, and $t^*$ is the time at which
the leaf removal procedure halts.

For small ratios $\a$,
${\cal S}'$ is empty, and so is $X'$. The halt time
$t^*=\a$ and integration of (\ref{sentxor}) with the help of (\ref{sol})
leads to the simple result $s(\a) =(1 -\a)\log 2$.
As we have reconstructed all possible solutions of ${\cal S}$ from the
empty $X'$, $s$ coincides with the total entropy $s_{tot}$ of solutions,
that in this case coincides with $\frac1N \log \overline{\NN}$. At such
small ratios the fluctuations of $\NN$ are not important.

On the contrary, for $\a > \a_d$
the leaf removal procedure stops at $t^* < \a$, and has 
not succeeded in eliminating all equations and variables. The 
average entropy associated to a cluster reconstructed from one
seed is
\begin{equation}\label{sentxor2}
\frac{s(\a)}{\log 2} = 1- \a-b^* - \a\, b^{*\, 2} (2 \,b^* -3) \quad ,
\end{equation}
where $b^*$ is, as before, the largest positive root of (\ref{backbone}).

To complete our description of clusters, some statistical knowledge
about their seeds is required. The number ${\cal N}'$ of solutions
of ${\cal S}'$ can be analyzed by means of the first and second moments 
method (\ref{ineq}), giving respectively some upper and lower bound to the
probability Prob[${\cal N}'\ge 1$] of existence of solutions.
The first moment is easy to calculate:
as the second members of the various equations are 
uncorrelated, we find
\begin{equation} \label{avxorp}
\overline {{\cal N}' }= 2^{N'} \times \left( \frac 12 \right) ^{M'}= 
2^{N'\left( 1-\a' \right)}
\quad .
\end{equation}
The overbar denotes here the unbiased expectation value over all
instances with $N'$ variables, $M'$ equations such that any variable
appears at least twice. The second moment calculation is made more
difficult by the existence of contraints on the minimal number (two)
of occurrences of variables in ${\cal S}'$. It requires a
combinatorial analysis of the number of systems ${\cal S}'$ {\em i.e.}
ways of choosing equations, having a given pair of configurations for
solutions. This calculation is beyond the scope of these notes and
was done by Dubois and Mandler~\cite{DM02}. The outcome is that, at large
$N$, the second moment $\overline{({\cal N'})^{2}}$ is asymptotically
equal to the first moment squared,$\big(\overline{{\cal N'}}\big)^2$,
when $\a'<1$ and exponentially larger when $\a'>1$. This result has 
two important consequences.

First, from (\ref{ineq}), we conclude that $\a'= 1$ is at the same 
time an upper
and a lower bound to the exact value of the threshold for ${\cal S}'$. 
Therefore $\a'=1$ is the location of the sat-unsat threhold for 
reduced subsystems. Using (\ref{reducedratioxor}), we see that $\a'=1$
is reached for an initial ratio equal to
\begin{equation}
\a_s \simeq 0.917936 ... \quad ,
\end{equation} 
as shown in Fig. \ref{algo}. Remarkably, while the moments method 
directly applied to ${\cal S}$ gave lower and upper bounds to the
threshold separated by a finite gap only, the outcome is much better
when applied to  ${\cal S}'$.
The concentration of ${\cal N}'$ constrasts with the (instance--to-instance)
fluctuations exhibited by ${\cal N}$, and suggests 
that the latter thus essentially comes from fluctuations in 
the numbers $N_0$ and $N_1$ of 0- and 1-vertices
eliminated by the leaf removal algorithm. This comes as no surprise since
variations of $N_0$ and $N_1$ induce 
drastic changes on the number of solutions \eg
the presence of a 0-vertex multiply the number of GS by two.
Conversely, in ${\cal S}'$, variables appear at least twice and are more 
interconnected, giving rise to weaker fluctuations for ${\cal N}'$. 

\begin{figure}
\includegraphics[width=8cm,angle=-90]{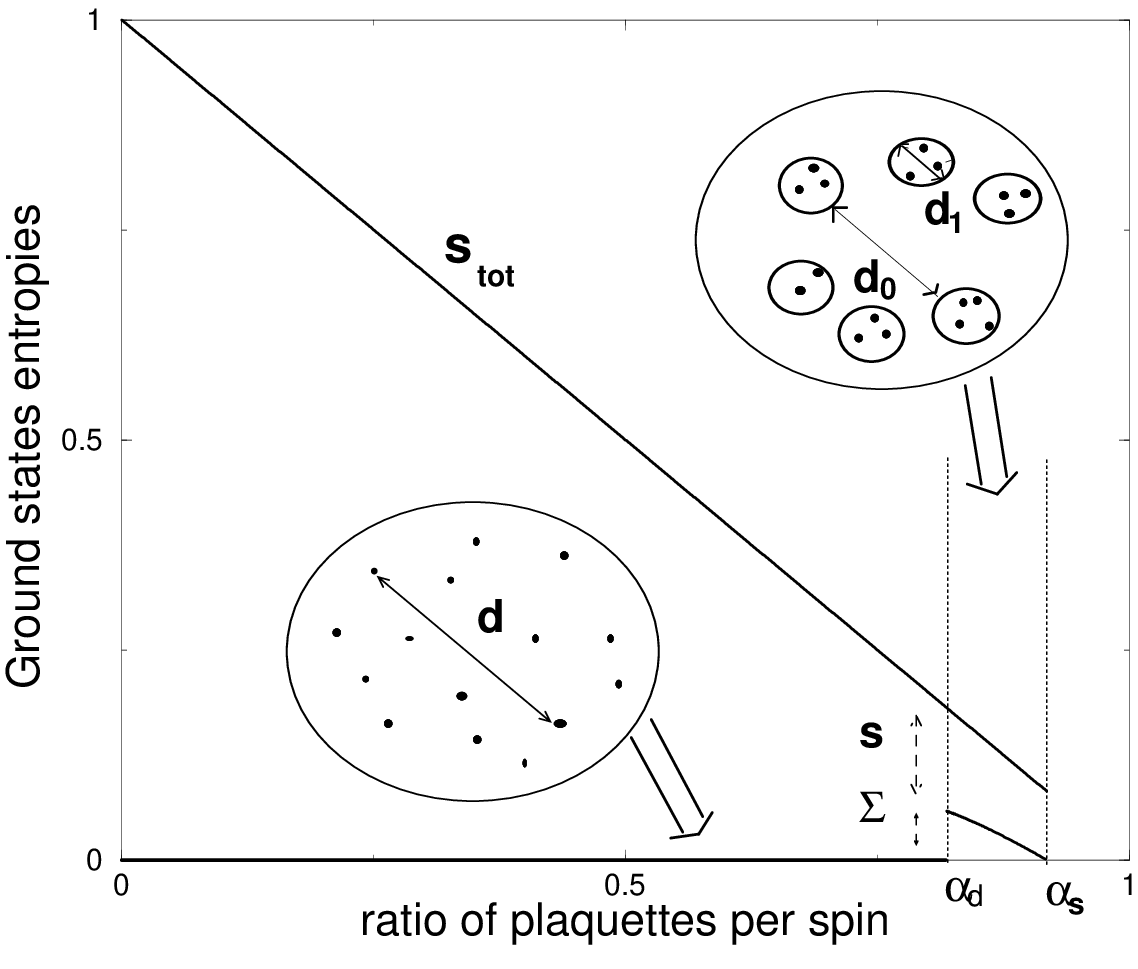}
\caption{
(From~\cite{xor_2}: {\it entropies are in units of $\log 2$})
Ground state structure and entropies as a function of the ratio $\a$
of equations per spin. The total entropy (logarithm of the number of unfrustrated
ground states per spin, or solutions) is $s_{tot}= (1-\a) \log 2$ for $\a<\a_s \simeq 0.918$. 
For $\a<\a_d\simeq 0.818$, solutions are uniformly
scattered on the $N$-dimensional hypercube, with a typical normalized 
Hamming distance $d=1/2$.  
At $\a_d$, the solution space discontinuously
breaks into disjoint clusters: the Hamming distance $d_1\simeq 0.14$ 
between solutions inside a cluster is much smaller than
the typical distance $d_0=1/2$ between two clusters
(RSB transition). The entropy of clusters, $s$,
and of solutions in each cluster, $\Si$, are such that $\Si+s=s_{tot}$. 
At $\a_s$, the number of clusters ceases to be exponentially large
($\Si=0$). Above $\a_s$, ground states are frustrated, \ie there are no solutions.}
\label{entroxor}
\end{figure}

Secondly, when $\a'<1$, the number of solutions of ${\cal S}'$ is,
with high probability, given by (\ref{avxorp})\footnote{This statement
comes, again, from the fact that the second moment is equal to the 
squared first moment, and therefore fluctuations of ${\cal N}'$ 
around the average value are extremely rare.}. This is the number of
seeds each of which gives a cluster of solutions for ${\cal S}$.
The number of clusters of solutions is therefore
\begin{equation}
{\cal N} _{clu} = 2^{N'\left( 1-\a' \right)} = e^{N\, \Si}
\quad ,
\end{equation}
where the entropy of clusters is defined as
\begin{equation} \label{resultxorentro}
\frac{\Si (\a)}{\log 2} = \frac {N'}N\, (1-\a') = b^*-3 \,\a\, b^{*\,2} + 2\, \a\, 
b^{* \,3} \quad .
\end{equation}
From the reconstruction process it is clear that two different seeds
cannot give twice the same solutions. Therefore the total entropy
of solutions of ${\cal S}$ is
\begin{equation}
s_{tot}(\a) = \Si(\a) + s(\a) = (1-\a)\log 2
\quad ,
\end{equation}
as obtained when summing (\ref{resultxorentro}) and (\ref{sentxor2}).
This is exactly the analytic continuation of the entropy of the unclustered
phase.
We have now established that $s_{tot}=(1-\a)\log 2$ for all ratios $\a<\a_s$.
The entropies of solutions in a cluster, $s$, and of clusters,
$\Si$, are shown in Fig.~\ref{entroxor}.

What we are left with is merely justifying the cluster denomination
used above. The reconstruction process allows a complete
characterization of solutions, in terms of an extensive number of
(possibly overlapping) blocks made of few variables, each block being
allowed to flip as a whole from a solution to another. When $\a<\a_d$,
with high probability, two randomly picked solutions differ over a
fraction $d=1/2$ of variables, but are connected through a sequence of
$O(N)$ successive solutions differing over $O(1)$ variables only. For
$\a_d<\a<\a_s$, flippable blocks are juxtaposed to a set of
seed-dependent frozen variables. For picture \ref{entroxor} to be
true, we must establish that different clusters do not overlap or,
more precisely, that any two solutions belonging to two different
clusters are far away from each other. A lower bound to the Hamming
distance $d$ between these two solutions is the Hamming distance
$d_s$ between the seeds of their respective clusters. 
A lower bound $d_s$, as a function of $\a$, can in turn be estimated 
using again the first moment method
{\em i.e.} from the vanishing condition of the expectation number 
$\overline{{\cal N'}^{2}  (d_s)}$ of solutions of ${\cal S}'$
lying apart at distance $\le d_s$. An explicit calculation
shows that $d_s>0$ for all $\a>\a_d$~\cite{xor_1,xor_2}.

\subsubsection{On backbones}

The reconstruction procedure helps us to interpret equation
(\ref{backbone}) as a self-consistent equation for the size of the
backbone of a cluster of solutions. Let $X'$ be a solution of ${\cal
S}'$, and ${\cal B}$ the set of variables taking the same value in all
solutions of the cluster associated to seed $X'$. The cardinality of
${\cal B}$ divided by $N$ is the backbone $b$. Obviously, the
variables assigned by $X'$ are elements of ${\cal B}$, thus
\begin{equation}
b \ge \sum _{\ell \ge 2} n_\ell (t^*) = b^* (1-3\, \a \,b^* (1-b^*))
\quad ,
\end{equation}
showing that the backbone is strictly positive.  So far no proof of
the equality between $b$ and $b^*$ has been obtained in literature,
though this could, in principle, be inferred from the analysis of the
removal and reconstruction procedures described above. The following
is a hand-waving argument supporting the interpretation of $b^*$ as
the cluster backbone~\cite{xor_replica}.

We want to estimate the probability that a variable, say, $x_1$ does
not belong to the backbone {\em i.e.} that $x_1$ takes value 0 on some
solutions and 1 on others.  $x_1$ appears in $\ell$ equations, where
$\ell$ is a Poisson variable with parameter $3\, \a$. Consider one of
these equations,
\begin{equation}
x_1 + x_j + x_k = v \quad .
\end{equation}
If either $x_j$ or $x_k$ does not belong to ${\cal B}$, neither does
$x_1$.  Thus the probability that the above equation does not
constrain $x_1$ to take the same value on all solutions is
$1-b^2$. This expression assumes that the events ``$x_j \in {\cal B}$''
and ``$x_k \in {\cal B}$'' are independent, a literally wrong assumption
which, however, apparently becomes asymptotically true for large
system sizes (this is the basis of the cavity method that we will discuss
in the next section).
Similarly, under the assumption that the above events
are independent from equation to equation, we obtain
\begin{equation}
\hbox{\rm Prob}[ x_1 \in {\cal B} | \ell ] = 1 - (1-b^2) ^\ell \quad .
\end{equation}
Summing over $\ell$, we have
\begin{equation}
\hbox{\rm Prob}[ x_1 \in {\cal B}] = \sum _{\ell = 0}^\infty
e^{-3\,\a}\frac {(3\,\a)^\ell}{\ell!} \big[ 1 - (1-b^2) ^\ell \big] = 1
- e^{-3\, \a\, b^2} \ .
\end{equation}
As the l.h.s. of the above equation is precisely the size of the
backbone, $b$, we see that $b$ fulfills (\ref{backbone}) and is equal
to $b^*$ as announced.

\subsubsection{Summary}

Unfortunately, the method discussed in this section can be applied
only to the XORSAT problem. In fact, the clusters in this problem
have a very simple structure: they are built around 
a seed (solution of the reduced system $\SS'$) by adding back the
leaves. This has a number of peculiar consequences:
\begin{enumerate}
\item {\it In all clusters there is a finite fraction of ``frozen'' variables
(backbone)}:
these are the variables in the seed 
that are constrained to take the same
value in all the solutions belonging to the cluster; and some of the leaves,
those that belong to equations where both other variables belong to the seed
(see the discussion in the last section).
\item {\it All the clusters have the same internal entropy}: this is
because adding back the leaves, the fraction of ``free'' variables
is determined only by the topology of the graph and not by the particular ``seed'' that defines the cluster.
Moreover, all these variables (those that are not in the backbone) are completely
free to be $\pm 1$ with probability $1/2$. Therefore, the internal entropy is determined only 
by the topology of the graph (the size of its backbone) and does not depend on the cluster.
\item {\it The SAT/UNSAT transition coincides with the point where $\Si$ vanishes}:
this is because when $\Si=0$ the reduced system becomes UNSAT.
\end{enumerate}
As we will see the structure of other optimization problems is more complicated:
the internal entropy may fluctuate from cluster to cluster, and a backbone is not
always present. In these cases we need a different method, that we explain in the
next section. Before turning to that, it is useful to reproduce the results obtained above
by mean of the replica method~\cite{xor_replica} ({\bf $\Rightarrow$ Ex.\ref{sec:optim}.\ref{ex:II.1}}).

\subsection{The replica symmetric cavity method}

The cavity method, initially invented to deal with the Sherrington Kirkpatrick
model of spin glasses \cite{MPV87}, is a powerful method to compute the
properties of ground states in many condensed matter and optimization
problems. It is in principle equivalent to the replica method, but it
turns out to have a much clearer and more direct interpretation, that allows
in practice to find solutions to some problems which remain rather difficult
to understand in the replica formalism: the replica approach is very elegant
and compact, but it is more difficult to get an intuitive feeling of what is
going on. Also, the cavity approach deals with usual probabilistic objects,
and can lend itself to rigorous studies \cite{Ta03}. We shall present
it here at two successive levels of approximation. The first one,
corresponding in replica language to the replica symmetric ({\sc RS}) solution, is an
easy one and has already been studied a lot. The one corresponding to one
step replica symmetry breaking ({\sc 1rsb}) is more involved and has been
fully understood only very recently~\cite{cavity,cavity_T0,KrMoRiSeZd,col2,MoSe2,rearr_csp}.

\subsubsection{Recursions on a finite tree}
\label{sec:treerec}

The key properties of random graphs that is exploited by the cavity method is that
loops are very large in the thermodynamic limit, as we discussed in the introduction
to this section. Therefore, locally random graphs look like trees. Before studying the
cavity method, we must understand how to solve statistical mechanics models defined
on trees. This can be done by a generalization of the transfer matrix methods that are
used to solve models in one dimension 
(which is indeed the special case of a regular tree with connectivity $c=2$). We will for the moment
restrict to consider models where the clauses have connectivity $k=2$: for these models, one does not
need a factor graph representation, since interactions can be
represented by a standard graph with vertices $i=1,\cdots,N$ 
representing variables $\s_i$ and links $\ij$ representing interactions $\psi_{ij}(\s_i,\s_j)$.
We shall repeatedly use in the following the same notation we used for factor graphs, specialized
to the $k=2$ case: therefore we use $\di$
for the set of vertices adjacent to a given vertex $i$, \ie for the sites
which interact with $i$, and $\dimj$ for those vertices around $i$ distinct
from $j$. 

\begin{figure}
\begin{center}
\includegraphics{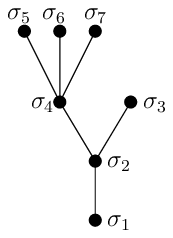}
\end{center}
\caption{Example of an Ising tree model on 7 vertices. The definition of
$Z_{2\to 1}$ and its recursive computation reads here:
$Z_{2\to 1}(\s_2)=\underset{\s_3,\dots,\s_7}{\sum} 
\psi_{23}(\s_2,\s_3)
\psi_{24}(\s_2,\s_4)
\psi_{45}(\s_4,\s_5)
\psi_{46}(\s_4,\s_6)
\psi_{47}(\s_4,\s_7)
= \underset{\s_3,\s_4}{\sum} Z_{3 \to
  2}(\s_3) Z_{4 \to 2}(\s_4) \psi_{23}(\s_2,\s_3)
\psi_{24}(\s_2,\s_4)
$. 
}
\label{fig_Ising_example}
\end{figure}

Let us then consider the case where the interaction graph is a finite
tree. In this case the computation can be organized in a very simple way,
taking benefit of the natural recursive structure of a tree. 
We define the
quantity $Z_{i \to j}(\s_i)$, for two adjacent sites $i$ and $j$, as the 
partial partition function for the subtree rooted at $i$, excluding the branch
directed towards $j$, with a fixed value of the spin variable on the site
$i$. We also introduce $Z_i(\s_i)$, the partition function of the whole tree
with a fixed value of $\s_i$. These quantities can be computed according to
the following recursion rules, see Fig.~\ref{fig_Ising_example} for an
example, 
\beq
Z_{i \to j}(\s_i) = 
\prod_{k \in \dimj} \left(\sum_{\s_k} Z_{k \to i}(\s_k) \psi_{ik}(\s_i,\s_k) 
\right) \ , \qquad
Z_i(\s_i) =  \prod_{j \in \di} \left( 
\sum_{\s_j} Z_{j \to i}(\s_j) \psi_{ij}(\s_i,\s_j)  \right) \ .
\eeq
It will be useful for the following discussion to rewrite these equations in
terms of normalized quantities which can be interpreted as probability laws
for the random variable $\s_i$, namely 
$\eta_{i \to j}(\s_i)=Z_{i \to j}(\s_i)/\sum_{\s'}Z_{i \to j}(\s')$ and
$\eta_i(\s_i)=Z_i(\s_i)/\sum_{\s'}Z_i(\s')$. The quantity $\eta_{i\to j}(\s_i)$ is the
marginal probability law of variable $\s_i$ in a modified system where the link 
$\ij$ has been removed.
The recursion equations read in
these notations
\beq
\eta_{i \to j}(\s_i) = \frac{1}{z_{i \to j}} 
\prod_{k \in \dimj} \left(\sum_{\s_k} \eta_{k \to i}(\s_k) 
 \psi_{ik}(\s_i,\s_k)  \right) \ , \qquad
\eta_i(\s_i) = \frac{1}{z_i}  \prod_{j \in \di} \left( 
\sum_{\s_j} \eta_{j \to i}(\s_j)  \psi_{ij}(\s_i,\s_j)  \right) \ ,
\label{eq_msg_Ising}
\eeq
where $z_{i \to j}$ and $z_i$ are normalization constants:
\beq
z_{i \to j} = \sum_{\s_i}
\prod_{k \in \dimj} \left(\sum_{\s_k} \eta_{k \to i}(\s_k) 
 \psi_{ik}(\s_i,\s_k)  \right) \ , \qquad
z_i = \sum_{\s_i}  \prod_{j \in \di} \left( 
\sum_{\s_j} \eta_{j \to i}(\s_j)  \psi_{ij}(\s_i,\s_j)  \right) \ ,
\label{eq_norm_Ising}
\eeq
Since the leaves are isolated when the link connecting them
is removed, one has $Z_{i \to j}(\s_i) =$const. and 
$\eta_{i \to j}(\s_i)=$const. for leaves. However, one can also choose to
put an arbitrary $\eta_{i \to j}(\s_i)$ on the leaves: this might represent
an external field acting on them, or the effect of a given boundary condition.
Moreover the quantity
$\eta_i(\s_i)$ is exactly the marginal probability law of the Gibbs-Boltzmann
distribution, hence the local magnetizations can be
computed as $m_i = \langle \s_i \rangle = \sum_{\s} \eta_i(\s) \s$.
Finally, it is useful to define the object
\beq
z_{ij} = \sum_{\s_i,\s_j} \eta_{j \to i}(\s_j)  \eta_{i \to j}(\s_i) \psi_{ij}(\s_i,\s_j) = \frac{z_j}{z_{j \to i}} = \frac{z_i}{z_{i \to j}} \ , 
\label{eq_norm_Ising2}
\eeq
where the last two equalities are easily derived using Eqs.~(\ref{eq_msg_Ising}).

We can now write the free energy of the system.
Clearly, for any spin $\s_i$
the total partition function is $Z = \sum_{\s_i} Z_i(\s_i)$.
Note that using Eqs.~(\ref{eq_msg_Ising}) and (\ref{eq_norm_Ising}), we obtain
\beq
z_i = \sum_{\s_i}  \prod_{j \in \di} \left( 
\sum_{\s_j} \eta_{j \to i}(\s_j)  \psi_{ij}(\s_i,\s_j)  \right) = \sum_{\s_i}  \prod_{j \in \di} \left( 
\sum_{\s_j} \frac{Z_{j \to i}(\s_j)}{\sum_{\s'} Z_{j \to i}(\s')}  \psi_{ij}(\s_i,\s_j)  \right) 
= \frac{\sum_{\s_i} Z_i(\s_i) }{\prod_{j \in \di} \sum_{\s_j} Z_{j\to i}(\s_j) } \ ,
\eeq
and along the same steps
\beq
 z_{j\to i} = \frac{ \sum_{\s_j} Z_{j\to i}(\s_j) }{\prod_{k \in \partial j \setminus i} \sum_{\s_k} Z_{k\to j}(\s_k) } \ .
\eeq
So we can start from an arbitrary spin $i$ and 
\beq
Z = \sum_{\s_i} Z_i(\s_i) = z_i \prod_{j \in \di} \left( \sum_{\s_j} Z_{j\to i}(\s_j) \right)= z_i  \prod_{j \in \di} \left( z_{j\to i} \prod_{k \in \partial j \setminus i} \sum_{\s_k} Z_{k\to j}(\s_k) \right) \ ,
\eeq
and we can continue to iterate this relation until we reach the leaves of the tree.
Using Eq.~(\ref{eq_norm_Ising2}), we finally obtain
\beq
Z = z_i \prod_{j \in \di} \left( z_{j \to i} \prod_{k \in \partial j \setminus i } z_{k\to j} \cdots \right)
= z_i \prod_{j \in \di} \left( \frac{z_j}{z_{ij}} \prod_{k \in \partial j \setminus i} \frac{z_k}{z_{jk}} \cdots \right)
= \frac{\prod_{i} z_i}{\prod_{\ij} z_{ij}}
\eeq
and the free energy is
\beq\label{f_Bethe}
\begin{split}
&F = -T \log Z = \sum_i f_i - \sum_{\ij} f_{ij} \ , \\
&f_i = -T \log z_i \ ,\\
&f_{ij} = -T \log z_{ij} \ .
\end{split}\eeq
The advantage of this expression of $F$ is that it does not depend on the arbitrary choice of the initial site $i$ we made above.

On a given tree, the equations (\ref{eq_msg_Ising}) for the $\eta_{i \to j}$ for all
directed edges of the graph have a single solution, which is easily found by
propagating the recursion from the leaves of the graph. From this solution, one can compute
the free energy using Eq.~(\ref{f_Bethe}). If the connectivity of the tree is bounded, then the number
of steps required to compute $F$ is proportional to the number of nodes $N$ in the tree when $N\to\io$.
Note that in the particular case of a ``tree'' with connectivity two (in other words, a one
dimensional chain with open boundaries) the above equations are exactly equivalent to the
transfer matrix method ({\bf $\Rightarrow$ Ex.\ref{sec:optim}.\ref{ex:transfer}}).

\begin{figure}
\begin{center}
\includegraphics{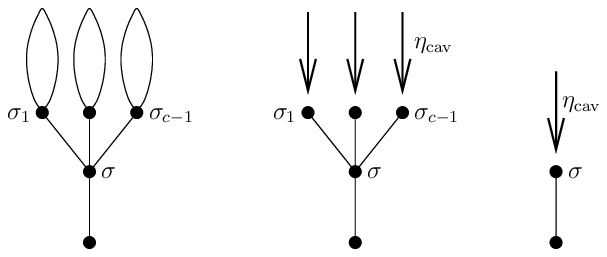}
\end{center}
\caption{Pictorial representation of Eq.~(\ref{eq_fpoint_Ising}). The bubbles
on the first panel represent subtrees of the graph; their effect on the spins
$\s_1,\dots\s_{c-1}$ is summarized by $\eta_{\rm cav}$, represented as a bold
arrow in the second panel. Tracing over these $c-1$ spins leads to the third
panel.}
\label{fig_Ising_iter}
\end{figure}

\subsubsection{From a tree to a random graph}

The reasoning of section~\ref{sec:treerec} was made under the assumption that the interaction graph
was a tree, and we arrived on the recursion (\ref{eq_msg_Ising}).
Suppose now that we are on a random graph, and we cut a link $\ij$; then we produce
two ``cavity'' variables on sites $i$ and $j$, whose connectivity has been decreased by one.
As before, we define the quantity $\eta_{i\to j}(\s_i)$ as the
marginal probability law of variable $\s_i$ in a modified system where the link 
$\ij$ has been removed.
We can write an exact equation:
\beq
\eta_{i \to j}(\s_i) = \frac{1}{z_{i \to j}} \sum_{\{\s_k\}, k \in \dimj} 
\eta_{(\dimj) \to i}(\{\s_k\}, k \in \dimj)
\prod_{k \in \dimj} \psi_{ik}(\s_i,\s_k)   \ , 
\eeq
where we introduced a multi-spin cavity field $\eta_{(\dimj) \to i}(\{\s_k\}, k \in \dimj)$ that
describes the joint distribution of the spins in $\dimj$ in absence of the links that connect them
to $i$. The problem obviously is that in these way the equations are not closed. We can
close the equations if we assume that the cavity variables are uncorrelated and write
\beq\label{cavity_decorr}
\eta_{(\dimj) \to i}(\{\s_k\}, k \in \dimj) = \prod_{k \in \dimj}  \eta_{k \to i}(\s_k) \ ,
\eeq
which gives back Eqs.~(\ref{eq_msg_Ising}).
Now, Eqs.~(\ref{eq_msg_Ising}) have to be interpreted as a set of coupled 
equations for the unknown $\eta_{i\to j}$. Contrary to the tree, on a graph with loops
they cannot be solved by recursion. Still the number of equations is clearly equal to the number
of unknown and we can hope for a unique solution.

{\it The replica symmetric cavity method corresponds to the use of the local recursion equations (\ref{eq_msg_Ising})
derived under the assumption (\ref{cavity_decorr})
to compute the free energy of models defined on sparse random graphs, that are
locally tree-like.} 

How can we justify assumption (\ref{cavity_decorr})? The key observation is that, as we already discussed,
random graphs converge locally to trees in the thermodynamic limit. Loops have typically length $\sim \log(N)$.
Therefore, if we cut the links between variables in $\dimj$ and variable $i$, in the modified system variables
in $\dimj$ are very far away. {\it If there is a single pure state}, then correlations in the Gibbs measure decay quickly
with distance and Eq.~(\ref{cavity_decorr}) becomes asymptotically correct for $N\to \io$.
In summary, the assumption of the existence of a single pure state 
implies that the effect of the loops
does not spoil the existence of a unique solution to the local recursions (\ref{eq_msg_Ising}).
Their presence
simply provides self-consistent boundary conditions.
We conclude that Eqs.~(\ref{eq_msg_Ising}) provide an exact description of
a disordered model on a random graph for $N\to\io$, in a phase where there is a single pure state.

Note that one can look for a solution of the recursion equations
(\ref{eq_msg_Ising}) on any graph, even in the presence of short loops. Although this procedure is not exact in general,
it might provide a good approximation to the true solution of the problem.
This approach is known as Belief Propagation in inference problems~\cite{fgraphs}, 
and corresponds to the Bethe approximation of statistical mechanics~\cite{Yedidia}. 

\subsubsection{Replica symmetric cavity equations in absence of local disorder}

To become more familiar with the cavity method, we now consider the simplest possible case.
We specialize to a model such that
\begin{enumerate}
\item The underlying graph is a random regular graph: the connectivity
of the constraints is $k=2$, so we do not need a factor graph representation, and the connectivity of the variables is fixed to $c$. 
We have therefore $N$ variables and $M=Nc/2$ links (constraints).
\item There is no disorder in the Hamiltonian, \ie the constraints are all equal to a
deterministic function of the involved variables.
\end{enumerate}
Simple example of models in this class are
\begin{enumerate}
\item The Ising ferromagnet/antiferromagnet: $H=-\sum_{(ij)} J S_i S_j$;
\item The Potts antiferromagnet $H=\sum_{(ij)} \d(\s_i,\s_j)$, that corresponds
to $q$-COL.
\item A model of hard spheres, defined by $n_i \in\{0,1\}$ (occupation numbers)
and local constraints that impose that if a site is occupied ($n_i = 1$) at least
some of the neighboring sites cannot be occupied~\cite{BM01}.
\end{enumerate}
Therefore this class is already rich enough to show some interesting features and
structures: actually, $q$-COL already contains the richest {\sc 1rsb} structure and
further complications do not add much to the physical picture~\cite{col2}.

The main simplification in these models is that {\it all sites are statistically equivalent},
\ie the local environment is tree-like (with probability $1$ for $N\to\io$) 
without fluctuations of the connectivity or of
the local interactions. The latter have the form $\psi_{ij}(\s_i,\s_j) = \psi(\s_i,\s_j)$ for a fixed function $\psi$.
Notice that in these systems, the frustration and the disorder (if any) are due to the
presence of loops, and thus occur only on large scales ($\sim \log N/\log (c-1)$). 

\begin{figure}
\begin{center}
\includegraphics{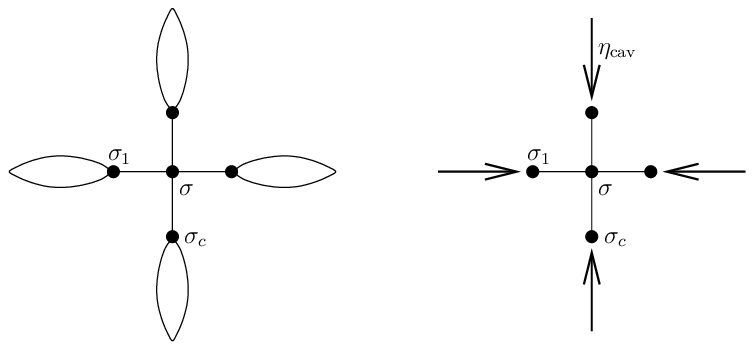}
\end{center}
\caption{Illustration of Eq.~(\ref{eq_Ising_final}); the local magnetization
  of one site is computed by taking into account all the $c$ neighbors.} 
\label{fig_Ising_final}
\end{figure}

Consider a site $i$ and the region around this site. For $N\to\io$, this region will be a tree with probability 1
and we can use the recursion (\ref{eq_msg_Ising}) to compute everything inside the tree. The problem is that
in this case the ``leaves'' are connected to the rest of the graph, which provides a boundary condition
that has to be determined. Suppose that we are in a phase that is not frustrated. Then, we expect the system to
be homogeneous. Suppose then that each of the leaves feel the same external field due to the rest of the graph.
We initiate then the recursion with the same $\eta_0(\s)$ for all the leaves. Then, at each iteration, the $\eta$'s 
remain identical and satisfy the recursion:
\beq
\eta_{g+1}(\s) = \frac{1}{z_{g}}
\sum_{\s_1,\dots,\s_{c-1}} 
\eta_{g}(\s_1)\dots \eta_{g}(\s_{c-1}) \psi(\s,\s_1) \cdots \psi(\s,\s_{c-1}) \ .
\eeq
Since the tree region around site $i$ can be arbitrarily large, we have to iterate this recursion for a very
large number of times, and it is natural to assume that we will converge to a fixed point\footnote{
In presence of spontaneous symmetry breaking, for instance for a ferromagnetic system at low temperature,
there might be several fixed points. In ordered systems, one can usually select one of them by adding a suitable
small external field to break explicitly the symmetry and obtain a system with a single pure state.}
$\eta_{\rm cav}$ which
satisfies the relation
\beq
\eta_{\rm cav}(\s) = \frac{1}{z_{\rm cav}}
\sum_{\s_1,\dots,\s_{c-1}} 
\eta_{\rm cav}(\s_1)\dots \eta_{\rm cav}(\s_{c-1}) 
 \psi(\s,\s_1) \cdots \psi(\s,\s_{c-1}) 
 = \frac1{z_{\rm cav}} \left( \sum_{\s'} \eta_{\rm cav}(\s') \psi(\s,\s') \right)^{c-1} \ , 
\label{eq_fpoint_Ising}
\eeq
with $z_{\rm cav}$ is the normalization constant. A pictorial representation of
this equation can be found in Fig.~\ref{fig_Ising_iter}. The local
magnetization is then computed as
\beq
\langle \s \rangle = \sum_\s \eta(\s) \s \ , \qquad
\eta(\s) = \frac{1}{z^{(s)}} 
\sum_{\s_1,\dots,\s_c} \eta_{\rm cav}(\s_1)\dots \eta_{\rm cav}(\s_c) 
 \psi(\s,\s_1) \cdots \psi(\s,\s_{c}) \ ,
\label{eq_Ising_final}
\eeq
including the $c$ neighbors of a central site as represented in
Fig.~\ref{fig_Ising_final}.
Although we fixed a particular site $i$ at the beginning, this reasoning is clearly
independent of the site in the thermodynamic limit, since in that limit all sites
have the same environment around them. Therefore, we expect that in this limit
all the fields $\eta_{i\to j}$ (or equivalently $\eta_g$) 
will converge to $\eta_{\rm cav}$, solution of (\ref{eq_fpoint_Ising}),
and all the fields $\eta_i(\s)$ will converge to $\eta(\s)$ defined
in (\ref{eq_Ising_final}). 

We can now compute the free energy of the system. 
We start from Eq.~(\ref{f_Bethe}) together with (\ref{eq_norm_Ising}) and
(\ref{eq_norm_Ising2}). 
We observe that, under the homogeneity assumption made above, 
all $z_i = z^{(s)}$ and $z_{ij} = z^{(l)}$ are equal with
\beq\label{zlinkRS}
\log z^{(l)} = 
\log \sum_{\s_1 \s_2} \eta_{\rm cav}(\s_1)\eta_{\rm cav}(\s_2) \psi(\s_1,\s_2) \ .
\eeq
and 
\beq\label{zsiteRS}
\log z^{(s)} = \log \sum_{\s,\s_1,\cdots,\s_c} 
\eta_{\rm cav}(\s_1) \cdots \eta_{\rm cav}(\s_{c}) \psi(\s,\s_1) \cdots \psi(\s,\s_{c}) 
=
\log \sum_{\s} 
\left(\sum_{\s'} \eta_{\rm cav}(\s') \psi(\s,\s')\right)^c \ .
\eeq
Recall that for a random regular graph
the number of links is $N c/2$. Then we get
\beq\label{freecavRS}
\begin{split}
&f = \frac{F}{N} = f^{(s)} - \frac{c}{2} f^{(l)} \ , \\
&f^{(s)} = -T \log z^{(s)} \ , \\
&f^{(l)} = -T \log z^{(l)} \ .
\end{split}\eeq
The cavity method (here presented for an homogeneous state of a model without disorder 
on a random regular graph) consists in solving Eq.(\ref{eq_fpoint_Ising}) and using then
Eq.~(\ref{freecavRS}) to compute the free energy. Obviously, any other observable can be computed
from the free energy by adding suitable external fields.

\subsubsection{An alternative derivation}
\label{sec:cavity2}

Before moving to more complicated cases, it is useful to present an alternative derivation
of the cavity method, that also explain the historical origin of its name. This derivation was
the original one of \cite{cavity} and this section is reprinted from that paper.

\begin{figure}
\includegraphics[width=8cm]{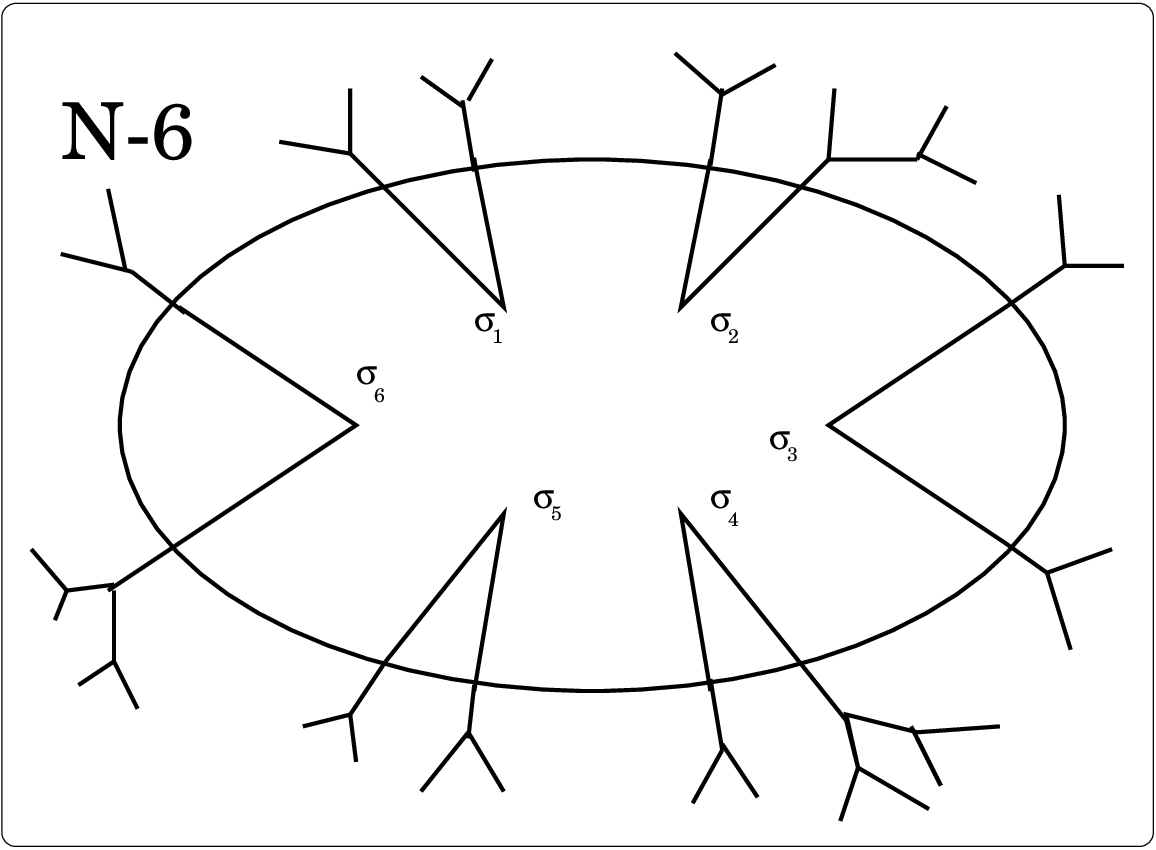}
\caption{An example, for the case $c=3$, of a  $\GG_{N,6}$ cavity
graph where
$q=6$ randomly chosen cavity variables have $c-1=2$ neighbours only. All the other 
$N-6$ spins  outside the cavity are connected through a random
graph such that every spin has $c=3$ neighbours.
}
\label{fignmoinssix}
\end{figure}

Let us introduce an intermediate object which is 
a model with $N$ variables, on a slightly different random lattice, where $q$ randomly chosen 
``cavity'' variables have only $c-1$ neighbours, while the other $N-q$ spins all have $c$ neighbours 
(see fig.\ref{fignmoinssix}).  We call such a graph a $\GG_{N,q}$ ``cavity graph''.
The cavity variables are characterized by some joint probability
distribution $\eta_{\rm cav}(\sigma_1,...,\sigma_{q})$.

While our primary interest is in $\GG_{N,0}$
graphs, the intermediate construction of $\GG_{N,q}$ is helpful.
The basic operations which one can perform on cavity graphs are the following:
\begin{itemize}
\item {\it Iteration:}
By adding a new spin $\sigma_0$ of into the cavity, connecting it
to $c-1$
of the cavity spins, say $\sigma_1,...,\sigma_{c-1}$,
one changes a $\GG_{N,q}$ into a  $\GG_{N+1,q-c+2}$ graph:
\beq
\D N =1, \ \ \ \ \ \  \D q =-c+2 \ .
\eeq
\item {\it Link addition}:
By adding a new link between two randomly chosen cavity spins $\sigma_1,\sigma_2$,
one changes a  $\GG_{N,q}$ into a  $\GG_{N,q-2}$ graph:
\beq
\D N =0, \ \ \ \ \ \  \D q =-2 \ .
\eeq
\item {\it Site addition}:
By adding a new spin $\sigma_0$ into the cavity, connecting it
to $c$
of the cavity spins say $\sigma_1,...,\sigma_{c}$,
one changes a $\GG_{N,q}$ into a  $\GG_{N+1,q-c}$ graph:
\beq
\D N =1, \ \ \ \ \ \  \D q =-c \ .
\eeq
\end{itemize}
In particular, if one starts from a $\GG_{N,2 c}$ cavity graph and perform
$c$ link additions, one gets a $\GG_{N,0}$ graph, \ie our original
problem with $N$ variables. Starting from the same $\GG_{N,2 c}$ cavity
graph and performing $2$ site additions, one gets a $\GG_{N+2,0}$ graph, \ie
our original problem with $N+2$ variables. Therefore the variation in
the free energy when going from $N$ to $N+2$ sites ($F_{N+2}-F_N$) is related
to the average free energy shifts $\Delta F^{(s)}$ for a site addition, and $\Delta
F^{(l)}$ for a link addition, through:  
\beq
F_{N+2}-F_N = 2 \Delta
F^{(s)}- c \Delta F^{(l)} \ .
\label{enetshift}
\eeq
Using the fact that the total free energy is asymptotically linear in $N$,
the free energy is finally 
\beq
f=\lim_{N \to \infty} {F_N}/N = \frac{F_{N+2}-F_N}{2}=\Delta F^{(s)}-\frac{c}{2}\Delta F^{(l)} \ .
\label{enetot}
\eeq
\begin{figure}
\includegraphics[width=8cm]{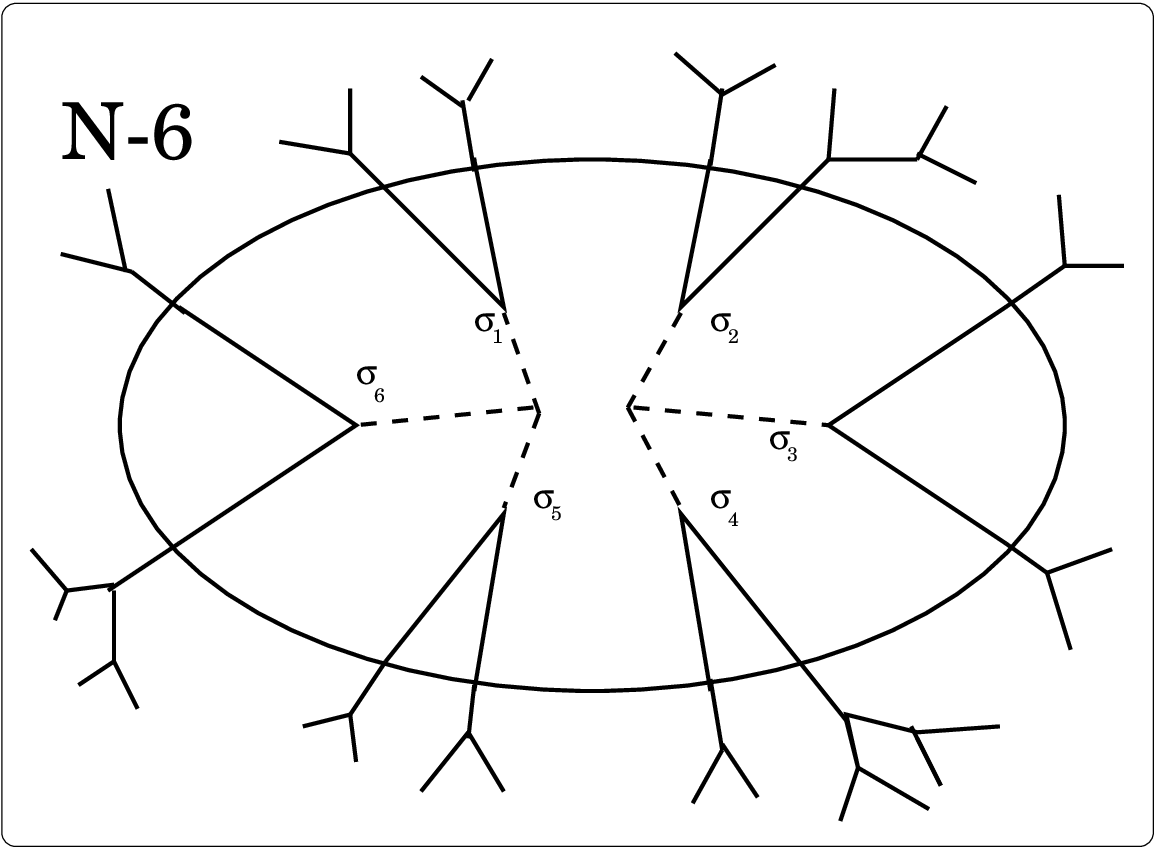} 
\includegraphics[width=8cm]{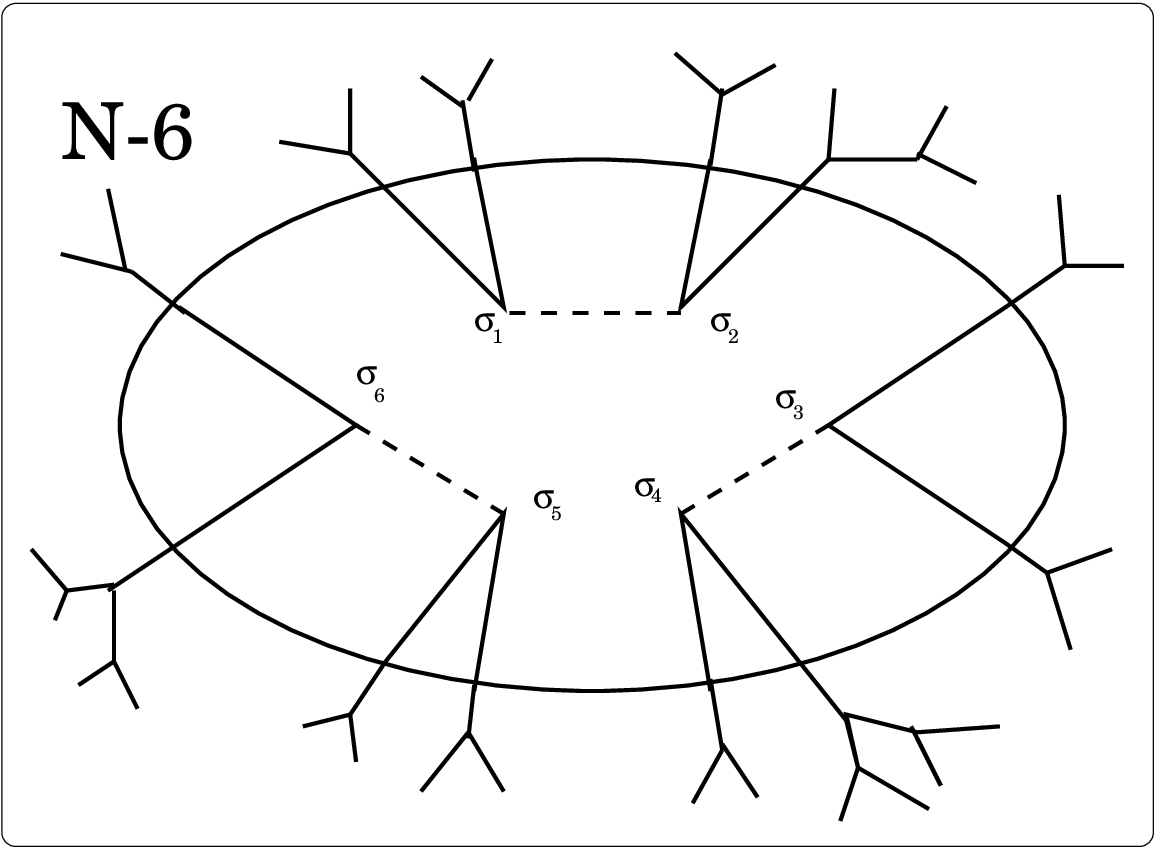}
\caption{Starting from the ${\GG}_{N,6}$ cavity graph,
one can either add two sites (left figure) and create a 
${\GG}_{N+2,0}$ graph, or add three links
(right figure) and create a 
${\GG}_{N,0}$ graph.}
\label{fignplusdeux}
\end{figure}
An intuitive interpretation of this result (for even $c$) is that in order
to go from $N$ to $N+1$ one should remove $c/2$ links (the energy for
removing a link is minus the energy for adding a link) and then add a site.

Now we should derive the equation for $\eta_{\rm cav}(\s)$.
When $q/N \ll 1$, generically, the distance on the lattice between two generic
cavity spins is large (it is of the order of the size of the loops, therefore 
diverges logarithmically in the large $N$ limit). 
We will therefore assume that:
\begin{enumerate}
\item Different cavity variables become
uncorrelated. This is true only if the Gibbs measure is a {\it pure state},
and the {\it clustering property}, stating that distant variables are uncorrelated,
holds.
\item The states of the system before and after any of the previous graph
operations (\eg iteration) are related. Equivalently one should assume that
the perturbation corresponding to the variation of one of the cavity spins
remains {\sl localized} and it does not propagate to the whole lattice.
This is clearly a decorrelation hypothesis that is very similar to the previous one.
\end{enumerate}
Under these hypotheses, which are reasonable for a non frustrated homogeneous phase,
it is very simple to compute the free energy shifts defined
above.

The important point is that our hypotheses imply that the joint probability of
the cavity variables {\it factorizes}; we have 
$\eta_{\rm cav}(\s_1,\cdots,\s_q) = \eta^1_{\rm cav}(\s_1) \cdots \eta^q_{\rm cav}(\s_q)$. 
To simplify the problem, let us assume that
the cavity distributions are identical,
$\eta^i_{\rm cav}(\s)=\eta_{\rm cav}(\s)$, for all cavity variables because of our assumption that
the local environment is the same everywhere in the graph (no fluctuations of
connectivity and interactions). This is not an harmless assumption, because in general
frustration might induce different biases on different variables; we will come back to
this in next section.

Consider now an iteration where a new cavity variable is added and connected to
$c-1$ cavity variables. The new cavity variable has a probability distribution
\beq\label{cavityeq}
\begin{split}
&\eta_{\rm cav}(\s) = \frac{1}{z_{\rm cav}} \sum_{\s_1,\cdots,\s_{c-1}} 
\eta_{\rm cav}(\s_1) \cdots \eta_{\rm cav}(\s_{c-1}) \psi(\s,\s_1) \cdots \psi(\s,\s_{c-1}) \ , \\
&z_{\rm cav} = \sum_{\s}  \left( \sum_{\s'} \eta_{\rm cav}(\s') \psi(\s,\s') \right)^{c-1} \ ,
\end{split}\eeq
that must again be identical to the old ones;
this is because we assumed that the local environment does not fluctuate and that, for large
$N$, the state of the system remains the same after any of the above operations.
Therefore (\ref{cavityeq}) is a self-consistent equation that can be solved to obtain
$\eta_{\rm cav}(\s)$. We got back Eq.~(\ref{eq_fpoint_Ising}).

Once this is done, we can compute the free energy shifts for a link and site addition.
Consider a link addition between two cavity spins $\s_1$ and $\s_2$. Before addition,
we have
\beq\begin{split}
&Z_{before} = \sum_{\s} \prod_a \psi_a(\s_a) = \sum_{\s_1 \s_2} Z(\s_1,\s_2) \ , \\
&\eta_{\rm cav}(\s_1,\s_2) = \eta_{\rm cav}(\s_1)\eta_{\rm cav}(\s_2) = \frac{Z(\s_1,\s_2)}{Z_{before}} \ ,
\end{split}\eeq
where $Z(\s_1,\s_2)$ is the trace of the Gibbs measure over all variables except the
two cavity ones. When we add a link (\ie a constraint) connecting $\s_1$ and $\s_2$,
we have
\beq
Z_{after} = \sum_{\s} \psi(\s_1,\s_2) \prod_a \psi_a(\s_a) =
\sum_{\s_1 \s_2} \psi(\s_1,\s_2) Z(\s_1,\s_2) = 
Z_{before} \sum_{\s_1 \s_2} \eta_{\rm cav}(\s_1)\eta_{\rm cav}(\s_2) \psi(\s_1,\s_2) \ ,
\eeq
and we obtain the final result
\beq
\log z^{(l)} = \log \frac{Z_{after}}{Z_{before}} =
\log \sum_{\s_1 \s_2} \eta_{\rm cav}(\s_1)\eta_{\rm cav}(\s_2) \psi(\s_1,\s_2) \ ,
\eeq
which coincides with (\ref{zlinkRS}).
With a very similar argument one can show that
\beq
\log z^{(s)} = \log \sum_{\s_0,\s_1,\cdots,\s_c} 
\eta_{\rm cav}(\s_1) \cdots \eta_{\rm cav}(\s_{c}) \psi(\s_0,\s_1) \cdots \psi(\s_0,\s_{c}) 
=
\log \sum_{\s} 
\left(\sum_{\s'} \eta_{\rm cav}(\s') \psi(\s,\s')\right)^c \ .
\eeq
which gives back (\ref{zlinkRS}). Putting these results in (\ref{enetot}) we get back
Eq.~(\ref{freecavRS}). 
Finally, note that from $\eta_{\rm cav}(\s)$ we can reconstruct the true {\it marginal probability}
of the variables $\eta(\s)$, defined as the trace of the Gibbs measure over all variables but
$\s$. To do this we consider a graph with $c$ cavity spin $\s_1,\cdots,\s_c$ 
and perform a site addiction to 
get the real graph; we get
\beq
P(\s) =  \frac{1}{z^{(s)}} \sum_{\s_1,\cdots,\s_{c}} 
\eta_{\rm cav}(\s_1) \cdots \eta_{\rm cav}(\s_{c}) \psi(\s,\s_1) \cdots \psi(\s,\s_{c}) 
= \frac1{z^{(s)}} \left( \sum_{\s'} \eta_{\rm cav}(\s') \psi(\s,\s') \right)^{c} \ . 
\eeq
which is (\ref{eq_Ising_final}).
This concludes our derivation; we have then presented two slightly different
but equivalent derivations of the cavity method.

A very important property of the free energy (\ref{freecavRS}) is that it is {\it variational}\footnote{
However, it cannot be proven that the extremum is a minimum in general.
}:
one can easily show ({\bf $\Rightarrow$ Ex.\ref{sec:optim}.\ref{ex:II.6}}) that the functional equation $\frac{d f}{d \eta_{\rm cav}(\s)}=0$
gives back Eq.~(\ref{cavityeq}).
In particular, this allows for a simple computation of the energy $e = \frac{d(\b f)}{d\b}$ and
entropy $s = -\frac{df}{dT}$, because it is enough to compute the explicit derivative with respect
to $T$ without deriving with respect to $\eta_{\rm cav}(\s)$. The result 
is\footnote{Recall that we defined $\psi(\s_1,\s_2) = \exp [-\b E(\s_1,\s_2)]$.}:
\beq\label{eneentroRS}
\begin{split}
&s = \log z^{(s)} - c/2 \log z^{(l)} + \b e \ , \\
&e = e^{(s)} - c/2 e^{(l)} \ , \\
&e^{(l)} = \frac1{z^{(l)}} \sum_{\s_1 \s_2} \eta_{\rm cav}(\s_1)\eta_{\rm cav}(\s_2) \psi(\s_1,\s_2) E(\s_1,\s_2) \ , \\
&e^{(s)} = \frac1{z^{(s)}}
\sum_{\s_0,\s_1,\cdots,\s_c} 
\eta_{\rm cav}(\s_1) \cdots \eta_{\rm cav}(\s_{c}) \psi(\s_0,\s_1) \cdots \psi(\s_0,\s_{c}) 
[ E(\s_0,\s_1) + \cdots + E(\s_0,\s_c)] \ .
\end{split}\eeq
The equations above can be applied to many problems on random graphs, such as
the Ising ferromagnet ({\bf $\Rightarrow$ Ex.\ref{sec:optim}.\ref{ex:II.2}}), the SK model  ({\bf $\Rightarrow$ Ex.\ref{sec:optim}.\ref{ex:II.2bis}})
and the $q$-COL problem
({\bf $\Rightarrow$ Ex.\ref{sec:optim}.\ref{ex:II.3}}).

In the case of un-frustrated ferromagnetic models it has
been shown rigorously~\cite{DM10} that the assumption we made on the way are correct,
the predictions of the cavity method being exact in the thermodynamic limit,
both for the local magnetizations and for the free-energy per site. 

\subsubsection{Fluctuations of the local environment: distributions of cavity probabilities}

We will now allow for the presence of local fluctuations, in the form of quenched random couplings, and fluctuations of the variable connectivity.
We show here how to take the average over this quenched disorder. 
We will still restrict to $k=2$ for simplicity.
In this more general setting, the cavity variables
are not equivalent: they do not have the same distribution, because of the
explicit disorder present in the Hamiltonian, or in the graph, or both. 

In principle we could take a given realization of the disorder (graph and couplings) for finite (but large) $N$,
and try to find a solution of (\ref{eq_msg_Ising}).
This can be done in some cases numerically, but it is not very practical.
If we are interested in computing
the free energy averaged over the disorder, we
can introduce a {\it distribution of cavity distributions},
$\PP[\eta_{\rm cav}(\s)]$, that gives the probability (over the links $i\to j$) that variable
$\s_i$ has a cavity probability distribution:
\beq
\PP[\eta_{\rm cav}(\s)] = \text{Prob}[ \eta_{i \to j}(\s_i) =\eta_{\rm cav}(\s_i) ] \ .
\eeq
If we take the average over the disorder (both the couplings and the random graph), then 
this distribution must be the same
for all sites, since all sites are statistically equivalent.

The equation for $\PP$ is deduced by using Eq.~(\ref{eq_msg_Ising}) and imposing that
the probability distribution of $\eta_{i\to j}$ is the same as that of the $\eta_{k\to i}$. 
Defining
\beq\label{cavityfluctRS}
\eta^0_{\rm cav}(\s_0) = \frac{1}{z_{\rm cav}} \sum_{\s_1,\cdots,\s_{c-1}} 
\eta^1_{\rm cav}(\s_1) \cdots \eta^{c-1}_{\rm cav}(\s_{c-1}) \psi_1(\s_0,\s_1) \cdots \psi_{c-1}(\s_0,\s_{c-1})
= \FF_J[\eta^1_{\rm cav},\cdots,\eta^{c-1}_{\rm cav}] \ ,
\eeq
where the interactions $\psi_1 \cdots \psi_{c-1}$ may contain one realization of the disorder,
one obtains
\beq\label{PPeqRS}
\PP[\eta^0_{\rm cav}] = \overline{ \int d\PP[\eta^1_{\rm cav}] \cdots d\PP[\eta^{c-1}_{\rm cav}] 
\ \ \d[\eta^0_{\rm cav} - \FF_J[\eta^1_{\rm cav},\cdots,\eta^{c-1}_{\rm cav}]] }^{\{c,J\}} \ ,
\eeq
where the overline denotes an average over the couplings $J$ and over the distribution
of $c$. This equation has the following interpretation: one must extract the $\eta^i_{\rm cav}$ from $\PP$, and produce
a new $\eta^0_{\rm cav}$ that must also be typical of $\PP$. The function $\FF_J$ may depend on
the couplings that appear in the constraints; these have to be extracted from their distribution
at each iteration. 

Moreover $c$ might also be a random variable, and an important remark on the distribution of $c$ is in order.
Suppose that $P(c)$ is the distribution of the connectivity of the variables of the random graph.
We {\it should not} use $P(c)$ to take the average in Eq.~(\ref{PPeqRS}). The reason is the following. 
The cavity distributions $\eta_{i \to j}$ are defined on the links of the graph. When we perform a cavity iteration
using Eq.~(\ref{eq_msg_Ising}), we fix an oriented link $i\to j$ and consider all the other 
links $k\to i$ that are connected to site $i$. Since $\PP[\eta_{\rm cav}(\s)]$ is the histogram over the links 
of cavity distributions, to derive Eq.~(\ref{PPeqRS}) we must take a random link, not a random site. The average 
in Eq.~(\ref{PPeqRS}) is taken over {\it the probability $\wt P(c)$ that, picking uniformly at random a link $i\to j$, the variable $i$
has connectivity $c$}. It is easy to show that
\beq\label{wtPc}
\begin{split}
& \wt P(c) = \frac{c P(c)}{\overline c } \ , \\
& \overline c  = \sum_{c=0}^{\io} c P(c) \ ,
\end{split}\eeq
and the average over $c$ in Eq.~(\ref{PPeqRS}) must be taken with the distribution $\wt P(c)$.

To compute the free energy one has to average (\ref{f_Bethe}) over the distribution
$\PP$ that solves the previous equation. It is easy to see that the result is
\beq\label{fRSdis}
f =   \overline{ f^{(s)} }^{\{\PP,c,J\}} - \frac{\overline c }2 \,  \overline{ f^{(l)} }^{\{\PP,J\}} \ ,
\eeq
where
\beq\begin{split}
& \overline{ f^{(s)} }^{\{\PP,c,J\}}  = -T \overline{ \int d\PP[\eta^1_{\rm cav}] \cdots d\PP[\eta^{c}_{\rm cav}] 
 \log \left[ \sum_{\s,\s_1,\cdots,\s_c} 
\eta^1_{\rm cav}(\s_1) \cdots \eta^c_{\rm cav}(\s_{c}) \psi_1(\s,\s_1) \cdots \psi_c(\s,\s_{c}) \right] }^{\{c,J\}} \\
& \overline{ f^{(l)} }^{\{\PP,J\}} =
-T \overline{  \int d\PP[\eta^1_{\rm cav}] d\PP[\eta^{2}_{\rm cav}]  \log \left[ \sum_{\s_1 \s_2} \eta^1_{\rm cav}(\s_1)\eta^2_{\rm cav}(\s_2) \psi(\s_1,\s_2)    \right] }^J
\ .
\end{split}\eeq
We note that here the site term must be averaged using $P(c)$, since it is an average over all sites, see Eq.~(\ref{f_Bethe}).

\subsubsection{The zero temperature limit}

As we already discussed there are two ways of taking the zero-temperature limit.
The first is to assume that the system is in the SAT phase and therefore the ground
state has zero energy. In this case a solution to the constraints exist and we can just
take infinitely hard contraints ($\b \to \io$). The energy is zero as can be seen from
(\ref{eneentroRS}), and the entropy is simply $\frac1N \log Z$ and is given by
the first of (\ref{eneentroRS}) without the last term $\b e$.

On the other hand we might be interested in a situation where not all the constraints
can be satisfied at the same time and the ground state energy is non-zero. In this case
the limit $\b \to \io$ of Eq.~(\ref{cavityeq}) does not make sense since the normalization
constant might vanish when a contradiction is met.

In this case one has to take care. For simplicity, let us focus on a spin glass Hamiltonian
$H=-\sum_{(ij)} J_{ij} S_i S_j$ for Ising spins~\cite{cavity_T0}. 
The denominator in (\ref{cavityfluctRS})
is given by $z_{\rm cav} \sim e^{-\b e_{\rm cav}}$ at leading order for large $\b$. We introduce
a {\it cavity field} $h_{i \to j}$ by the parametrization, again at leading order for $\b\to\io$:
\beq
\eta_{i \to j}(S_i) \sim e^{\b (h_{i \to j} S_i - |h_{i \to j}|)} \ .
\eeq
The cavity equation becomes, substituting the previous parametrizations and taking the
leading order for $\b\to \io$:
\beq
h_0 S_0 - |h_0| =  e_{\rm cav} + \sum_{i=1}^{c-1} ( | h_i + J_{0i} S_0| - |h_i|) \ .
\eeq
We can write
\beq\begin{split}
&| h_i + J_{0i} S_0| = a(h_i,J_{0i}) + S_0 u(h_i,J_{0i}) \ , \\
& a(h_i,J_{0i}) = \frac12 (| h_i + J_{0i} | + | h_i - J_{0i} |) \ ,\\
& u(h_i,J_{0i}) = \frac12 (| h_i + J_{0i} | - | h_i - J_{0i} |) \ .
\end{split}\eeq
Then we get
\beq\begin{split}
&h_0 = \sum_{i=1}^{c-1} u(h_i,J_{0i}) \ , \\
&e_{\rm cav}=-|h_0| + \sum_{i=1}^{c-1} [|h_i| - a(h_i,J_{0i})] \ .
\end{split}\eeq
Similarly we get
\beq\label{esel}
\begin{split}
&e^{(l)} = |h_1| + |h_2| - \max_{S_1,S_2} [ h_1 S_1 + h_2 S_2 + J_{12} S_1 S_2 ] \ , \\
&e^{(s)} =   \sum_{i=1}^{c} [|h_i| - a(h_i,J_{0i})] + | \sum_{i=1}^{c} u(h_i,J_{0i} |    \ .
\end{split}\eeq

If we assume that the $J_{ij}$ are drawn from a given distribution $P(J)$, and we introduce
the distribution of cavity fields $\PP(h)$, we obtain \cite{cavity_T0} 
the zero-temperature limit of (\ref{PPeqRS})
\beq
\PP(h) =\overline{ \int d h_i \PP(h_i) \, 
\d\left(h -   \sum_{i=1}^{c-1} u(h_i,J_{0i}) \right) }^{\{c,J\}} \ ,
\eeq
where the overline is an average over $P(J)$ and, if one wishes, also on fluctuations of $c$.
The ground state energy is
\beq
e_0 = \overline{ e^{(s)} }^{\{ \PP,c,J\}  } - \frac{ \overline c }2 \,  \overline{ e^{(l)} }^{\{\PP, J\}  }  \ ,
\eeq
\ie the site and link energies (\ref{esel}) have to be averaged over the distribution
of $J$ and $c$, and over $\PP(h)$, as in Eq.~(\ref{fRSdis}).

The solution to these equations for the case $J=\pm 1$ with probability $1/2$ can be
found in \cite{cavity_T0}. Many other problems have been studied using this zero-temperature
``energetic'' formalism, including $q$-COL~\cite{BMPWZ03} and $k$-SAT~\cite{MeZe,MeMeZe}.
In the following we will focus more on the ``entropic'' limit in which one assumes to be
in the SAT phase and studies the structure of the solutions.

\subsubsection{On a factor graph}
\label{sec:factor}

Let us conclude by generalizing the previous equations to the case of a factor graph.
We will only briefly sketch the derivation, more details can be found in \cite{MRS08}.

In this case we have two type of nodes, variables and constraints. It is convenient
to introduce {\it cavity variables} as before: they are variables connected only
to $c-1$ constraints. We also introduce {\it cavity constraints}: they are constraints
that are connected only to $k-1$ variables.

To each cavity variable we associate its cavity distribution $\eta_{\rm cav}(\s)$. To each
cavity contraint, we associate a distribution $\eta_{\rm test}(\s)$ that is defined as follows:
imagine to add a $k$-th missing variable $\s$ to the cavity constraint and {\it only to it}. Then
$\eta_{\rm test}(\s)$ is the distribution of this variable.

Recall that the number of constraints is $M = c N /k$. We denote by $\GG_{N,M,p,q}$ a graph
with $N$ variables and $M$ constraints, of which $p$ are cavity variables and $q$ are 
cavity constraints.

We have five possible operations:
\begin{enumerate}
\item {\it Variable iteration}: We add a new cavity variable and connect it to
$c-1$ cavity constraints.
\item {\it Constraint iteration}: We add a new cavity constraint and connect it to
$k-1$ cavity variables.
\item {\it Site addition}: We add a new variable and connect it to
$c$ cavity constraints.
\item {\it Constraint addition}: We add a new constraint and connect it to
$k$ cavity variables.
\item {\it Link addition}: We add a new link connecting a cavity constraint and a cavity
variable.
\end{enumerate}

We can start with a $\GG_{N,M,0,0}$ graph and delete $c k$ links. Now we get a graph
$\GG_{N,M,c k,c k}$ as each link deletion produces a cavity variable and a cavity constraint.
To the latter graph we add $k$ new variables, that have to be connected to the $c k$ cavity
constraints, and $c$ new constraints, that have to be connected to the $c k$ cavity variables.
We thus produce a graph with $N+k$ variables and $M+c$ constraint (note that the relation
$M = c N/k$ still holds), $\GG_{N+k,M+c,0,0}$. Then we get
\beq\label{free-factor-rs}
F(N+k)-F(N) = k f = k \D F^{(s)} + c \D F^{(c)} - ck \D F^{(l)} \ .
\eeq

The iteration equation are the following. When we add a new cavity variable $\s$, we connect
it to $c-1$ constraints. The influence of each constraint on the new variable is independent
from the others and given by $\eta_{\rm test}(\s)$. Then we have
\beq
\eta^0_{\rm cav}(\s_0) = \frac{1}{z_{\rm cav}} \prod_{i=0}^{c-1} \eta_{\rm test}^i(\s_0)
=\FF_{\rm cav}[\eta_{\rm test}^1 \cdots 
\eta_{\rm test}^{c-1}]
 \ ,
\eeq
where in principle each constraint can produce a different distribution.

When we add a cavity constraint $\psi_a$, we connect it to $k-1$ cavity variables $\s_1,\cdots,\s_{k-1}$.
To compute the cavity distribution we need to add also a fictitious variable $\s$ that is 
connected only to the new constraint. Its distribution is then
\beq
\eta^0_{\rm test}(\s_0) = \frac{1}{z_{\rm test}} \sum_{\s_1,\cdots,\s_{k-1}} \eta_{\rm cav}^1(\s_1) \cdots 
\eta_{\rm cav}^{k-1}(\s_{k-1}) \psi_a(\s_0,\s_1,\cdots,\s_{k-1}) = \FF_{\rm test}[\eta_{\rm cav}^1 \cdots 
\eta_{\rm cav}^{k-1}]   \ .
\eeq

Finally we need to compute the free energy shifts. When we add a new constraint, $k$ cavity variables
become connected and
\beq
\log z^{(c)} = \log \sum_{\s_1,\cdots,\s_{k}} \eta_{\rm cav}^1(\s_1) \cdots 
\eta_{\rm cav}^{k}(\s_{k}) \psi_a(\s_1,\cdots,\s_{k}) \ .
\eeq
When we add a new variable, $c$ constraints are connected to it and give independent
influence, then
\beq
\log z^{(s)} = \log \sum_\s \prod_{i=1}^{c} \eta_{\rm test}^i(\s) \ .
\eeq
When we add a link, we connect a cavity variable $\s$ with distribution $\eta_{\rm cav}(\s)$ with
a cavity constraint whose influence on $\s$ is $\eta_{\rm test}(\s)$. Then
\beq
\log z^{(l)} = \log \sum_\s \eta_{\rm cav}(\s)\eta_{\rm test}(\s) \ .
\eeq
It is easy to check that the equations above reduce to the ones for a normal graph for $k=2$
({\bf $\Rightarrow$ Ex.\ref{sec:optim}.\ref{ex:II.4}}).

In presence of site fluctuations all the considerations we made in the $k=2$ case can be easily
generalized. The same holds for the zero temperature limit.
The generalization of the formalism to the factor graph allows to discuss for instance the case of
XORSAT ({\bf $\Rightarrow$ Ex.\ref{sec:optim}.\ref{ex:II.5}}).

\subsubsection{Summary}

Before turning to the {\sc 1rsb} equations, let's summarize this discussion. The replica symmetric cavity method works when the Gibbs measure
is a single pure state. Under this assumption, cavity variables are uncorrelated for $N\to\io$ since loops
are very long. The equations derived on a tree can then be used to describe a factor graph:
\begin{itemize}
\item On a given system of finite (large) size $N$, one can use Eqs.~(\ref{eq_msg_Ising}) and (\ref{f_Bethe}).
\item For a homogeneous system (such that locally there is no disorder in the couplings and the graph), in the thermodynamic
limit all the cavity fields are equal and the {\sc rs} cavity equations reduce to Eqs.~(\ref{eq_fpoint_Ising}) and (\ref{freecavRS}).
\item In presence of local disorder, one must introduce a distribution of cavity fields over the disorder; the cavity equation for this object
is given by Eq.~(\ref{PPeqRS}), and the free energy can be obtained from Eq.~(\ref{fRSdis}).
\end{itemize}
Similar equations are obtained for systems defined on factor graphs, and in the zero temperature limit.

\subsection{1-step replica symmetry breaking}

In the one-step replica symmetry breaking scenario, we will assume that the Gibbs state
is split in a large number of states, as we obtained for the spherical $p$-spin model.
Each state has a weight $w_\a \propto \exp(-\b N f_\a)$ in the partition function.
As discussed in section \ref{sec:realreplica}, in such a situation we wish to compute
\beq\label{partmcavity}
Z_m = \sum_a e^{-\b N m f_\a} = \int df e^{N [\Si(f) - \b m f]} \ ,
\eeq
from which the thermodynamics of the system can be reconstructed.

Recall that the central hypothesis of the {\sc rs} cavity method is that (distant) 
cavity spins are uncorrelated and the Gibbs state is stable (for $N\to\io$)
under the operations on the graph defined above.
Both these properties are false in the presence of many states, because {\it i)} the
Gibbs state is not a pure state and the decorrelation (clustering) property does not
hold, and {\it ii)} the free energy of the states are shifted when operating on the
graph, and this might change the relative weight of the states in the partition function,
therefore changing the nature of the Gibbs state.
The treatment of the free energy shift is quite complicated~\cite{cavity}. Therefore
in the following we will not use the derivation of the {\sc 1rsb} cavity equations 
based on the graph operation of section~\ref{sec:cavity2}. This derivation can be found
in \cite{cavity}.
We will present instead a derivation based on the idea of first writing the recurrence equations
on the tree and then justify its use on the random graph by a decorrelation assumption~\cite{MM09}.
This derivation has also the advantage that it is formulated for a single graph and choice of
the couplings, so one does not need to take the average over the disorder.

In the {\sc rs} treatment, the key property is the factorization of the joint distribution
of the cavity variables, expressed by Eq.~(\ref{cavity_decorr}), that leads to the closed
equations (\ref{eq_msg_Ising}) for the cavity messages. Based on the discussion above,
in presence of many pure states we can only assume that Eq.~(\ref{cavity_decorr}) is true
for the messages $\eta_{i\to j}^\a$ restricted to one pure state.
A precise formalization of this hypothesis is the following. 
By definition, we must be able to select one state by acting on each spin with infinitesimal
field $h_i^\a$. If we take {\it first} the limit $N\to \io$ and {\it then} the limit $h_i^\a \to 0$,
we will end up in the state $\a$. 
For $N\to \io$ in presence of $h_i^\a$, we have a single pure state and we can use the
{\sc rs} cavity equations to obtain a set of messages $\eta_{i\to j}^\a$. Then we can take the
limit $h_i^\a$, and the resulting messages will describe the state $\a$ in absence of the external
field. Therefore the $\eta_{i\to j}^\a$ are a fixed point of the {\sc rs} cavity equation (\ref{eq_msg_Ising}).
The free energy of a state is given by equation (\ref{f_Bethe}), calculated in the fixed point $\eta_{i\to j}^\a$.
While finding the fixed points analytically is not possible, 
we could hope to determine them numerically. The problem is that often, in presence of many
fixed points, an iterative solution of equation (\ref{eq_msg_Ising}) is not possible because the recursion
will not converge if one starts with random messages. Still, in many cases we do not need to know
the full set of fixed points. We only want to count how many solutions have a given free energy $f$, to
compute the complexity, and this can be done by mean of Eq.~(\ref{partmcavity}).

\subsubsection{The auxiliary model}

Let's summarize the situation, and rewrite once again the relevant equations (\ref{eq_msg_Ising}) and (\ref{f_Bethe}), 
in the case $k=2$ (simple graph) for simplicity. 
We should find
all the solutions of the {\sc rs} equations
\beq\label{eqrs1}
\eta_{i \to j}(\s_i) = \frac{\psi_i(\s_i)}{z_{i \to j}} 
\prod_{k \in \dimj} \left(\sum_{\s_k} \eta_{k \to i}(\s_k) 
 \psi_{ik}(\s_i,\s_k)  \right) = \FF_{i \to j}[\eta_{k \to i}, k\in \dimj] \ , 
\eeq
and make use of the Bethe free energy to obtain the free energy of each state:
\beq\label{eqrs2}
-\b F_{\rm Bethe}[\{\eta_{i\to j}\}] =  \sum_i \log z_i - \sum_{\ij} \log z_{ij}  \ , \\
\eeq
where 
\beq\label{eqrs3}
\begin{split}
&z_{i \to j} = \sum_{\s_i} \psi_i(\s_i)
\prod_{k \in \dimj} \left(\sum_{\s_k} \eta_{k \to i}(\s_k) 
 \psi_{ik}(\s_i,\s_k)  \right) \ , \\
&z_i = \sum_{\s_i} \psi_i(\s_i) \prod_{j \in \di} \left( 
\sum_{\s_j} \eta_{j \to i}(\s_j)  \psi_{ij}(\s_i,\s_j)  \right) \ , \\
&z_{ij} = \sum_{\s_i,\s_j} \eta_{j \to i}(\s_j)  \eta_{i \to j}(\s_i) \psi_{ij}(\s_i,\s_j) = \frac{z_j}{z_{j \to i}} = \frac{z_i}{z_{i \to j}} \ . \\
\end{split}\eeq
In the equations above, 
we also added a local term $\psi_i(\s_i)$, which may represent
a local field $\psi_i(\s_i) = e^{\b h_i \s_i}$. One can easily understand how to place this factor
by doing the calculation on the tree. The solution of the equations above gives the free energy of a generic statistical mechanics model
having the partition function
\beq
Z = \sum_{\{\s_i\}} \prod_i \psi_i(\s_i) \prod_{\ij} \psi_{\ij}(\s_i,\s_j) \ .
\eeq

Note that on random graphs, the cavity messages $\eta_{i\to j}^\a$ play the role of the local magnetizations
in fully connected models: they fully specify a given state $\a$ of the system.
We can write the partition function (\ref{partmcavity}) as an integral\footnote{
For continuous messages, it is not completely clear what is the correct integration measure $\DD\eta_{i \to j}$ in 
Eq.~(\ref{Zmmess}). In principle, the delta functions require the introduction
of a determinant of the second derivatives since one should write
\beq\begin{split}
Z_m & = \sum_\a e^{-\b m F_{\rm Bethe}[\eta^\a_{i\to j}]} =
\int d\eta_{i \to j} e^{-\b m F_{\rm Bethe}[\eta_{i\to j}]}  \sum_\a  \prod_{i,j} \d\left[ \eta_{i\to j} - \eta^\a_{i\to j} \right] \\
& = \int d\eta_{i \to j}  
e^{-\b m F_{\rm Bethe}[\eta_{i\to j}]}
\left(\prod_{i,j} \d\left[ \eta_{i\to j} - \FF_{i\to j} \right] \right)
\det \left[ \frac{\partial ( \eta_{i\to j} - \FF_{i\to j} )}{\partial \h_{k \to l}} \right]
\end{split}\eeq
Here we don't discuss this issue in details and just neglect the determinant. 
A discussion, based on a discretization of the cavity distributions, can be found
in \cite[pag.436]{MM09} (thanks to P.Urbani for pointing out this problem).
}
over the messages $\eta_{i \to j}$:
\beq\label{Zmmess}
Z_m = \sum_\a e^{-\b m F_{\rm Bethe}[\eta^\a_{i\to j}] } =
\int \DD\eta_{i \to j} \ e^{-\b m F_{\rm Bethe}[\eta_{i\to j}] } \prod_{\ij} \d\left[ \eta_{i\to j} - \FF_{i\to j} \right]
\d\left[ \eta_{j\to i} - \FF_{j\to i} \right]
\eeq
where on each link $\ij$ we have two messages, $\eta_{i \to j}$ and $\eta_{j \to i}$, and the two delta functions
enforce the {\sc rs} cavity equations (\ref{eqrs1}).
Using (\ref{eqrs2}), we get
\beq\label{Zm_eta}
\begin{split}
Z_m & = 
\int \DD\eta_{i \to j} \prod_i z_i^m \prod_{\ij} z_{ij}^{-m} \d\left[ \eta_{i\to j} - \FF_{i\to j} \right]
\d\left[ \eta_{j\to i} - \FF_{j\to i} \right] \\
& = 
\int \DD\eta_{i \to j} \prod_i \left( z_i^m \prod_{j \in \di}\d\left[ \eta_{i\to j} - \FF_{i\to j} \right] \right)
 \prod_{\ij} z_{ij}^{-m}  \\
 & = \int \DD\eta_{i \to j} \prod_i \Psi_i( \{ \eta_{i\to j}, \eta_{j\to i} \}_{j \in \di}) \prod_{\ij} \Psi_{ij}( \eta_{i\to j}, \eta_{j\to i} )
 \end{split}\eeq
To simplify the notations we will often omit the arguments of the different functions. 
There are two interaction terms in the last equation above. The first term is a product over all the sites
$i$ of a term $\Psi_i$ that depends on all the messages involving site $i$. The second term is a product over
all the links of a term $\Psi_{ij}$ that depends only on the messages living on that link.

We can therefore interpret Eq.~(\ref{Zm_eta}) as the partition function of a statistical mechanics
model, where the variables are the messages. On each link of the original graph, 
we have a variable made by the two
messages $\{ \eta_{i\to j}, \eta_{j\to i}\}$ with a local field $z_{ij}^{-m}$; 
on each site of the original graph, we have a {\it many-body} 
interaction $z_i^m \prod_{j \in \di}\d\left[ \eta_{i\to j}  - \FF_{i\to j} \right] $.
We can therefore re-interpret the original graph as a factor graph, where on each link there is a variable
node, and the original variable nodes act as interaction nodes.

\subsubsection{RS cavity equations for the auxiliary model: the 1RSB equations}

We can write the {\sc rs} cavity equations for the auxiliary model in a compact way by introducing
the message $Q_{i\to j}( \eta_{i\to j}, \eta_{j\to i})$, which is the probability distribution of the two
messages sitting on $\ij$ when the connection between this link and node $j$ is absent.

A quite straightforward calculation, similar to the one of section~\ref{sec:treerec}, leads to
\beq
Q_{i \to j}( \eta_{i\to j}, \eta_{j\to i}) = \frac{ \Psi_{ij}( \eta_{i\to j}, \eta_{j\to i}  )}{\ZZ_{i\to j}}
\sum_{ \{  \eta_{i\to k}, \eta_{k\to i} \}_{ k \in \dimj }} \Psi_i( \{ \eta_{i\to l}, \eta_{l\to i} \}_{l \in \di})
\prod_{ k \in \dimj} Q_{k \to i}( \eta_{k\to i}, \eta_{i\to k}) 
\eeq
Using the explicit expressions of $\Psi_i$ and $\Psi_{ij}$ we obtain
\beq\begin{split}
Q_{i \to j}&( \eta_{i\to j}, \eta_{j\to i}) = \\
& = \frac{ z_{ij}(  \eta_{i\to j}, \eta_{j\to i} )^{-m} }{\ZZ_{i\to j}}
\sum_{ \{  \eta_{i\to k}, \eta_{k\to i} \}_{ k \in \dimj }}
z_i(\{\eta_{k\to i}\}_{ k \in \di})^m \prod_{l \in \di}\d\left[ \eta_{i\to l} - \FF_{i\to l} \right]
\prod_{ k \in \dimj} Q_{k \to i}( \eta_{k\to i}, \eta_{i\to k}) \\
& = \frac{ z_{ij}(  \eta_{i\to j}, \eta_{j\to i}  )^{-m} }{\ZZ_{i\to j}}
\sum_{ \{ \eta_{k\to i} \}_{ k \in \dimj }}
z_i(\{\eta_{k\to i}\}_{k \in \di})^m \d\left[ \eta_{i\to j} - \FF_{i\to j}[ \{ \eta_{k\to i} \}_{ k \in \dimj }] \right]
\prod_{ k \in \dimj} Q_{k \to i}( \eta_{k\to i}, \FF_{i\to k}) \\
& = \frac{ 1 }{\ZZ_{i\to j}}
\sum_{ \{ \eta_{k\to i} \}_{ k \in \dimj }}
z_{i\to j}(\{\eta_{k\to i}\}_{k \in \dimj})^m \d\left[ \eta_{i\to j} - \FF_{i\to j}[ \{ \eta_{k\to i} \}_{ k \in \dimj }] \right]
\prod_{ k \in \dimj} Q_{k \to i}( \eta_{k\to i}, \FF_{i\to k}) \\
\end{split}\eeq
where in the second step we used the delta functions to integrate over the $\eta_{k \to i}$, and in the third
step we used the identity $z_i / z_{ij} = z_{i\to j}$, see Eq.~(\ref{eqrs3}). This last simplification makes 
$\eta_{j \to i}$ disappear from the last line of the equation above. 
The only point where $\eta_{j\to i}$ appears in the right hand side of the above equations is in the argument
of $\FF_{i\to k}$ inside the function $Q$. Therefore, a consistent choice is to assume that 
$Q_{i\to j}$ does not depend
on $\eta_{j \to i}$, or in other word $Q$ does not depend on its second argument. Using this assumption, we 
finally obtain
\beq\label{eq_msg_1rsb}
Q_{i \to j}( \eta_{i\to j}) = \frac{ 1 }{\ZZ_{i\to j}}
\sum_{ \{ \eta_{k\to i} \}_{ k \in \dimj }}
z_{i\to j}(\{\eta_{k\to i}\}_{k \in \dimj})^m \d\left[ \eta_{i\to j} - \FF_{i\to j}[ \{ \eta_{k\to i} \}_{ k \in \dimj }] \right]
\prod_{ k \in \dimj} Q_{k \to i}( \eta_{k\to i}) \ .
\eeq
Note that $Q_{i\to j}$ is the probability distribution of $\eta_{i\to j}$ over the states, with weight $e^{-\b m F_\a}$,
or in formula:
\beq
Q_{i\to j}( \eta_{i\to j}) = \sum_\a e^{-\b m F_{\rm Bethe}[\eta^\a_{i\to j}]} \d[ \eta_{i \to j} - \eta^\a_{i \to j}] \ .
\eeq

Following the same steps as in section~\ref{sec:treerec} (the details are slightly different because of the different graph
structure of the auxiliary partition function), we find that the ``replicated" free energy is
\beq\label{free_1rsb}
\begin{split}
-\b N & \Phi(m,T) = \log Z_m =  \sum_i \log \ZZ_i - \sum_{ij} \log \ZZ_{ij} \ , \\
& \ZZ_i =  \sum_{ \{ \eta_{k \to i} \}_{k \in \di } } z_i( \{ \eta_{k \to i} \}_{k \in \di })^m \prod_{k \in \di} Q_{k \to i}(\eta_{k \to i})  = \la  z_i^m \ra_\a \ , \\
& \ZZ_{ij} =   \sum_{\eta_{i\to j}, \eta_{j\to i} } 
z_{ij}( \eta_{i\to j}, \eta_{j\to i} )^m Q_{i\to j}(\eta_{i \to j}) Q_{j\to i}(\eta_{j\to i}) = \la z_{ij}^m \ra_\a \ , \\
\end{split}\eeq

Eq.~(\ref{eq_msg_1rsb}) and Eq.~(\ref{free_1rsb}) constitute the set of {\sc 1rsb} cavity equations for a given graph and choice of the couplings.
The reader has probably already guessed that {\sc 2rsb} equations could in principle be obtained by performing a {\sc 1rsb} calculation
on the auxiliary model, and so on. Unfortunately the complexity of the calculation is already prohibitive at {\sc 2rsb} so we will not explore
further this possibility.

\subsubsection{Homogeneous 1RSB equations}

As we did in the {\sc rs} case, we can now consider a system where there are no spatial fluctuations: the graph
is regular and the coupling are all equal (an important example is the coloring of random regular graphs, that as we
already said corresponds to the Potts antiferromagnetic model).

In that case, it is natural to assume that the distribution $Q_{i\to j}$ does not depend on the particular site.
Even if in a given glass state $\a$ the local fields are different, once we take the average over the states, all the sites
must be statistically equivalent in the thermodynamic limit.
Then we have $Q_{i\to j}(\eta_{i \to j}) \to Q_{\rm cav}(\eta_{\rm cav})$ for $N\to \io$, and Eq.~(\ref{eq_msg_1rsb}) becomes:
\beq\begin{split}\label{cavity1rsb}
Q_{\rm cav}[\eta_{\rm cav}] &= \frac{1}{\ZZ_{\rm cav}} \int dQ[\eta^1_{\rm cav}] \cdots dQ[\eta^{c-1}_{\rm cav}] \ 
\d\big[\eta_{\rm cav} - \FF[\eta^1_{\rm cav},\cdots,\eta^{c-1}_{\rm cav}]\big] \ \{ z_{\rm cav}[\eta^1_{\rm cav},\cdots,\eta^{c-1}_{\rm cav}] \}^m \\
& = \FFF\big\{ Q[\eta^1_{\rm cav}] \cdots Q[\eta^{c-1}_{\rm cav}]\big\}
\ ,
\end{split}\eeq
where the function $\FF$ is defined in (\ref{cavityfluctRS}). 

Now we can compute the free energy.
This is done by noting that in Eq.~(\ref{free_1rsb}) all the terms are equal if $Q_{i \to j} \to Q$.
We get
\beq\begin{split}
& \ZZ^{(l)} = \int dQ[\eta^1_{\rm cav}] dQ[\eta^2_{\rm cav}] \ \ \{ z^{(l)}[\eta^1_{\rm cav},\eta^2_{\rm cav}] \}^m \ , \\
& z^{(l)}  = \sum_{\s_1 \s_2} \eta^1_{\rm cav}(\s_1)\eta^2_{\rm cav}(\s_2) \psi(\s_1,\s_2) \ . \\
\end{split}\eeq
and similarly
\beq
\ZZ^{(s)} = \int dQ[\eta^1_{\rm cav}] \cdots dQ[\eta^{c}_{\rm cav}] \ \ 
\{ z^{(s)}[\eta^1_{\rm cav},\cdots,\eta^{c}_{\rm cav}] \}^m \ ,
\eeq
with $z^{(s)}$ given in (\ref{zsiteRS}). 
The free energy $\Phi(m,T) = -\frac{T}N \log Z_m$ is given by
\beq
\Phi(m,T) = -T \big[ \log \ZZ^{(s)}  -\frac{c}{2} \log \ZZ^{(l)} \big] \ .
\label{enetot1rsb}
\eeq
The equation above is still variational; differentiating it with respect to $Q[\h_{\rm cav}]$ leads to the self-consistency equation
(\ref{cavity1rsb}).

A final remark is in order to conclude this discussion.
Eq.~(\ref{cavity1rsb}) is dangerously reminiscent of equation (\ref{PPeqRS}), that corresponds to the {\sc rs} case in presence of 
local fluctuations (indeed, the two equations are formally equivalent at $m=0$). These two equations should not be confused.
There is a deep physical difference between the two distributions $Q[\eta]$ and $\PP[\eta]$: the former
describes the fluctuations over the many states for a given sample, while the latter describes the fluctuations over the samples
of a single pure state. For this reason, in (\ref{PPeqRS}) the weight $z_{\rm cav}^m$ is absent. The physical difference between these equations
is better seen if we compare the free energy (\ref{fRSdis}) with the one we derived here, Eq.~(\ref{enetot1rsb}). In the former case, the average
over $\PP[\eta]$ represents an average over the disorder and is taken {\it outside} the logarithm. In the latter case, the average over $Q[\h]$ is
over states for a given sample, and therefore it is taken {\it inside} the logarithm.

\subsubsection{Complications: spatial fluctuations, factor graphs}

Now it should be clear how to introduce spatial fluctuations. Instead of considering a single $Q[\h_{\rm cav}]$,
we must introduce a distribution {\it over the sites}, $\PP\big[Q[\h_{\rm cav}]\big]$, defined as the probability
that a cavity variable has distribution of cavity fields $Q[\h_{\rm cav}]$ {\it over the states}.
The equation (\ref{cavity1rsb}) now depends on the disorder, \ie $\FFF_{J}$ depends on the coupling and the number
$c$ might fluctuate. Similarly to (\ref{PPeqRS}) we get
\beq\label{PPeq1RSB}
\PP\big[Q \big] = \overline{\int d\PP\big[Q^1 \big] \cdots d\PP\big[Q^{c-1} \big]
\delta\big[Q - \FFF_J[Q^1,\cdots,Q^{c-1}] \big]}^{\{c,J\}} \ .
\eeq
The free energy has now to be computed according to (\ref{enetot1rsb}) taking an external average over $c,J,\PP$.

In the case of a factor graph all these considerations are easily generalized along the lines of section \ref{sec:factor}.
We must introduce a $Q_{\rm cav}[\h_{\rm cav}]$ and a $Q_{\rm test}[\h_{\rm test}]$, with recursions
\beq\begin{split}\label{cavity1rsb-factor}
Q_{\rm cav}[\eta_{\rm cav}] &= \frac{1}{\ZZ_{\rm cav}} \int dQ_{\rm test}[\eta^1_{\rm test}] \cdots dQ_{\rm test}[\eta^{c-1}_{\rm test}] \ 
\d[\eta_{\rm cav} - \FF_{\rm cav}[\eta^1_{\rm test},\cdots,\eta^{c-1}_{\rm test}]] \ \{ z_{\rm cav}[\eta^1_{\rm test},\cdots,\eta^{c-1}_{\rm test}] \}^m \\
& = \FFF_{\rm cav}\big\{ Q_{\rm test}[\eta^1_{\rm test}] \cdots Q_{\rm test}[\eta^{c-1}_{\rm test}]\big\}
\ , \\
Q_{\rm test}[\eta_{\rm test}] &= \frac{1}{\ZZ_{\rm test}} \int dQ_{\rm cav}[\eta^1_{\rm cav}] \cdots dQ_{\rm cav}[\eta^{k-1}_{\rm cav}] \ 
\d[\eta_{\rm test} - \FF_{\rm test}[\eta^1_{\rm cav},\cdots,\eta^{k-1}_{\rm cav}]] \ \{ z_{\rm test}[\eta^1_{\rm cav},\cdots,\eta^{k-1}_{\rm cav}] \}^m \\
& = \FFF_{\rm test}\big\{ Q_{\rm cav}[\eta^1_{\rm cav}] \cdots Q_{\rm cav}[\eta^{k-1}_{\rm cav}]\big\}
\ .
\end{split}\eeq
The free energy $\Phi(m,T)$ is computed as in (\ref{free-factor-rs}) by replacing $z \to \ZZ$ as in the $k=2$ case:
\beq\label{phi1rsbfactor}
\Phi(m,T) =\D F^{(s)} + \frac{c}k \D F^{(c)} - c \D F^{(l)} =
-T \big[ \log \ZZ^{(s)}  +\frac{c}{k} \log \ZZ^{(c)} - c  \log \ZZ^{(l)} \big] \ .
\eeq

In presence of external disorder one introduces distributions $\PP_{\rm cav}\big[Q[\h_{\rm cav}]\big]$,
$\PP_{\rm test}\big[Q[\h_{\rm test}]\big]$,
of these fields and equations like (\ref{PPeq1RSB}),
\beq\begin{split}\label{PPeq1RSB-factor}
&\PP_{\rm cav}\big[Q_{\rm cav} \big] = \overline{\int d\PP_{\rm test}\big[Q^1_{\rm test} \big] \cdots d\PP_{\rm test}\big[Q^{c-1}_{\rm test} \big]
\delta\big[Q_{\rm cav} - \FFF_{J,{\rm cav}}[Q^1_{\rm test},\cdots,Q^{c-1}_{\rm test}] \big]}^{\{J, c\}} \ , \\
&\PP_{\rm test}\big[Q_{\rm test} \big] = 
\overline{\int d\PP_{\rm cav}\big[Q^1_{\rm cav} \big] \cdots d\PP_{\rm cav}\big[Q^{k-1}_{\rm cav} \big]
\delta\big[Q_{\rm test} - \FFF_{J,{\rm test}}[Q^1_{\rm cav},\cdots,Q^{k-1}_{\rm cav}] \big]}^{\{J\}} \ ,
\end{split}\eeq
and the free energy (\ref{phi1rsbfactor}) has to be averaged over $J,c$, and $\PP$.

The explicit solution of the {\sc 1rsb} cavity equations is possible for the case of $k$-XORSAT ({\bf $\Rightarrow$ Ex.\ref{sec:optim}.\ref{ex:II.7}}).
This is a very useful exercise that will allow to familiarize with {\sc 1rsb} cavity equations, and to check that we can obtain in this way the same result
that we already obtained by mean of the leaf removal algorithm and the replica method ({\bf $\Rightarrow$ Ex.\ref{sec:optim}.\ref{ex:II.1}}).

\subsection{Phase transitions in $q$-COL}

We wish to conclude the discussion of the cavity method by presenting the spectacular results that have been obtained for
the $q$-coloring of random graphs. These results are particularly interesting because,
at variance with XORSAT, $q$-COL has a nontrivial $\Si(s)$ which
seems to be common also to other optimization problems \cite{KrMoRiSeZd,KZ08,MRS08}.
In this case the {\sc 1rsb} equations have to be solved numerically. As we already discussed, 
$q$-COL corresponds to an antiferromagnetic Potts model. There is no disorder in the 
coupling and one has only two-body interactions, hence there is no need to introduce a factor
graph representation. For regular random graphs, there is no local disorder at all.
At the {\sc 1rsb} level, one can use therefore Eq.~(\ref{cavity1rsb}) and (\ref{enetot1rsb}) to compute
the free energy. For Erd\H{o}s-R\'enyi graphs, there is local disorder due to the fluctuations of the connectivity
and one needs to solve Eq.~(\ref{PPeq1RSB}). There are several tricks that help the numerical resolution
of these complicated equations.
We do not discuss here the details and we refer to the original paper \cite{KZ08}, from which the following discussion
is reprinted.

\begin{figure}
\begin{center}
\includegraphics[width=\linewidth]{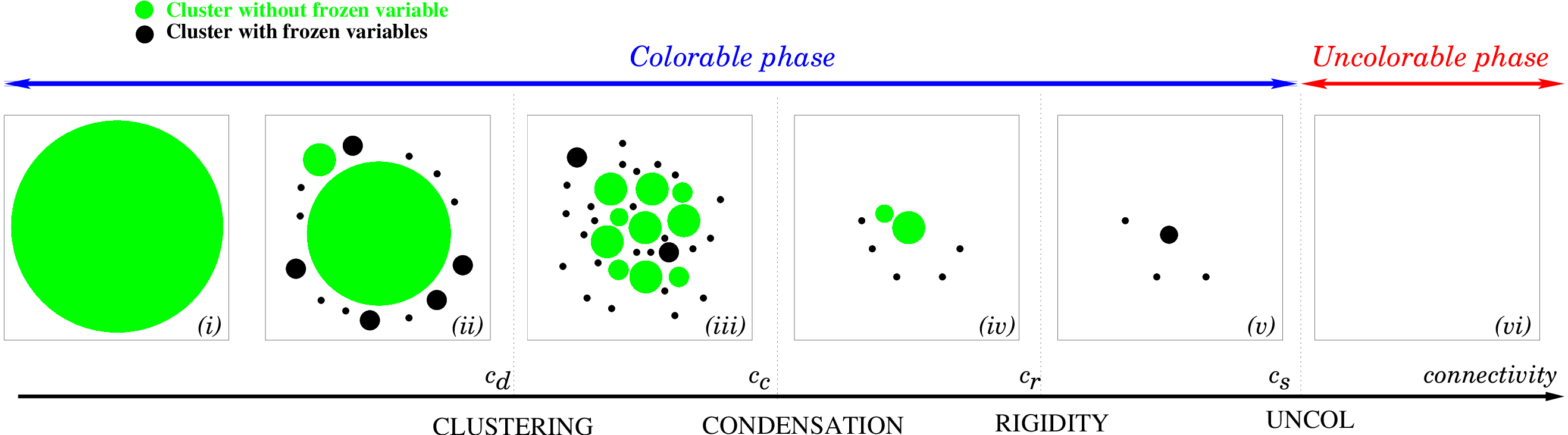}
\end{center}
\caption{(From \cite{KZ08}; {\it replace $c_c$ by $c_K$ in the figure})
Sketch of the space of solutions ---colored
  points in this representation--- in the $q$-coloring problem on random
  graphs when the connectivity $c$ is increased.  (i) At low $c$, all
  solutions belong to a single cluster. (ii) For larger $c$, other clusters of
  solutions appear but a giant cluster still contains almost all solutions.  (iii)
  At the clustering transition $c_d$, it splits into an exponentially large
  number of clusters.  (iv) At the condensation transition $c_K$, most
  colorings are found in the few largest of them. (v) The rigidity transition
  $c_r$ ($c_r<c_K$ and $c_r>c_K$ are both possible depending on $q$) arises
  when typical solutions belong to clusters with frozen variables (that are
  allowed only one color in the cluster). (vi) No proper coloring exists beyond the
  COL/UNCOL threshold $c_s$.}
\label{fig1} 
\end{figure}

Consider that we have $q\ge 4$ colors (the $q=3$ case being a bit particular
\cite{KrMoRiSeZd,KZ08}, as we shall see) and a large Erd\H{o}s-R\'enyi  
random graph whose average
connectivity $c$ we shall increase continuously. Different phases are encountered that we
will now describe (and enumerate) in order of appearance (the corresponding
phase diagram is depicted in figure \ref{fig1}).

\begin{itemize}
\item[(i)]{\bf A unique cluster exists}: For low enough connectivities, all the
  proper colorings are found in a single cluster, where it is easy to ``move''
  from one solution to another. The {\sc 1rsb} equation reduce to the {\sc rs}
ones and the cavity fields are uniform. The entropy can
  be computed and reads in the large graph size $N$ limit
\beq
s_{tot} = \frac{\log{\cal{N}}} N = \log{q} + \frac {c}{2}
\log{(1-\frac{1}{q})}\, .\label{S_RS} 
\eeq
This corresponds to the region $T>T_{TAP}$ in figure~\ref{fig1:TAP}.

\item[(ii)] {\bf Some (irrelevant) clusters appear}: As the connectivity is
  slightly increased, a {\sc 1rsb} solution appears and 
the phase space of solutions decomposes into a large
  (exponential) number of different clusters. It is tempting to identify that
  as the clustering transition, but it happens that all (but one) of these
  clusters contain relatively very few solutions ---as compare to the whole set---
  and that almost all proper colorings still belong to one single giant
  cluster.  Clearly, this is not a proper clustering phenomenon and in fact,
  for all practical purpose, there is still only one single cluster.
  Equation (\ref{S_RS}) still gives the correct entropy at this stage. This corresponds
to $T_{TAP} > T >T_d$ in figure~\ref{fig1:TAP}.

\item[(iii)] {\bf The clustered phase}: For larger connectivities, the large
  single cluster also decomposes into an exponential number of smaller ones:
  this now defines the genuine clustering threshold $c_d$. 
  Beyond this threshold, a local algorithm that
  tries to move in the space of solutions will remain trapped in a cluster of
  solutions \cite{MoSe2}.
Interestingly, as in the case of the fully connected $p$-spin model and of $k$-XORSAT, 
it can be shown that the total
  number of solutions is still given by equation (\ref{S_RS}) in this phase.  This
  is because, as we already discussed for the $p$-spin, the free energy has no
  singularity at the dynamical transition (which is therefore not a true
  transition, but rather a dynamical or geometrical transition in
  the space of solutions). This region corresponds to $T_K < T < T_d$ in figure~\ref{fig1:TAP}.

\item[(iv)] {\bf The condensed phase}: As the connectivity is further
  increased, a new sharp phase transition arises at the condensation threshold
  $c_K$ where most of the solutions are found in a finite number of clusters (the largest). 
  From this point, equation (\ref{S_RS}) is not valid anymore 
  and becomes just an upper bound.  The entropy is non-analytic at $c_K$, therefore
  this is a genuine static phase transition. This correspond to $T<T_K$ in figure~\ref{fig1:TAP}.

\item[(v)] {\bf The rigid phase}: 
Recall that in XORSAT there was a finite fraction of frozen variables (the backbone)
in each cluster. Here the situation is different and two
  different types of clusters exist: in the first type, that we shall call the
  {\it unfrozen} ones, all spins can take at least two different colors.  In
  the second type, however, a finite fraction of spins are allowed only one
  color within the cluster and are thus ``frozen'' into this color. 
It follows that a transition exists, that we call {\it rigidity},
  when frozen variables appear inside the dominant clusters (those that
  contain most colorings).  If one takes a proper coloring at
  random beyond $c_r$, it will belong to a cluster where a finite fraction of
  variables is frozen into the same color.  Depending on the value of $q$,
  this transition may arise before or after the condensation transition.

\item[(vi)] {\bf The UNCOL phase}: Eventually, the connectivity $c_s$ is
  reached beyond which no more solutions exist. The ground state energy
  (as sketched in figure \ref{fig:Tc}) is zero for $c<c_{s}$ and then grows
  continuously for $c>c_{s}$. The values $c_s$ computed within the cavity
  formalism are in perfect agreement with the rigorous bounds
  derived using probabilistic methods and are widely believed to be exact
  (although this remains to be rigorously proven).
\end{itemize}
Precise values of all threshold connectivities corresponding to all
these transitions are reported in \cite{KZ08} for the regular and the Poissonian
({\it i.e.~}Erd\H{o}s-R\'enyi) random graphs ensembles. The peculiarity of
$3$-coloring is that $c_d=c_K$ so that the clustered phase is
always condensed in this case. 

\subsection{Exercises}

\begin{enumerate}
\item \label{ex:II.1}
{\bf XORSAT with replicas}:
The aim of this (long) exercise is to re-obtain the results on the clustering in XORSAT discussed
in section~\ref{sec:XORSAT} by means of the replica method discussed in section~\ref{sec:realreplica}.
Consider $k$-XORSAT on an Erd\"os-R\'eny graph of mean connectivity $c = \a k$.
Recall that this means just that the variables entering in each equation are taken independently at random.
It is convenient here to use boolean variables, $X = (x_1,\cdots,x_N)$, and the form of the constraints
is $x_{i_1} + \cdots + x_{i_k} = b_i$, with $b_i$ a random boolean variable.
In the following we discuss the problem for general $k$ but one might first try to do the exercise for $k=3$ for
simplicity.

In the replica method (section~\ref{sec:realreplica}) we wish to compute the entropy of $m$ coupled replicas $\SS(m)$. 
The partition function of one replica is $Z = \NN=\sum_X \I(X)$, \ie the number of solutions. 
Based on the discussion of sections~\ref{sec:realreplicapspin} and~\ref{sec:1rsbpspin}, we have: 
\beq
\SS(m) = \frac{1}{N} \overline{ \log \NN_m} = \frac{1}{N} \lim_{n\to 0} \partial_n \overline{(\NN_m)^n} = \frac{1}{N}\lim_{n\to 0}  \partial_n \overline{\NN^{m n}} \ ,
\eeq
where the $\nu = m n$ replicas are divided in blocks of $m$ coupled replicas.
Then we want to compute all the moments of $\NN$. Note that the computation of the first two moments
has already been discussed in section~\ref{sec:XORSAT}.

\begin{itemize}
\item Following the same route than for the second moment, show that
\beq
\overline{\NN^\nu} = \sum_{X^1,\cdots,X^\nu} \overline{\prod_{a=1}^M \I(X^1)\cdots \I(X^\nu)} =
\sum_{X^1,\cdots,X^\nu} [p(X^1,\cdots,X^\nu)]^M \ , 
\eeq
where $p(X^1,\cdots,X^\nu)$ is the probability that all configurations $X^a$ are solutions of a randomly
drawn equation.
\item Denote $\vec X = (X^1,\cdots,X^\nu)$ and $\vec x=(x^1,\cdots,x^\nu)$; denote by $\vec 0$ and $\vec 1$ the
vectors of all $0$ and $1$ respectively. Define $\vec x + \vec y = (x^1 + y^1,\cdots , x^\nu + y^\nu)$. 
Show that, for large $N$,
\beq\begin{split}
&p(\vec X) \sim \frac{1}{N^k} \sum_{i_1,\cdots,i_k}^{1,N} \EE(\vec x_{i_1}, \cdots, \vec x_{i_k} ) \ , \\
&\EE(\vec x_1, \cdots, \vec x_k )  = \frac12 \left[ \delta(\vec x_1 + \cdots + \vec x_k = \vec 0) +
 \delta(\vec x_1 + \cdots + \vec x_k = \vec 1) \right] \ .
\end{split}\eeq
Introduce the function
\beq
\r(\vec x | \vec X) = \frac{1}{N} \sum_{i=1}^N \delta(\vec x = \vec x_i) \ ,
\eeq
note that it is normalized to 1 when summed over $\vec x$, and show that
\beq
p(\vec X) = \sum_{\vec x_1,\cdots, \vec x_k} \r(\vec x_1 | \vec X) \cdots \r(\vec x_k | \vec X) \EE(\vec x_1, \cdots, \vec x_k ) \ .
\eeq
\item Using the previous results, denote by $r(\vec x)$ a generic normalized function,
and show that
\beq
\overline{\NN^\nu} = \int Dr(\vec x) \MM[r(\vec x)] \left[\sum_{\vec x_1,\cdots, \vec x_k} r(\vec x_1) \cdots r(\vec x_k) \EE(\vec x_1, \cdots, \vec x_k ) \right]^M \ ,
\eeq
where $\MM[r(\vec x)]$ is the number of replicated configurations $\vec X$ giving rise to the same $\r(\vec x | \vec X) = r(\vec x)$. Show that the latter
is given by the multinomial factor
\beq
\MM[r(\vec x)] = \frac{N !}{\prod_{\vec x} (N r(\vec x) !)} \ .
\eeq
\item Take the large $N$ limit with $M=\a N$ and deduce that
\beq\label{ZnXORSATrep}
\overline{\NN^{\nu}} =\exp \left\{ N \max_{r(\vec x)} \left[      - \sum_{\vec x} r(\vec x) \log r(\vec x) + 
\a \log \sum_{\vec x_1,\cdots, \vec x_k} r(\vec x_1) \cdots r(\vec x_k) \EE(\vec x_1, \cdots, \vec x_k )  \right] \right\}
\eeq
\item
Make the following ansatz for $r(\vec x)$:
\beq
r(\vec x) = \frac{1-b}{2^\nu} + b \prod_{k=1}^n \left[ \frac12 \left( \d(\vec x_k = \vec 0) + \d(\vec x_k = \vec 1) \right) \right] \ ,
\eeq
where $\vec x_k = (x_{1+ m (k-1)}, \cdots, x_{mk})$ is the vector of the replicas in the $k$-th block.
Compute the overlap between two replicas; using spin notations, $S = (-1)^x$,
\beq
Q_{ab} = \la (-1)^{x_a} (-1)^{x_b} \ra = \sum_{\vec x} r(\vec x) (-1)^{x_a} (-1)^{x_b} \ .
\eeq
Show that it is equal to $b$ if the two replicas are in the same block, and zero otherwise.
Based on the results of section~\ref{sec:XORSAT}, give a justification of this ansatz and
interpret $b$ as the fraction of variables in the backbone.
\item Substitute this ansatz in (\ref{ZnXORSATrep}); show that
$\overline{\NN^{m n}} = \exp \{N \max_b S(m,n; b)\}$ with 
\beq\begin{split}
S(m,n ; b) & = - 2^n \left( \frac{b}{2^n} + \frac{1-b}{2^{m n}} \right) \log\left( \frac{b}{2^n} + \frac{1-b}{2^{m n}} \right)
- (2^{m n} - 2^n) \frac{1-b}{2^{m n}} \log \left( \frac{1-b}{2^{m n}} \right) \\
&+ \a \log \left( \frac{1}{2^{m n}} (1-b^k) + \frac{1}{2^n} b^k \right)
\end{split}\eeq
and deduce that (for the interesting case $m<1$):
\beq\begin{split}
\SS(m) &= \min_b \lim_{n\to 0} \partial_n S(m,n ; b) \\ & =
\log( 2 ) \min_b \left\{  b + m (1 - b) + b^k (m -1) \a - 
   m \a - (m -1) (1 -b ) \log(1 - b) \right\} \ .
\end{split}\eeq
\item Write the equation for $b$ and check that it does not depend on $m$.
Using Eq.~(\ref{fentropy}), deduce the expressions of $s(m)$ and $\Si(m)$. 
Note that $\SS(m)$ is linear in $m$, therefore for each value of $\a$, $s(m)$ and $\Si(m)$
do not depend on $m$.
Show that this gives back
Eq.(\ref{resultxorentro}) and (\ref{sentxor2}).
\end{itemize}

\item \label{ex:transfer}
{\bf The Bethe equations and the transfer matrix method}:
Adapt the reasoning of section~\ref{sec:treerec} to the case of a one dimensional chain with open
boundaries, and show that it is equivalent to the transfer matrix method.

\item \label{ex:II.2}
{\bf The Ising ferromagnet}: Consider the Ising model on the fixed connectivity
random lattice. The cavity probability can be parametrized as 
$\eta_{\rm cav}(S) = \frac{1 + m_c S}2$. Show that the recurrence equation (\ref{cavityeq})
in terms of the magnetization $m_c$ becomes
\beq
m_c = \tanh[(c-1) \text{atanh}[ m_c \tanh(\b J) ] ] \ .
\eeq
Show that there is a phase transition from a
paramagnetic to a ferromagnetic phase; compute the critical temperature $T_c$.
Express the real magnetization $m$ that enters in $P(S) = \frac{1+m S}2$ in terms of $m_c$;
compute the critical exponent $\b$ associated to $m \sim |T-T_c|^\b$.

Now, in order to mimic a finite-dimensional system of dimension $d$, choose $c=2d$ and 
$J = 1/(2d)$. Show that for $d\to\io$ one recovers the mean field equation of the fully-connected model and that $T_c=1$ in this limit.
Compute numerically $T_c$ from the cavity method as a function of $d$. Find in the literature 
data for the exact $T_c$ in dimension $d = 2, 3, 4, 5, 6, 7, 8$ (at least) and compare it with the cavity method result and from the fully-connected
result.

\item \label{ex:II.2bis}
{\bf A cavity derivation of the reaction term}: Consider the SK model with $N$ spins 
and think about it as a spin glass model on the fully connected graph.
\begin{itemize}
\item Show, using the same parametrization of Exercise~\ref{ex:II.2}, that the cavity equations can be written as
\beq\label{ex2bisA}
\begin{split}
m^{i\to j}_c &= \tanh\left[ \sum_{k \neq \{i, j \}} \text{atanh}[ m^{k\to i}_c \tanh(\b J_{ik}) ] \right] \ , \\
m_i &= \tanh\left[ \sum_{k \neq i} \text{atanh}[ m^{k\to i}_c \tanh(\b J_{ik}) ] \right] \ , \\
\end{split}\eeq
where $m_c^{i\to j}$ are the cavity magnetizations and $m_i$ are the true magnetizations.
\item
Recall that in the SK model $J_{ij} \sim N^{-1/2}$ and therefore the couplings are small. Using this,
show that at the leading order
\beq
m_c^{i\to j} - m_i \sim - m_j \b J_{ij} ( 1 - m_i^2 ) \ .
\eeq
\item Plug this result in the second Eq.~\eqref{ex2bisA} and show that
\beq
m_i = \tanh\left[ \sum_{k\neq i} \b J_{ik} m_k - m_i \sum_{k\neq i} (\b J_{ik})^2 (1-m_k^2) \right] \ .
\eeq
Show that this is exactly the same result obtained in Eq.~\eqref{TAP_SK} through the high temperature expansion.
\end{itemize}
Discuss why this derivation justifies the name ``reaction term'' that is used for the correction term.

\item \label{ex:II.3}
{\bf RS solution of $q$-COL}: 
Consider the $q$-coloring of fixed connectivity random graphs 
at zero temperature.
\begin{itemize}
\item 
Assume that the replica symmetric solution is uniform over all possible colors,
$\h_{\rm cav}(\s) = 1/q$; show that this is indeed a solution. Compute the entropy 
$s(c,q) = \log q + c/2 \log((q-1)/q)$; note that it vanishes for a given value of 
$c \sim q \log q$ for large $q$.
\item 
Consider solving the cavity equation by iteration, at each step 
using the right hand side to compute a new estimate to the solution until convergence.
Study the stability of the uniform solution under this process. In other words,
consider a small perturbation of the uniform solution, $\h_{\rm cav}(\s) = 
1/q + \d \h_{\rm cav}(\s)$, $\sum_\s \d \h_{\rm cav}(\s) = 0$.
Linearize the cavity
equation to obtain a linear equation for $\d \h_{\rm cav}(\s)$. Show that the perturbation
decays exponentially under iteration for $c < q$ while it grow exponentially for $c>q$.
Then the uniform solution is unstable for $c \geq q$. See \cite{col2} for an interpretation of
this instability.
\end{itemize}
\item \label{ex:II.4}
{\bf A consistency check}: Show that for $k=2$ the factor graph cavity equations
reduce to (\ref{cavityeq}) and the free energy reduces to (\ref{freecavRS}).
\item \label{ex:II.5}
{\bf XORSAT}: Consider the $3$-XORSAT problem with fixed or fluctuating connectivity,
at zero temperature in the SAT phase (``entropic'' cavity method).
Show that the uniform solution $\eta_{\rm cav}(S)=\eta_{\rm test}(S)=1/2$ is indeed a solution of the
iteration equations on a factor graph. Deduce that the zero-temperature
entropy is $s = (1-\a) \log 2$ as found using the leaf removal algorithm.
\item \label{ex:II.6}
{\bf Variational principle}: Show that differentiation of (\ref{freecavRS})
with respect to $\eta_{\rm cav}(\s)$ gives back (\ref{cavityeq}). Keep in mind that $\eta_{\rm cav}(\s)$
must be normalized! Repeat the calculation also in the factor graph case.

\item\label{ex:II.7}
{\bf Explicit solution of {\sc 1rsb} equations for XORSAT}:
Consider $k$-XORSAT on an Erd\"os-R\'enyi graph of mean connectivity $c = \a k$.
Variables are represented by Ising spins and the form of the constraint is 
$\psi_a(S_1,\cdots,S_k) = \d(J_a S_1 \cdots S_k = 1)$, where $J_a = \pm 1$ with uniform probability.

Solve the iteration equation as follows:
\begin{itemize}
\item For a given graph, variables may be in the backbone or not.
According to the analysis of section \ref{sec:XORSAT}, this depends only on the topological
structure of the graph.
Then the cavity fields $\h$ (for both cavity
spins and cavity tests) can be of three different types: $\h(S) = 1/2$, if the variable is not
in the backbone; $\h(S) = \d_{S,1}$ or $\h(S) = \d_{S,-1}$ if the variable is in the backbone.

\item If the variable is not in the backbone, then for all states $\a$ it is free; the distribution
\beq
Q[\h(S)] = \d\big[ \h(S) = 1/2 \big] \equiv \D_{1/2} \ .
\eeq
If the variable is in the backbone, then it is frozen to $\pm 1$ with uniform probability,
\beq
Q[\h(S)] = \frac12 \d\big[ \h(S) = \d_{S,1} \big]+\frac12 \d\big[ \h(S) = \d_{S,-1} \big] \equiv \D_{\pm1} \ .
\eeq
Show that the latter statement is a consequence of the symmetries of the problem.

\item Assume that the {\sc 1rsb} distribution over the sites of $Q[\h_{\rm cav}]$ has the form
\beq
\PP\big[ Q[\h_{\rm cav}] \big] = b \d\big[  Q[\h_{\rm cav}] = \D_{\pm1}   \big]+ (1-b) \d\big[  Q[\h_{\rm cav}] = \D_{1/2}   \big] \ ,
\eeq
where $b$ is the probability (over the sites) that a variable is in the backbone.

\item Plug the equation above into the second Eq.(\ref{PPeq1RSB-factor}); show that
\beq
\PP\big[ Q[\h_{\rm test}] \big] = b^{k-1} \d\big[  Q[\h_{\rm test}] = \D_{\pm1}   \big]+ 
(1-b^{k-1}) \d\big[  Q[\h_{\rm test}] = \D_{1/2}   \big] \ ,
\eeq

\item Now plug the latter expression in the first Eq.(\ref{PPeq1RSB-factor}); show that for a given $c$ one has
\beq
\PP\big[ Q[\h_{\rm cav}] \big] =(1-(1-b^{k-1})^{c-1}) 
\d\big[  Q[\h_{\rm cav}] = \D_{\pm1}   \big]+ (1-b^{k-1})^{c-1}  \d\big[  Q[\h_{\rm cav}] = \D_{1/2}   \big] \ .
\eeq
Show that the average over $\ell=c-1 \geq 0$ must be taken using a Poissonian of average $\a k$; for this follow the
the reasoning before Eq.~(\ref{wtPc}).
Take the average over $c$ and show that
one gets back
\beq
\PP\big[ Q[\h_{\rm cav}] \big] = (1-f(b)) \d\big[  Q[\h_{\rm cav}] = \D_{\pm1}   \big]+ f(b) \d\big[  Q[\h_{\rm cav}] = \D_{1/2}   \big] \ ,
\eeq
with $f(b) = e^{-\a k b^{k-1}}$. Then $b$ must satisfy the equation
\beq
1-b = f(b) = e^{-\a k b^{k-1}} \ ,
\eeq
that gives back Eq.~(\ref{backbone}) obtained with the leaf removal. Show that for $\a \leq \a_d$ only
the solution $b=0$ exist, and show that it gives back the {\sc RS} solution. 
For $\a > \a_d$, there is a solution $b^* \neq 0$.
Note that $b^*$ does not depend on $m$.

\end{itemize}

Compute the free entropy (\ref{fentropy}) from Eq.(\ref{phi1rsbfactor}) (just drop the $-T$ to get the free entropy,
and take the average of the different terms over $J,c,\PP$.
The result is
\beq\begin{split}
&\SS^{(l)} =  \log \ZZ^{(l)} = -(\log 2) \big[ b^k + m (1 - b^k) \big] \ , \\
&\SS^{(c)} =\log \ZZ^{(c)} = -(\log 2) \big[ b^k + m (1 - b^k) \big] \ , \\
&\SS^{(s)} =\log \ZZ^{(s)} = -(\log 2) \big[ m(\a k -1 + b - \a k b^{k-1}) + \a k b^{k-1} - b \big] \ .
\end{split}\eeq
Then $\SS(m) = - \lim_{T\to 0} \b \Phi(m,T)$ is linear in $m$ and one gets a single value of $\Si$ and $s$ for all $m$. This gives back
Eq.(\ref{resultxorentro}) and (\ref{sentxor2}).

\end{enumerate}

\clearpage

\section{Conclusions and perspectives}

The main message of these notes was that {\it mean field spin glasses are characterized by
the existence of many pure states}, among which some are stable equilibrium states
and others are metastable. We discussed different models (spherical $p$-spin, SK model,
optimization problems such as XORSAT and $q$-COL) that share this feature, and some
methods (the replica and cavity method, and in the case of XORSAT the rigorous leaf-removal
method) to compute the complexity and other properties of these states.

This particular property of spin glasses (that is {\it not} shared by all disordered systems)
raises many intriguing questions. For instance, what is the influence of the presence of
these states on the dynamics? This question is also relevant for the analysis of search algorithms in optimization,
that can be regarded as (non-equilibrium) dynamics for the corresponding physical system.

And what about non mean-field models? Are so many states present also in finite
dimensional models? And how can they be described?

These questions did not receive a complete and satisfactory answer at present and are active research
topics. Still, many important
advances have been made. Here it follows a list of references that may be consulted to go deeper into
these fascinating problems. The list is very incomplete and strongly biased towards mean-field inspired 
work; it is intended only to stimulate the curiosity, and
the reader is strongly encouraged to look for further references. The articles cited below have also been chosen
because they are useful sources of more references on the same subject.
\begin{enumerate}
\item The {\it equilibrium} dynamics of the spherical $p$-spin model can be completely solved in the paramagnetic
phase, and it can be shown that a {\it dynamical transition} takes place at the temperature $T_d$ where
states first appear. The transition is charaterized by the divergence of the relaxation time. These
results are reviewed in \cite{BCKM97,Cu02,CC05}.
The same can be proven for the equilibrium 
dynamics of XORSAT, see \cite{MoSe,MoSe2}.
\item In optimization problems one wants to find the ground state of the system. The simplest way to
do that is to consider a dynamics satisfying detailed balance at temperature $T_i$, and then reduce
the temperature down to $T=0$ (or a low temperature $T=T_f$) at a given rate $\g$ 
({\it classical annealing}). In the limit $\g\to\io$ one just istantaneously
{\it quenches} the system from $T_i$ to $T_f$. For the spherical $p$-spin one can show that if $T_f < T_d$,
the system falls out of equilibrium and starts to {\it age}; its energy decreases but approaches
the energy of the threshold states. Therefore classical annealing cannot find the ground state for
this system, as it is always trapped by higher energy metastable states. The {\it aging} dynamics of
this and many other models is reviewed in \cite{Cu02}; for the specific case of the
SK model see \cite{KC08}. In some cases it is not at all obvious
to understand which states dominate the aging dynamics, see \cite{MR04} and in particular \cite{ZK10} 
for a very detailed discussion of this point using the cavity method.
\item More generally, one can consider dynamics that do not satisfy detailed balance. In the case
of optimization problems, many algorithms designed to search for solutions falls in this class.
These algorithms are known to undergo {\it algorithmic transitions}: the probability (over formulas
and randomness built in the algorithm) to find a solution decreases abruptly from 1 to 0 ($N\to\io$)
when the density of constraints is increased over a value $\a_a$. Is $\a_a$ related to some property
of the equilibrium states (their existence, the presence of frozen variables,~$\cdots$)?
This is mainly an open problem. See \cite{AMSZ09}
for a review of algorithms that have been studied with methods borrowed from physics, and
\cite{KK07} for an original perspective on the general
connection between (free)energy landscapes and algorithms.
\item The {\sc 1rsb} spin glass transition has been conjectured to describe the glass transition in
finite-dimensional particle systems. This is based on the following observations:
\begin{itemize}
\item The equations that describe the equilibrium dynamics of the spherical $p$-spin closely resemble
the {\it Mode-Coupling equations} that describe liquids close to the glass transition. See \eg~\cite{Cu02}.
\item The existence of an exponential number of states between $T_d$ and $T_K$ at the mean field level
is impossible in finite-dimensional systems. Therefore one has to re-discuss the definition of these
states. This leads to very important ideas, such as {\it entropic-driven nucleation}, that might explain
the dynamics of liquids at temperatures below the Mode-Coupling regime.
A pedagogical discussion of the definition of states in finite dimension can be found in~\cite{BB04}
and \cite{MP00}, as well as in Appendix A of \cite{PZ10}.
A very successful 
theory of glasses based on the adaptation of the mean-field scenario has been developed by Wolynes and goes under
the name of {\it Random First Order Theory}; a review is \cite{LW07}, see also \cite{BB09,Ca09}.
Quantitative replica calculations of the thermodynamics are reviewed in~\cite{MP99,PZ10}.
\item A detailed investigation of the dynamics of liquids close to the glass transition revealed the existence
of {\it dynamical heterogeneities}, 
namely of regions of the sample that are more mobile than others.
This defines a {\it dynamical correlation length}, related to the typical size of the heterogeneities, that
seems to diverge at the glass transition.
Different theories account for heterogeneity; in particular mean-field like equations predict the existence
of such a correlation length and its divergence at $T_d$~\cite{FP00,BBBKMR07a,BBBKMR07b}.
\item The aging dynamics of glasses is very similar to the mean-field one \cite{BCKM97,Cu02}.
\item This mean-field-like scenario has been derived also for Kac versions of {\sc 1rsb} spin glasses in finite
dimension; see in particular~\cite{Fr05,Fr07,FM07}.
\end{itemize}
It is important to keep in mind that, as the transition is first-order in mean-field, its understanding in 
finite dimension is mostly related to nucleation phenomena.
\item On the contrary, the f{\sc RSB} transition of the SK model is a true second-order critical point.
A natural question is whether finite-dimensional spin glass models on cubic lattices with nearest-neighbor
two body interactions, like the Edward-Anderson model, also undergo a f{\sc rsb} spin glass transition. 
This is very much debated; a classical alternative picture is presented in \cite{FH88}.
Recent reviews of the status of the mean-field approach are in \cite{MPRRZ00,Pa08,Pa07c}.
In this case one would like to apply to the problem the whole machinery of standard second order phase
transition, like scaling, renormalization group, computation of upper/lower critical dimensions, etc.
Unfortunately, this is very difficult, see \eg~\cite{DG06}, and for the
moment most of the results come from numerical simulations.
\item It is very important to understand how glassy systems respond to
external drives. For instance, one can shear a liquid close to the glass transition, or consider
a granular driven by an external tapping, etc. This situations are often met in experiments and in
many practical applications. A seminal paper in this respect is \cite{BBK00},
where it was shown that the structure of states of the spherical $p$-spin 
model gives rise to a complex behavior of the system when subject to an external drive. This led
to the prediction of a ``complex rheology'' in liquids close to the glass transition and colloidal
systems \cite{BB02} that is able to explain most of the
phenomenology of these materials subject to external drives.
\item Finally, the role of quantum fluctuations for the glass transition has to be elucidated.
Is it possible to have many pure states in a quantum system? Quantum versions of the spherical
$p$-spin model have been solved in \cite{CGSC01} using the
replica method, and quantum TAP equations have been discussed in \cite{BC01}. 
It was shown that the spin glass transition
becomes first order at low temperature (and down to $T=0$) as a function of the quantum fluctuations
parameter (\eg a transverse field). However, much less is known for non-mean field models; already
on random graphs, the development of a quantum version of the cavity method is a very recent 
achievement~\cite{LSS08,KRSZ08}.
\end{enumerate}

\bibliographystyle{mioaps}
\bibliography{FR,HS,SG,SAT}

\end{document}